%
%
%
\def\unredoffs{} \def\redoffs{\voffset=-.31truein\hoffset=-.48truein}
\def\speclscape{}
%
%
%
%
%
\newbox\leftpage \newdimen\fullhsize \newdimen\hstitle \newdimen\hsbody
\tolerance=1000\hfuzz=2pt
\catcode`\@=11 
\ifx\hyperdef\UNd@FiNeD\def\hyperdef#1#2#3#4{#4}\def\hyperref#1#2#3#4{#4}\fi
\def\bigans{b }
\def\answ{b }
%
\ifx\answ\bigans\message{(This will come out unreduced.}
\magnification=1200\unredoffs\baselineskip=16pt plus 2pt minus 1pt
\hsbody=\hsize \hstitle=\hsize 
\else\message{(This will be reduced.} \let\l@r=L
\magnification=1000\baselineskip=16pt plus 2pt minus 1pt \vsize=7truein
\redoffs \hstitle=8truein\hsbody=4.75truein\fullhsize=10truein\hsize=\hsbody
\output={\ifnum\pageno=0 
  \shipout\vbox{\speclscape{\hsize\fullhsize\makeheadline}
    \hbox to \fullhsize{\hfill\pagebody\hfill}}\advancepageno
  \else
  \almostshipout{\leftline{\vbox{\pagebody\makefootline}}}\advancepageno
  \fi}
\def\almostshipout#1{\if L\l@r \count1=1 \message{[\the\count0.\the\count1]}
      \global\setbox\leftpage=#1 \global\let\l@r=R
 \else \count1=2
  \shipout\vbox{\speclscape{\hsize\fullhsize\makeheadline}
      \hbox to\fullhsize{\box\leftpage\hfil#1}}  \global\let\l@r=L\fi}
\fi
%
\newcount\yearltd\yearltd=\year\advance\yearltd by -1900

\def\Title#1#2{\nopagenumbers\abstractfont\hsize=\hstitle\rightline{#1}%
\vskip 1in\centerline{\titlefont #2}\abstractfont\vskip .5in\pageno=0}
\def\Date#1{\vfill\leftline{#1}\tenpoint\supereject\global\hsize=\hsbody%
\footline={\hss\tenrm\hyperdef\hypernoname{page}\folio\folio\hss}}%
%

\def\draftmode{\message{ DRAFTMODE }\def\draftdate{{\rm preliminary draft:
\number\month/\number\day/\number\yearltd\ \ \hourmin}}%
\headline={\hfil\draftdate}\writelabels\baselineskip=20pt plus 2pt minus 2pt
 {\count255=\time\divide\count255 by 60 \xdef\hourmin{\number\count255}
  \multiply\count255 by-60\advance\count255 by\time
  \xdef\hourmin{\hourmin:\ifnum\count255<10 0\fi\the\count255}}}
\def\nolabels{\def\wrlabeL##1{}\def\eqlabeL##1{}\def\reflabeL##1{}}
\def\writelabels{\def\wrlabeL##1{\leavevmode\vadjust{\rlap{\smash%
{\line{{\escapechar=` \hfill\rlap{\sevenrm\hskip.03in\string##1}}}}}}}%
\def\eqlabeL##1{{\escapechar-1\rlap{\sevenrm\hskip.05in\string##1}}}%
\def\reflabeL##1{\noexpand\llap{\noexpand\sevenrm\string\string\string##1}}}
\nolabels
%
\global\newcount\secno \global\secno=0
\global\newcount\meqno \global\meqno=1
\def\s@csym{}
\def\newsec#1{\global\advance\secno by1%
{\toks0{#1}\message{(\the\secno. \the\toks0)}}%
\global\subsecno=0\eqnres@t\let\s@csym\secsym\xdef\secn@m{\the\secno}\noindent
{\bf\hyperdef\hypernoname{section}{\the\secno}{\the\secno.} #1}%
\writetoca{{\string\hyperref{}{section}{\the\secno}{\the\secno.}} {#1}}%
\par\nobreak\medskip\nobreak}
\def\eqnres@t{\xdef\secsym{\the\secno.}\global\meqno=1\bigbreak\bigskip}
\def\sequentialequations{\def\eqnres@t{\bigbreak}}\xdef\secsym{}
\global\newcount\subsecno \global\subsecno=0
\def\subsec#1{\global\advance\subsecno by1%
{\toks0{#1}\message{(\s@csym\the\subsecno. \the\toks0)}}%
\ifnum\lastpenalty>9000\else\bigbreak\fi
\noindent{\it\hyperdef\hypernoname{subsection}{\secn@m.\the\subsecno}%
{\secn@m.\the\subsecno.} #1}\writetoca{\string\quad
{\string\hyperref{}{subsection}{\secn@m.\the\subsecno}{\secn@m.\the\subsecno.}}
{#1}}\par\nobreak\medskip\nobreak}
\def\appendix#1#2{\global\meqno=1\global\subsecno=0\xdef\secsym{\hbox{#1.}}%
\bigbreak\bigskip\noindent{\bf Appendix \hyperdef\hypernoname{appendix}{#1}%
{#1.} #2}{\toks0{(#1. #2)}\message{\the\toks0}}%
\xdef\s@csym{#1.}\xdef\secn@m{#1}%
\writetoca{\string\hyperref{}{appendix}{#1}{Appendix {#1.}} {#2}}%
\par\nobreak\medskip\nobreak}
%
%
\def\checkm@de#1#2{\ifmmode{\def\f@rst##1{##1}\hyperdef\hypernoname{equation}%
{#1}{#2}}\else\hyperref{}{equation}{#1}{#2}\fi}
\def\eqnn#1{\DefWarn#1\xdef #1{(\noexpand\relax\noexpand\checkm@de%
{\s@csym\the\meqno}{\secsym\the\meqno})}%
\wrlabeL#1\writedef{#1\leftbracket#1}\global\advance\meqno by1}
\def\f@rst#1{\c@t#1a\em@ark}\def\c@t#1#2\em@ark{#1}
\def\eqna#1{\DefWarn#1\wrlabeL{#1$\{\}$}%
\xdef #1##1{(\noexpand\relax\noexpand\checkm@de%
{\s@csym\the\meqno\noexpand\f@rst{##1}}{\hbox{$\secsym\the\meqno##1$}})}
\writedef{#1\numbersign1\leftbracket#1{\numbersign1}}\global\advance\meqno by1}
\def\eqn#1#2{\DefWarn#1%
\xdef #1{(\noexpand\hyperref{}{equation}{\s@csym\the\meqno}%
{\secsym\the\meqno})}$$#2\eqno(\hyperdef\hypernoname{equation}%
{\s@csym\the\meqno}{\secsym\the\meqno})\eqlabeL#1$$%
\writedef{#1\leftbracket#1}\global\advance\meqno by1}
\def\xeqn{\expandafter\xe@n}\def\xe@n(#1){#1}
\def\xeqna#1{\expandafter\xe@n#1}
\def\eqns#1{(\e@ns #1{\hbox{}})}
\def\e@ns#1{\ifx\UNd@FiNeD#1\message{eqnlabel \string#1 is undefined.}%
\xdef#1{(?.?)}\fi{\let\hyperref=\relax\xdef\next{#1}}%
\ifx\next\em@rk\def\next{}\else%
\ifx\next#1\xeqn#1\else\def\n@xt{#1}\ifx\n@xt\next#1\else\xeqna#1\fi
\fi\let\next=\e@ns\fi\next}

\def\DefWarn#1{\ifx\UNd@FiNeD#1\else
\immediate\write16{*** WARNING: the label \string#1 is already defined ***}\fi}
%
\newskip\footskip\footskip14pt plus 1pt minus 1pt 
\def\footnotefont{\ninepoint}\def\f@t#1{\footnotefont #1\@foot}
\def\f@@t{\baselineskip\footskip\bgroup\footnotefont\aftergroup\@foot\let\next}
\setbox\strutbox=\hbox{\vrule height9.5pt depth4.5pt width0pt}
\global\newcount\ftno \global\ftno=0
\def\foot{\global\advance\ftno by1\def\foot@rg{\hyperref{}{footnote}%
{\the\ftno}{\the\ftno}\xdef\foot@rg{\noexpand\hyperdef\noexpand\hypernoname%
{footnote}{\the\ftno}{\the\ftno}}}\footnote{$^{\foot@rg}$}}
%
\newwrite\ftfile
\def\footend{\def\foot{\global\advance\ftno by1\chardef\wfile=\ftfile
\hyperref{}{footnote}{\the\ftno}{$^{\the\ftno}$}%
\ifnum\ftno=1\immediate\openout\ftfile=\jobname.fts\fi%
\immediate\write\ftfile{\noexpand\smallskip%
\noexpand\item{\noexpand\hyperdef\noexpand\hypernoname{footnote}
{\the\ftno}{f\the\ftno}:\ }\pctsign}\findarg}%
\def\footatend{\vfill\eject\immediate\closeout\ftfile{\parindent=20pt
\centerline{\bf Footnotes}\nobreak\bigskip\input \jobname.fts }}}
\def\footatend{}
%
%
\global\newcount\refno \global\refno=1
\newwrite\rfile
\def\ref{[\hyperref{}{reference}{\the\refno}{\the\refno}]\nref}
\def\nref#1{\DefWarn#1%
\xdef#1{[\noexpand\hyperref{}{reference}{\the\refno}{\the\refno}]}%
\writedef{#1\leftbracket#1}%
\ifnum\refno=1\immediate\openout\rfile=\jobname.refs\fi
\chardef\wfile=\rfile\immediate\write\rfile{\noexpand\item{[\noexpand\hyperdef%
\noexpand\hypernoname{reference}{\the\refno}{\the\refno}]\ }%
\reflabeL{#1\hskip.31in}\pctsign}\global\advance\refno by1\findarg}
\def\findarg#1#{\begingroup\obeylines\newlinechar=`\^^M\pass@rg}
{\obeylines\gdef\pass@rg#1{\writ@line\relax #1^^M\hbox{}^^M}%
\gdef\writ@line#1^^M{\expandafter\toks0\expandafter{\striprel@x #1}%
\edef\next{\the\toks0}\ifx\next\em@rk\let\next=\endgroup\else\ifx\next\empty%
\else\immediate\write\wfile{\the\toks0}\fi\let\next=\writ@line\fi\next\relax}}
\def\striprel@x#1{} \def\em@rk{\hbox{}}
\def\lref{\begingroup\obeylines\lr@f}
\def\lr@f#1#2{\DefWarn#1\gdef#1{\let#1=\UNd@FiNeD\ref#1{#2}}\endgroup\unskip}

\def\addref#1{\immediate\write\rfile{\noexpand\item{}#1}} 
\def\listrefs{\footatend\vfill\supereject\immediate\closeout\rfile\writestoppt
\baselineskip=\footskip\centerline{{\bf References}}\bigskip{\parindent=20pt%
\frenchspacing\escapechar=` \input \jobname.refs\vfill\eject}\nonfrenchspacing}
\def\startrefs#1{\immediate\openout\rfile=\jobname.refs\refno=#1}
\def\xref{\expandafter\xr@f}\def\xr@f[#1]{#1}
\def\refs#1{\count255=1[\r@fs #1{\hbox{}}]}
\def\r@fs#1{\ifx\UNd@FiNeD#1\message{reflabel \string#1 is undefined.}%
\nref#1{need to supply reference \string#1.}\fi%
\vphantom{\hphantom{#1}}{\let\hyperref=\relax\xdef\next{#1}}%
\ifx\next\em@rk\def\next{}%
\else\ifx\next#1\ifodd\count255\relax\xref#1\count255=0\fi%
\else#1\count255=1\fi\let\next=\r@fs\fi\next}
%

%
\newwrite\ffile\global\newcount\figno \global\figno=1
\def\fig{fig.~\hyperref{}{figure}{\the\figno}{\the\figno}\nfig}
\def\nfig#1{\DefWarn#1%
\xdef#1{fig.~\noexpand\hyperref{}{figure}{\the\figno}{\the\figno}}%
\writedef{#1\leftbracket fig.\noexpand~\xfig#1}%
\ifnum\figno=1\immediate\openout\ffile=\jobname.figs\fi\chardef\wfile=\ffile%
{\let\hyperref=\relax
\immediate\write\ffile{\noexpand\medskip\noexpand\item{Fig.\ %
\noexpand\hyperdef\noexpand\hypernoname{figure}{\the\figno}{\the\figno}. }
\reflabeL{#1\hskip.55in}\pctsign}}\global\advance\figno by1\findarg}
\def\listfigs{\vfill\eject\immediate\closeout\ffile{\parindent40pt
\baselineskip14pt\centerline{{\bf Figure Captions}}\nobreak\medskip
\escapechar=` \input \jobname.figs\vfill\eject}}
\def\xfig{\expandafter\xf@g}\def\xf@g fig.\penalty\@M\ {}
\def\figs#1{figs.~\f@gs #1{\hbox{}}}
\def\f@gs#1{{\let\hyperref=\relax\xdef\next{#1}}\ifx\next\em@rk\def\next{}\else
\ifx\next#1\xfig #1\else#1\fi\let\next=\f@gs\fi\next}
\def\figin{\epsfcheck\figin}\def\figins{\epsfcheck\figins}
\def\epsfcheck{\ifx\epsfbox\UNd@FiNeD
\message{(NO epsf.tex, FIGURES WILL BE IGNORED)}
\gdef\figin##1{\vskip2in}\gdef\figins##1{\hskip.5in}
\else\message{(FIGURES WILL BE INCLUDED)}%
\gdef\figin##1{##1}\gdef\figins##1{##1}\fi}
\def\DefWarn#1{}
\def\figinsert{\goodbreak\midinsert}
\def\ifig#1#2#3{\DefWarn#1\xdef#1{fig.~\noexpand\hyperref{}{figure}%
{\the\figno}{\the\figno}}\writedef{#1\leftbracket fig.\noexpand~\xfig#1}%
\figinsert\figin{\centerline{#3}}\medskip\centerline{\vbox{\baselineskip12pt
\advance\hsize by -1truein\noindent\wrlabeL{#1=#1}\footnotefont%
{\bf Fig.~\hyperdef\hypernoname{figure}{\the\figno}{\the\figno}:} #2}}
\bigskip\endinsert\global\advance\figno by1}
\newwrite\lfile
{\escapechar-1\xdef\pctsign{\string\%}\xdef\leftbracket{\string\{}
\xdef\rightbracket{\string\}}\xdef\numbersign{\string\#}}
\def\writedefs{\immediate\openout\lfile=\jobname.defs \def\writedef##1{%
{\let\hyperref=\relax\let\hyperdef=\relax\let\hypernoname=\relax
 \immediate\write\lfile{\string\def\string##1\rightbracket}}}}%
\def\writestop{\def\writestoppt{\immediate\write\lfile{\string\pageno
 \the\pageno\string\startrefs\leftbracket\the\refno\rightbracket
 \string\def\string\secsym\leftbracket\secsym\rightbracket
 \string\secno\the\secno\string\meqno\the\meqno}\immediate\closeout\lfile}}
\def\writestoppt{}\def\writedef#1{}
\def\seclab#1{\DefWarn#1%
\xdef #1{\noexpand\hyperref{}{section}{\the\secno}{\the\secno}}%
\writedef{#1\leftbracket#1}\wrlabeL{#1=#1}}
\def\subseclab#1{\DefWarn#1%
\xdef #1{\noexpand\hyperref{}{subsection}{\secn@m.\the\subsecno}%
{\secn@m.\the\subsecno}}\writedef{#1\leftbracket#1}\wrlabeL{#1=#1}}
\def\applab#1{\DefWarn#1%
\xdef #1{\noexpand\hyperref{}{appendix}{\secn@m}{\secn@m}}%
\writedef{#1\leftbracket#1}\wrlabeL{#1=#1}}
\newwrite\tfile \def\writetoca#1{}
\def\leaderfill{\leaders\hbox to 1em{\hss.\hss}\hfill}
\def\writetoc{\immediate\openout\tfile=\jobname.toc
   \def\writetoca##1{{\edef\next{\write\tfile{\noindent ##1
   \string\leaderfill {\string\hyperref{}{page}{\noexpand\number\pageno}%
                       {\noexpand\number\pageno}} \par}}\next}}}
\newread\ch@ckfile
\def\listtoc{\immediate\closeout\tfile\immediate\openin\ch@ckfile=\jobname.toc
\ifeof\ch@ckfile\message{no file \jobname.toc, no table of contents this pass}%
\else\closein\ch@ckfile\centerline{\bf Contents}\nobreak\medskip%
{\baselineskip=12pt\footnotefont\parskip=0pt\catcode`\@=11\input\jobname.toc
\catcode`\@=12\bigbreak\bigskip}\fi}
\catcode`\@=12 
%
\edef\tfontsize{\ifx\answ\bigans scaled\magstep3\else scaled\magstep4\fi}
\font\titlerm=cmr10 \tfontsize \font\titlerms=cmr7 \tfontsize
\font\titlermss=cmr5 \tfontsize \font\titlei=cmmi10 \tfontsize
\font\titleis=cmmi7 \tfontsize \font\titleiss=cmmi5 \tfontsize
\font\titlesy=cmsy10 \tfontsize \font\titlesys=cmsy7 \tfontsize
\font\titlesyss=cmsy5 \tfontsize \font\titleit=cmti10 \tfontsize
\skewchar\titlei='177 \skewchar\titleis='177 \skewchar\titleiss='177
\skewchar\titlesy='60 \skewchar\titlesys='60 \skewchar\titlesyss='60
\def\titlefont{\def\rm{\fam0\titlerm}
\textfont0=\titlerm \scriptfont0=\titlerms \scriptscriptfont0=\titlermss
\textfont1=\titlei \scriptfont1=\titleis \scriptscriptfont1=\titleiss
\textfont2=\titlesy \scriptfont2=\titlesys \scriptscriptfont2=\titlesyss
\textfont\itfam=\titleit \def\it{\fam\itfam\titleit}\rm}
 \ifx\answ\bigans\else scaled\magstep1\fi
\ifx\answ\bigans\def\abstractfont{\tenpoint}\else
\font\absit=cmti10 scaled \magstep1
\font\abssl=cmsl10 scaled \magstep1
\font\absrm=cmr10 scaled\magstep1 \font\absrms=cmr7 scaled\magstep1
\font\absrmss=cmr5 scaled\magstep1 \font\absi=cmmi10 scaled\magstep1
\font\absis=cmmi7 scaled\magstep1 \font\absiss=cmmi5 scaled\magstep1
\font\abssy=cmsy10 scaled\magstep1 \font\abssys=cmsy7 scaled\magstep1
\font\abssyss=cmsy5 scaled\magstep1 \font\absbf=cmbx10 scaled\magstep1
\skewchar\absi='177 \skewchar\absis='177 \skewchar\absiss='177
\skewchar\abssy='60 \skewchar\abssys='60 \skewchar\abssyss='60
\def\abstractfont{\def\rm{\fam0\absrm}
\textfont0=\absrm \scriptfont0=\absrms \scriptscriptfont0=\absrmss
\textfont1=\absi \scriptfont1=\absis \scriptscriptfont1=\absiss
\textfont2=\abssy \scriptfont2=\abssys \scriptscriptfont2=\abssyss
\textfont\itfam=\absit \def\it{\fam\itfam\absit}\def\footnotefont{\tenpoint}%
\textfont\slfam=\abssl \def\sl{\fam\slfam\abssl}%
\textfont\bffam=\absbf \def\bf{\fam\bffam\absbf}\rm}\fi
\def\tenpoint{\def\rm{\fam0\tenrm}
\textfont0=\tenrm \scriptfont0=\sevenrm \scriptscriptfont0=\fiverm
\textfont1=\teni  \scriptfont1=\seveni  \scriptscriptfont1=\fivei
\textfont2=\tensy \scriptfont2=\sevensy \scriptscriptfont2=\fivesy
\textfont\itfam=\tenit \def\it{\fam\itfam\tenit}\def\footnotefont{\ninepoint}%
\textfont\bffam=\tenbf \def\bf{\fam\bffam\tenbf}\def\sl{\fam\slfam\tensl}\rm}
\font\ninerm=cmr9 \font\sixrm=cmr6 \font\ninei=cmmi9 \font\sixi=cmmi6
\font\ninesy=cmsy9 \font\sixsy=cmsy6 \font\ninebf=cmbx9
\font\nineit=cmti9 \font\ninesl=cmsl9 \skewchar\ninei='177
\skewchar\sixi='177 \skewchar\ninesy='60 \skewchar\sixsy='60
\def\ninepoint{\def\rm{\fam0\ninerm}
\textfont0=\ninerm \scriptfont0=\sixrm \scriptscriptfont0=\fiverm
\textfont1=\ninei \scriptfont1=\sixi \scriptscriptfont1=\fivei
\textfont2=\ninesy \scriptfont2=\sixsy \scriptscriptfont2=\fivesy
\textfont\itfam=\ninei \def\it{\fam\itfam\nineit}\def\sl{\fam\slfam\ninesl}%
\textfont\bffam=\ninebf \def\bf{\fam\bffam\ninebf}\rm}
%
%
\def\noblackbox{\overfullrule=0pt}
\hyphenation{anom-aly anom-alies coun-ter-term coun-ter-terms}
\def\inv{^{\raise.15ex\hbox{${\scriptscriptstyle -}$}\kern-.05em 1}}

\def\Dsl{\,\raise.15ex\hbox{/}\mkern-13.5mu D} 
\def\dsl{\raise.15ex\hbox{/}\kern-.57em\partial}

 \def\Tr{{\rm Tr}}
\def\lspace{\ifx\answ\bigans{}\else\qquad\fi}
\def\lbspace{\ifx\answ\bigans{}\else\hskip-.2in\fi} 
\def\boxeqn#1{\vcenter{\vbox{\hrule\hbox{\vrule\kern3pt\vbox{\kern3pt
	\hbox{${\displaystyle #1}$}\kern3pt}\kern3pt\vrule}\hrule}}}
\def\mbox#1#2{\vcenter{\hrule \hbox{\vrule height#2in
		\kern#1in \vrule} \hrule}}  
%
 \def\CO{{\cal O}} 
\def\CA{{\cal A}} \def\CC{{\cal C}}  
\def\CL{{\cal L}} \def\CH{{\cal H}}  
   \def\CT{{\cal T}}

\def\darr#1{\raise1.5ex\hbox{$\leftrightarrow$}\mkern-16.5mu #1}

\def\half{{\textstyle{1\over2}}} 
\def\roughly#1{\raise.3ex\hbox{$#1$\kern-.75em\lower1ex\hbox{$\sim$}}}


\lref\SonXQA{
  D.~T.~Son,
  ``Is the Composite Fermion a Dirac Particle?,''
Phys.\ Rev.\ X {\bf 5}, no. 3, 031027 (2015).
[arXiv:1502.03446 [cond-mat.mes-hall]].
}

\lref\FradkinTT{
  E.~H.~Fradkin and F.~A.~Schaposnik,
  ``The Fermion - boson mapping in three-dimensional quantum field theory,''
Phys.\ Lett.\ B {\bf 338}, 253 (1994).
[hep-th/9407182].
}

\lref\PolyakovMD{
  A.~M.~Polyakov,
  ``Fermi-Bose Transmutations Induced by Gauge Fields,''
Mod.\ Phys.\ Lett.\ A {\bf 3}, 325 (1988).
}

\lref\ShajiIS{
  N.~Shaji, R.~Shankar and M.~Sivakumar,
  ``On Bose-fermi Equivalence in a U(1) Gauge Theory With {Chern-Simons} Action,''
Mod.\ Phys.\ Lett.\ A {\bf 5}, 593 (1990).
}

\lref\WangQMT{
  C.~Wang and T.~Senthil,
  ``Dual Dirac Liquid on the Surface of the Electron Topological Insulator,''
Phys.\ Rev.\ X {\bf 5}, no. 4, 041031 (2015). [arXiv:1505.05141 [cond-mat.str-el]].
}

\lref\MetlitskiEKA{
  M.~A.~Metlitski and A.~Vishwanath,
  ``Particle-vortex duality of 2d Dirac fermion from electric-magnetic duality of 3d topological insulators,''
[arXiv:1505.05142 [cond-mat.str-el]].
}
\lref\GurAriXFF{
  G.~Gur-Ari, S.~A.~Hartnoll and R.~Mahajan,
  ``Transport in Chern-Simons-Matter Theories,''
JHEP {\bf 1607}, 090 (2016).
[arXiv:1605.01122 [hep-th]].
}

\lref\FidkowskiJUA{
  L.~Fidkowski, X.~Chen and A.~Vishwanath,
  ``Non-Abelian Topological Order on the Surface of a 3D Topological Superconductor from an Exactly Solved Model,''
Phys.\ Rev.\ X {\bf 3}, no. 4, 041016 (2013).
[arXiv:1305.5851 [cond-mat.str-el]].
}

\lref\WangFQL{
  C.~Wang and T.~Senthil,
  ``Half-filled Landau level, topological insulator surfaces, and three-dimensional quantum spin liquids,''
Phys.\ Rev.\ B {\bf 93}, no. 8, 085110 (2016). [arXiv:1507.08290 [cond-mat.st-el]].
}

\lref\GeraedtsPVA{
  S.~D.~Geraedts, M.~P.~Zaletel, R.~S.~K.~Mong, M.~A.~Metlitski, A.~Vishwanath and O.~I.~Motrunich,
  ``The half-filled Landau level: the case for Dirac composite fermions,''
Science {\bf 352}, 197 (2016).
[arXiv:1508.04140 [cond-mat.str-el]].
}

\lref\MrossIDY{
  D.~F.~Mross, J.~Alicea and O.~I.~Motrunich,
  ``Explicit derivation of duality between a free Dirac cone and quantum electrodynamics in (2+1) dimensions,''
[arXiv:1510.08455 [cond-mat.str-el]].
}

\lref\FradkinTT{
  E.~H.~Fradkin and F.~A.~Schaposnik,
  ``The Fermion - boson mapping in three-dimensional quantum field theory,''
Phys.\ Lett.\ B {\bf 338}, 253 (1994).
[hep-th/9407182].
}

\lref\FradkinWY{
  E.~Fradkin and A.~Lopez,
  ``Fractional Quantum Hall effect and Chern-Simons gauge theories,''
Phys.\ Rev.\ B {\bf 44}, 5246 (1991).
}

\lref\RoscherWOX{
  D.~Roscher, E.~Torres and P.~Strack,
  ``Dual QED$_3$ at "$N_F = 1/2$" is an interacting CFT in the infrared,''
[arXiv:1605.05347 [cond-mat.str-el]].
}

\lref\WangCTO{
  C.~Wang and T.~Senthil,
  ``Time-Reversal Symmetric $U(1)$ Quantum Spin Liquids,''
Phys.\ Rev.\ X {\bf 6}, no. 1, 011034 (2016).
}

\lref\IntriligatorLCA{
  K.~Intriligator and N.~Seiberg,
  ``Aspects of 3d N=2 Chern-Simons-Matter Theories,''
JHEP {\bf 1307}, 079 (2013).
[arXiv:1305.1633 [hep-th]].
}

\lref\MetlitskiYQA{
  M.~A.~Metlitski,
  ``$S$-duality of $u(1)$ gauge theory with $\theta =\pi$ on non-orientable manifolds: Applications to topological insulators and superconductors,''
[arXiv:1510.05663 [hep-th]].
}

\lref\MulliganGLM{
  M.~Mulligan, S.~Raghu and M.~P.~A.~Fisher,
  ``Emergent particle-hole symmetry in the half-filled Landau level,''
[arXiv:1603.05656 [cond-mat.str-el]].
}

\lref\WangGQJ{
  C.~Wang and T.~Senthil,
  ``Composite fermi liquids in the lowest Landau level,''
[arXiv:1604.06807 [cond-mat.str-el]].
}

\lref\WilczekDU{
  F.~Wilczek,
  ``Magnetic Flux, Angular Momentum, and Statistics,''
Phys.\ Rev.\ Lett.\  {\bf 48}, 1144 (1982).
}
\lref\JainTX{
  J.~K.~Jain,
  ``Composite fermion approach for the fractional quantum Hall effect,''
Phys.\ Rev.\ Lett.\  {\bf 63}, 199 (1989).
}

\lref\BorokhovCG{
  V.~Borokhov, A.~Kapustin and X.~k.~Wu,
  ``Monopole operators and mirror symmetry in three-dimensions,''
JHEP {\bf 0212}, 044 (2002).
[hep-th/0207074].
}

\lref\AharonyBX{
  O.~Aharony, A.~Hanany, K.~A.~Intriligator, N.~Seiberg and M.~J.~Strassler,
  ``Aspects of N=2 supersymmetric gauge theories in three-dimensions,''
Nucl.\ Phys.\ B {\bf 499}, 67 (1997).
[hep-th/9703110].
}

\lref\PeskinKP{
  M.~E.~Peskin,
  ``Mandelstam 't Hooft Duality in Abelian Lattice Models,''
Annals Phys.\  {\bf 113}, 122 (1978).
}
\lref\DasguptaZZ{
  C.~Dasgupta and B.~I.~Halperin,
  ``Phase Transition in a Lattice Model of Superconductivity,''
Phys.\ Rev.\ Lett.\  {\bf 47}, 1556 (1981).
}

\lref\JainGZA{
  S.~Jain, S.~Minwalla and S.~Yokoyama,
  ``Chern Simons duality with a fundamental boson and fermion,''
JHEP {\bf 1311}, 037 (2013).
[arXiv:1305.7235 [hep-th]].
}

\lref\ClossetVP{
  C.~Closset, T.~T.~Dumitrescu, G.~Festuccia, Z.~Komargodski and N.~Seiberg,
  ``Comments on Chern-Simons Contact Terms in Three Dimensions,''
JHEP {\bf 1209}, 091 (2012).
[arXiv:1206.5218 [hep-th]].
}

\lref\GurPCA{
  G.~Gur-Ari and R.~Yacoby,
  ``Three Dimensional Bosonization From Supersymmetry,''
JHEP {\bf 1511}, 013 (2015).
[arXiv:1507.04378 [hep-th]].
}
\lref\RadicevicYLA{
  D.~Radicevic,
  ``Disorder Operators in Chern-Simons-Fermion Theories,''
JHEP {\bf 1603}, 131 (2016).
[arXiv:1511.01902 [hep-th]].
}

\lref\NaculichPA{
  S.~G.~Naculich, H.~A.~Riggs and H.~J.~Schnitzer,
  ``Group Level Duality in {WZW} Models and {Chern-Simons} Theory,''
Phys.\ Lett.\ B {\bf 246}, 417 (1990).
}
\lref\MlawerUV{
  E.~J.~Mlawer, S.~G.~Naculich, H.~A.~Riggs and H.~J.~Schnitzer,
  ``Group level duality of WZW fusion coefficients and Chern-Simons link observables,''
Nucl.\ Phys.\ B {\bf 352}, 863 (1991).
}

\lref\BorokhovIB{
  V.~Borokhov, A.~Kapustin and X.~k.~Wu,
  ``Topological disorder operators in three-dimensional conformal field theory,''
JHEP {\bf 0211}, 049 (2002).
[hep-th/0206054].
}

\lref\NakanishiHJ{
  T.~Nakanishi and A.~Tsuchiya,
  ``Level rank duality of WZW models in conformal field theory,''
Commun.\ Math.\ Phys.\  {\bf 144}, 351 (1992).
}

\lref\GiveonZN{
  A.~Giveon and D.~Kutasov,
  ``Seiberg Duality in Chern-Simons Theory,''
Nucl.\ Phys.\ B {\bf 812}, 1 (2009).
[arXiv:0808.0360 [hep-th]].
}
\lref\BeniniMF{
  F.~Benini, C.~Closset and S.~Cremonesi,
  ``Comments on 3d Seiberg-like dualities,''
JHEP {\bf 1110}, 075 (2011).
[arXiv:1108.5373 [hep-th]].
}
\lref\AharonyDHA{
  O.~Aharony, S.~S.~Razamat, N.~Seiberg and B.~Willett,
  ``3d dualities from 4d dualities,''
JHEP {\bf 1307}, 149 (2013).
[arXiv:1305.3924 [hep-th]].
}

\lref\AharonyJZ{
  O.~Aharony, G.~Gur-Ari and R.~Yacoby,
  ``d=3 Bosonic Vector Models Coupled to Chern-Simons Gauge Theories,''
JHEP {\bf 1203}, 037 (2012).
[arXiv:1110.4382 [hep-th]].
}
\lref\GiombiKC{
  S.~Giombi, S.~Minwalla, S.~Prakash, S.~P.~Trivedi, S.~R.~Wadia and X.~Yin,
  ``Chern-Simons Theory with Vector Fermion Matter,''
Eur.\ Phys.\ J.\ C {\bf 72}, 2112 (2012).
[arXiv:1110.4386 [hep-th]].
}
\lref\AharonyNH{
  O.~Aharony, G.~Gur-Ari and R.~Yacoby,
  ``Correlation Functions of Large N Chern-Simons-Matter Theories and Bosonization in Three Dimensions,''
JHEP {\bf 1212}, 028 (2012).
[arXiv:1207.4593 [hep-th]].
}

\lref\VasilievVF{
  M.~A.~Vasiliev,
  ``Holography, Unfolding and Higher-Spin Theory,''
J.\ Phys.\ A {\bf 46}, 214013 (2013).
[arXiv:1203.5554 [hep-th]].
}

\lref\KarchTong{A.~Karch and D.~Tong, ``Particle Vortex Duality from 3d Bosonization,'' to appear.}

\lref\MuruganNastase{J. Murugan and H. Nastase, to appear.}

\lref\FradkinTT{
  E.~H.~Fradkin and F.~A.~Schaposnik,
  ``The Fermion - boson mapping in three-dimensional quantum field theory,''
Phys.\ Lett.\ B {\bf 338}, 253 (1994).
[hep-th/9407182].
}

\lref\PolyakovMD{
  A.~M.~Polyakov,
  ``Fermi-Bose Transmutations Induced by Gauge Fields,''
Mod.\ Phys.\ Lett.\ A {\bf 3}, 325 (1988).
}

\lref\ShajiIS{
  N.~Shaji, R.~Shankar and M.~Sivakumar,
  ``On Bose-fermi Equivalence in a U(1) Gauge Theory With {Chern-Simons} Action,''
Mod.\ Phys.\ Lett.\ A {\bf 5}, 593 (1990).
}

\lref\AlvarezGaumeNF{
  L.~Alvarez-Gaume, S.~Della Pietra and G.~W.~Moore,
  ``Anomalies and Odd Dimensions,''
Annals Phys.\  {\bf 163}, 288 (1985).
}

\lref\AharonyMJS{
  O.~Aharony,
  ``Baryons, monopoles and dualities in Chern-Simons-matter theories,''
JHEP {\bf 1602}, 093 (2016).
[arXiv:1512.00161 [hep-th]].
}

\lref\DaiKQ{
  X.~z.~Dai and D.~S.~Freed,
  ``eta invariants and determinant lines,''
J.\ Math.\ Phys.\  {\bf 35}, 5155 (1994), Erratum: [J.\ Math.\ Phys.\  {\bf 42}, 2343 (2001)].
[hep-th/9405012].
}

\lref\WittenABA{
  E.~Witten,
  ``Fermion Path Integrals And Topological Phases,''
[arXiv:1508.04715 [cond-mat.mes-hall]].
}

\lref\SeibergRSG{
  N.~Seiberg and E.~Witten,
  ``Gapped Boundary Phases of Topological Insulators via Weak Coupling,''
[arXiv:1602.04251 [cond-mat.str-el]].
}

\lref\WittenYA{
  E.~Witten,
  ``SL(2,Z) action on three-dimensional conformal field theories with Abelian symmetry,''
In *Shifman, M. (ed.) et al.: From fields to strings, vol. 2* 1173-1200.
[hep-th/0307041].
}

\lref\HalperinMH{
  B.~I.~Halperin, P.~A.~Lee and N.~Read,
  ``Theory of the half filled Landau level,''
Phys.\ Rev.\ B {\bf 47}, 7312 (1993).
}

\lref\PotterCDN{
  A.~C.~Potter, M.~Serbyn and A.~Vishwanath,
  ``Thermoelectric transport signatures of Dirac composite fermions in the half-filled Landau level,''
[arXiv:1512.06852 [cond-mat.str-el]].
}

\lref\WangUKY{
  C.~Wang, A.~C.~Potter and T.~Senthil,
  ``Gapped symmetry preserving surface state for the electron topological insulator,''
Phys.\ Rev.\ B {\bf 88}, no. 11, 115137 (2013).
[1306.3223].
}

\lref\MetlitskiBPA{
  M.~A.~Metlitski, C.~L.~Kane and M.~P.~A.~Fisher,
  ``Symmetry-respecting topologically ordered surface phase of three-dimensional electron topological insulators,''
Phys.\ Rev.\ B {\bf 92}, no. 12, 125111 (2015).
}

\lref\ChenJHA{
  X.~Chen, L.~Fidkowski and A.~Vishwanath,
  ``Symmetry Enforced Non-Abelian Topological Order at the Surface of a Topological Insulator,''
Phys.\ Rev.\ B {\bf 89}, no. 16, 165132 (2014).
[arXiv:1306.3250 [cond-mat.str-el]].
}

\lref\BondersonPLA{
  P.~Bonderson, C.~Nayak and X.~L.~Qi,
  ``A time-reversal invariant topological phase at the surface of a 3D topological insulator,''
J.\ Stat.\ Mech.\  {\bf 2013}, P09016 (2013).
}

\lref\WittenGF{
  E.~Witten,
  ``On S duality in Abelian gauge theory,''
Selecta Math.\  {\bf 1}, 383 (1995).
[hep-th/9505186].
}

\lref\MetlitskiYQA{
  M.~A.~Metlitski,
  ``$S$-duality of $u(1)$ gauge theory with $\theta =\pi$ on non-orientable manifolds: Applications to topological insulators and superconductors,''
[arXiv:1510.05663 [hep-th]].
}

\lref\IntriligatorEX{
  K.~A.~Intriligator and N.~Seiberg,
  ``Mirror symmetry in three-dimensional gauge theories,''
Phys.\ Lett.\ B {\bf 387}, 513 (1996).
[hep-th/9607207].
}

\lref\KapustinHA{
  A.~Kapustin and M.~J.~Strassler,
  ``On mirror symmetry in three-dimensional Abelian gauge theories,''
JHEP {\bf 9904}, 021 (1999).
[hep-th/9902033].
}

\lref\GaiottoAK{
  D.~Gaiotto and E.~Witten,
  ``S-Duality of Boundary Conditions In N=4 Super Yang-Mills Theory,''
Adv.\ Theor.\ Math.\ Phys.\  {\bf 13}, no. 3, 721 (2009).
[arXiv:0807.3720 [hep-th]].
}

\lref\WittenYA{
  E.~Witten,
  ``SL(2,Z) action on three-dimensional conformal field theories with Abelian symmetry,''
In *Shifman, M. (ed.) et al.: From fields to strings, vol. 2* 1173-1200.
[hep-th/0307041].
}
\lref\SeibergPQ{
  N.~Seiberg,
  ``Electric - magnetic duality in supersymmetric nonAbelian gauge theories,''
Nucl.\ Phys.\ B {\bf 435}, 129 (1995).
[hep-th/9411149].
}

\lref\MPS{C. Wang, A. C. Potter, and T. Senthil,
``Classification Of Interacting Electronic Topological Insulators In Three Dimensions,'' Science {\bf 343} (2014) 629, arXiv:1306.3238.}

\lref\McG{S. M. Kravec, J. McGreevy, and B. Swingle, ``All-Fermion Electrodynamics And Fermion Number Anomaly Inflow,'' arXiv:1409.8339.}

\lref\BarkeshliIDA{
  M.~Barkeshli and J.~McGreevy,
  ``Continuous transition between fractional quantum Hall and superfluid states,''
Phys.\ Rev.\ B {\bf 89}, no. 23, 235116 (2014).
}

\lref\ChenCD{
  W.~Chen, M.~P.~A.~Fisher and Y.~S.~Wu,
  ``Mott transition in an anyon gas,''
Phys.\ Rev.\ B {\bf 48}, 13749 (1993).
[cond-mat/9301037].
}

\lref\WangLCA{
  C.~Wang and T.~Senthil,
  ``Interacting fermionic topological insulators/superconductors in three dimensions,''
Phys.\ Rev.\ B {\bf 89}, no. 19, 195124 (2014), Erratum: [Phys.\ Rev.\ B {\bf 91}, no. 23, 239902 (2015)].
[arXiv:1401.1142 [cond-mat.str-el]].
}

\lref\NakaharaNW{
  M.~Nakahara,
  ``Geometry, topology and physics,''
Boca Raton, USA: Taylor and Francis (2003) 573 p.
}

\lref\NguyenZN{
  A.~K.~Nguyen and A.~Sudbo,
  ``Topological phase fluctuations, amplitude fluctuations, and criticality in extreme type II superconductors,''
Phys.\ Rev.\ B {\bf 60}, 15307 (1999).
[cond-mat/9907385].
}

\lref\KajantieVY{
  K.~Kajantie, M.~Laine, T.~Neuhaus, A.~Rajantie and K.~Rummukainen,
  ``Duality and scaling in three-dimensional scalar electrodynamics,''
Nucl.\ Phys.\ B {\bf 699}, 632 (2004).
[hep-lat/0402021].
}

\lref\ZhangWY{
  S.~C.~Zhang, T.~H.~Hansson and S.~Kivelson,
  ``An effective field theory model for the fractional quantum hall effect,''
Phys.\ Rev.\ Lett.\  {\bf 62}, 82 (1988).
}


\input epsf

\def\CSg{{\rm{CS}}_g}
\def\spinc{\rm{spin}_c}
\def\ICS{I_{\rm{CS}}}
\def\Tr{{\rm{Tr}}}
\def\L{{\cal L}}


\def\bb{
\font\tenmsb=msbm10
\font\sevenmsb=msbm7
\font\fivemsb=msbm5
\textfont1=\tenmsb
\scriptfont1=\sevenmsb
\scriptscriptfont1=\fivemsb
}




\def\tilde{\widetilde}
\def\t{\tilde}
\def\hat{\widehat}

\def\bar{\overline}
\def\b{\bar}
\def\bsq#1{{{\b{#1}}^{\lower 2.5pt\hbox{$\scriptstyle 2$}}}}
\def\bexp#1#2{{{\b{#1}}^{\lower 2.5pt\hbox{$\scriptstyle #2$}}}}
\def\dotexp#1#2{{{#1}^{\lower 2.5pt\hbox{$\scriptstyle #2$}}}}

\def\overleftrightarrow#1{\buildrel{\leftrightarrow}\over {#1}}


\def\rt2{\sqrt{2}}
\def\half {{1 \over 2}}

\def\mod{{\rm mod}}
\def\det{\mathop{\rm det}}

\def\Tr{\mathop{\rm Tr}}

\def\sign{\mathop{\rm sgn}}



\def\CA{{\cal A}}

\def\CC{{\cal C}}

\def\CH{{\cal H}}

\def\CL{{\cal L}}
\def\CM{{\cal M}}
\def\CN{{\cal N}}
\def\CO{{\cal O}}
\def\CP{{\cal P}}

\def\CT{{\cal T}}


\def\1{{\ds 1}}

\def\Z{\hbox{$\bb Z$}}

\def\em{{\rm{em}}}


\noblackbox

\def\unit{\relax{\rm 1\kern-.26em I}}
\def\nada{\relax{\rm 0\kern-.30em l}}
\def\tilde{\widetilde}
\def\t{\tilde}

\def\mod{{\rm mod}}
\def\CP{{\cal P}}
\noblackbox
\def\IL{\relax{\rm I\kern-.18em L}}
\def\IH{\relax{\rm I\kern-.18em H}}
\def\IR{\relax{\rm I\kern-.18em R}}
\def\IC{\relax\hbox{$\inbar\kern-.3em{\rm C}$}}
\def\IZ{\relax\ifmmode\mathchoice
{\hbox{\cmss Z\kern-.4em Z}}{\hbox{\cmss Z\kern-.4em Z}} {\lower.9pt\hbox{\cmsss Z\kern-.4em Z}}
{\lower1.2pt\hbox{\cmsss Z\kern-.4em Z}}\else{\cmss Z\kern-.4em Z}\fi}
\def\CM {{\cal M}}

\def\CN {{\cal N}}

\def\partialslash{\not{\hbox{\kern-2pt $\partial$}}}
\def\CP {{\cal P }}
\def\CL {{\cal L}}

\def\CO {{\cal O}}

\def\CH {{\cal H}}
\def\CC {{\cal C}}

\def\CA{{\cal A}}

\def\CM {{\cal M}}
\def\CN {{\cal N}}

\def\CO {{\cal O}}

\def\CP {{\cal P }}

\def\CSi{{\rm{CS}}}

\def\Tr{{\rm Tr}}

\font\manual=manfnt \def\dbend{\lower3.5pt\hbox{\manual\char127}}

\def\IZ{\relax\ifmmode\mathchoice
{\hbox{\cmss Z\kern-.4em Z}}{\hbox{\cmss Z\kern-.4em Z}} {\lower.9pt\hbox{\cmsss Z\kern-.4em Z}}
{\lower1.2pt\hbox{\cmsss Z\kern-.4em Z}}\else{\cmss Z\kern-.4em Z}\fi}
\def\half {{1\over 2}}

\def\bar{\overline}

\def\CH{{\cal H}}

\def\rt2{\sqrt{2}}
\def\irt2{{1\over\sqrt{2}}}

\def\t{\tilde}
\def\hat{\widehat}
\def\slashchar#1{\setbox0=\hbox{$#1$}           
   \dimen0=\wd0                                 
   \setbox1=\hbox{/} \dimen1=\wd1               
   \ifdim\dimen0>\dimen1                        
      \rlap{\hbox to \dimen0{\hfil/\hfil}}      
      #1                                        
   \else                                        
      \rlap{\hbox to \dimen1{\hfil$#1$\hfil}}   
      /                                         
   \fi}


\def\figcaption#1#2{\DefWarn#1\xdef#1{Figure~\noexpand\hyperref{}{figure}%
{\the\figno}{\the\figno}}\writedef{#1\leftbracket Figure\noexpand~\xfig#1}%
\medskip\centerline{{\footnotefont\bf Figure~\hyperdef\hypernoname{figure}{\the\figno}{\the\figno}:}  #2 \wrlabeL{#1=#1}}%
\global\advance\figno by1}

\Title {\vbox{}}
{\vbox{\centerline{A Duality Web in $2+1$ Dimensions}
\centerline{and Condensed Matter Physics}
\vskip7pt
}}

\centerline{Nathan Seiberg$^a$, T. Senthil$^b$, Chong Wang$^c$, and Edward Witten$^a$ }
\bigskip
\centerline{${}^a${\it School of Natural Sciences, Institute for Advanced Study, Princeton, NJ 08540, USA}}
\centerline{${}^b${\it Department of Physics, Massachusetts Institute of Technology, Cambridge, MA 02139, U. S. A.}}
\centerline{$^c${\it Department of Physics, Harvard University,
Cambridge, MA 02138, USA}}

\bigskip
\vskip.1in \vskip.1in
\centerline{\bf Abstract}
\noindent
Building on earlier work in the high energy and  condensed matter communities, we present a web of dualities  in $2+1$ dimensions
that generalize the known particle/vortex duality.  Some of the dualities relate theories of fermions to theories of bosons.  Others relate different theories of fermions.  For example, the long distance behavior of the $2+1$-dimensional analog of QED with a single Dirac fermion (a theory known as $U(1)_{\half}$) is identified with the $O(2)$ Wilson-Fisher fixed point.  The gauged version of that fixed point with a Chern-Simons coupling at level one is identified as a free Dirac fermion.  The latter theory also has a dual version as a fermion interacting with some gauge fields.  Assuming some of these dualities, other dualities can be derived.  Our analysis resolves a number of confusing issues in the literature including how time reversal is realized in these theories.  It also has many applications in condensed matter physics like the theory of topological insulators (and their gapped boundary states) and the problem of electrons in the lowest Landau level at half filling.  (Our techniques also clarify some points in the fractional Hall effect and its description using flux attachment.) In addition to presenting several consistency checks, we also present plausible (but not rigorous) derivations of the dualities and relate them to $3+1$-dimensional $S$-duality.

\vfill

\Date{June 2016}

\newsec{Introduction}

Duality in quantum field theory refers to two related but distinct phenomena. In one case we consider two or more presentations of the same theory.  Here the different dual presentations lead to identical physics.  Examples of that include the duality between two free field theories like the $2+1$ dimensional compact boson\foot{ We use here the high energy physics terminology.  A boson $\varphi$ is compact if it parameterizes a compact target space such as a circle.  In  condensed matter, it is common to refer to a free massless boson as non-compact and to call a scalar field ``compact''  if  a codimension two defect with nonzero winding can have finite action or (depending on the spacetime dimension)  energy or energy density.
In $1+1$ dimensions, this can be achieved by adding to the action a twist field with nonzero vorticity; these is no close analog of this in higher dimensions, although of course there
are many models that do have defects of the appropriate sort. There is an analogous difference in terminology
for gauge fields.  In high energy physics, to call an abelian gauge field ``compact'' means that the gauge group is compact.  For example, in $2+1$ dimensions this means that monopole operators exist
(they may or may not be included in the action).  In the condensed matter literature, in $2+1$ dimensions, an abelian gauge field is usually called compact if a monopole operator is included in the action, so that the corresponding
magnetic flux is not conserved.}
 and the $2+1$ dimensional free photon.  Typical interacting examples are dual $\CN=4$ supersymmetric theories in $3+1$ dimensions.  The second class of duality is IR duality.  Here two or more different quantum field theories flow to the same IR theory.  This latter theory could be free or interacting.  A typical example of IR duality that we will soon review is the $2+1$ dimensional particle/vortex duality.  Other examples use supersymmetry and include $\CN=1$ and $\CN=2$ supersymmetric theories in $3+1$ dimensions and mirror symmetry in $\CN=2$ and $\CN=4$ supersymmetric theories in $2+1$ dimensions.  These $2+1$ dimensional examples use particle/vortex duality.

In this paper, we will discuss  a number of non-supersymmetric examples in $2+1$ dimensions.  We will explore the relations between them and will discuss their applications in condensed matter physics.

\subsec{Review of the Bosonic Particle/Vortex Duality}

Particle/vortex duality relates the $2+1$ dimensional $O(2)$ Wilson-Fisher fixed point to a gauged version of that theory \refs{\PeskinKP,\DasguptaZZ}.  We write it as
\eqn\particlevortex{
|D_B\phi|^2 -|\phi|^4 \qquad \longleftrightarrow  \qquad |D_{\hat b} \hat\phi|^2 -|\hat\phi|^4+{1\over 2\pi} \hat b dB~.}
Here $\phi$ and $\hat \phi$ are complex scalar fields, $\hat b$ is a dynamical $U(1)$ gauge field and $B$ is a background $U(1)$ gauge field.  $D_B$ is the covariant derivative acting on charge $+1$ fields.  Our notation with the $|\phi|^4$ and $|\hat \phi|^4$ interaction means that the duality is valid only in the IR as the coefficient of this interaction flows to the IR fixed point.  In both sides of the duality we tune the coefficient $r$ in $r |\phi|^2$ and $\hat r$ in $\hat r |\hat \phi|^2$ to the fixed point and define them to vanish there.  As is well known, the theories in the two sides of the duality have a global $U(1)$ symmetry, which we identify.  On the left, the conserved current is $i\bar\phi\overleftrightarrow{\partial}_\mu\phi$ and
on the right it is $\epsilon_{\mu\nu\lambda }\partial^\nu \hat b^\lambda/2\pi$.  We have coupled these conserved currents to
 a classical $U(1)$ gauge field $B$ and we refer to the associated symmetry as $U(1)_B$.

The operator $\phi$ in the left hand side is charged under $U(1)_B$. It is mapped by the duality to a monopole operator $\CM_{\hat b}$ on the other side of the duality.

If we perturb the left hand side of \particlevortex\ by the relevant operator $r |\phi|^2$, the resulting theory depends on the sign of $r$.  For positive $r$ the theory becomes gapped by giving $\phi$ a mass and the global symmetry $U(1)_B$ is unbroken.  For negative $r$, the global symmetry $U(1)_B$ is spontaneously broken and the spectrum includes a massless Nambu-Goldstone boson, which is the phase of $\phi$.  The same physics is reproduced on the other side of the duality but with $\hat r = -r$.  Positive $r$ is mapped to negative $\hat r$ and the gauge symmetry of $\hat b$, which we call $U(1)_{\hat b}$, is Higgsed.  As before, the spectrum is gapped.  It includes vortex excitations, which carry charge $1$ under the unbroken global symmetry $U(1)_B$. They are identified as the $\phi$ particles of the left hand side of \particlevortex. Hence the name particle/vortex duality.  Negative $r$ is mapped to positive $\hat r$.  In this phase, $\hat \phi$ is massive and the low energy spectrum includes a massless gauge field $\hat b$. Its dual is a compact scalar (see the above footnote), which is identified as the Nambu-Goldstone boson of the broken $U(1)_B$ global symmetry.

 It is useful to consider the fate of time reversal and charge conjugation symmetries under the duality. We will actually keep track of the two anti-unitary symmetries $\CT$ and $\CC\CT$. For the left hand side of \particlevortex, we implement these\foot{We use a notation where for gauge fields, the $\CT$ or $\CC\CT$ action is indicated for the spatial components; the time components will transform with opposite sign} as
\eqn\TCTboson{\eqalign{
& \CT(\phi) = \phi,~~\CT(B) = -B\cr
& \CC\CT(\phi) = {\phi}^\dagger, ~~\CC\CT(B) = B  }}
On the dual side these symmetries are then implemented as
\eqn\TCTvortex{\eqalign{
& \CT(\hat \phi) = {\hat \phi}^\dagger,~~\CT(b) = -b\cr
& \CC\CT(\hat \phi) = \hat \phi, ~~\CC\CT(b) = b  }}
Note that under the duality the action of $\CT$ and $\CC\CT$ on $\phi$ and $\hat \phi$ are interchanged.

This  bosonic particle-vortex duality has been successfully tested by numerical simulations of lattice versions of both sides  (see e.g, \refs{\NguyenZN,\KajantieVY}).

\subsec{Quantum Field Theory Perspective}

The duality web that we will explore is actually part of a richer picture that has emerged  in recent studies of relativistic field theory.

Several different lines of research have influenced this development.  One of them originated from level/rank duality in $1+1$-dimensional conformal field theories.  Roughly, it relates $SU(N)_k$ and $SU(k)_{-N}$ Kac-Moody algebras and their corresponding WZW models.  (We say `roughly' because, as we state below, one needs to add certain $U(1)$ factors to one or the two sides of the duality.)  The same level-rank duality has a $2+1$-dimensional Topological Quantum Field Theory counter part, where the relation is between different Chern-Simons gauge theories \refs{\NaculichPA,\MlawerUV,\NakanishiHJ}.  These dualities have been rigorously established in the sense that all the observables of the dual theories were shown to be identical.

$\CN=2$ supersymmetric theories in $2+1d$ exhibit analogs of particle/vortex duality. The first examples were found in \AharonyBX.  These were later extended in various directions and in particular the authors of \refs{\GiveonZN\BeniniMF\IntriligatorLCA-\AharonyDHA} found such dualities, which relate different Chern-Simons matter theories.  When the matter fields are massive and are integrated out these dualities go over to the level-rank dualities of the topological theories. Also, it was shown in \AharonyDHA\ that these $2+1d$ dualities follow upon compactification on a circle from the previously found dualities  \SeibergPQ\ in $3+1$ dimensions.  Unlike the topological theories, where the duality is rigorously established, the dualities in these quantum field theories cannot be proven.  However, using the power of supersymmetry many observables in these theories can be computed exactly in the two sides of the duality and  shown to match.  And renormalization group flows between these dualities lead to additional consistency checks.

Once one has a duality between supersymmetric theories, one can attempt to break supersymmetry by turning on corresponding relevant operators on the two sides of the duality to flow to a duality between non-supersymmetric theories.  Since supersymmetry is broken, there is less control over the renormalization group flow.  Assuming that the flow is smooth, a new duality can be found.  It relates a theory of bosons coupled to a Chern-Simons gauge theory to a theory of fermions coupled to a Chern-Simons theory \refs{\JainGZA,\GurPCA}.  Since these theories are not supersymmetric, this duality cannot be subject to most of the tests of the supersymmetric dualities.

\lref\MaldacenaJN{
  J.~Maldacena and A.~Zhiboedov,
  ``Constraining Conformal Field Theories with A Higher Spin Symmetry,''
J.\ Phys.\ A {\bf 46}, 214011 (2013).
[arXiv:1112.1016 [hep-th]].
}

\lref\AharonyNS{
  O.~Aharony, S.~Giombi, G.~Gur-Ari, J.~Maldacena and R.~Yacoby,
  ``The Thermal Free Energy in Large N Chern-Simons-Matter Theories,''
JHEP {\bf 1303}, 121 (2013).
[arXiv:1211.4843 [hep-th]].
}
\lref\InbasekarTSA{
  K.~Inbasekar, S.~Jain, S.~Mazumdar, S.~Minwalla, V.~Umesh and S.~Yokoyama,
  ``Unitarity, crossing symmetry and duality in the scattering of $ {N}=1 $ susy matter Chern-Simons theories,''
JHEP {\bf 1510}, 176 (2015).
[arXiv:1505.06571 [hep-th]].
}

\lref\MinwallaSCA{
  S.~Minwalla and S.~Yokoyama,
  ``Chern Simons Bosonization along RG Flows,''
JHEP {\bf 1602}, 103 (2016).
[arXiv:1507.04546 [hep-th]].
}
\lref\GiombiMS{
  S.~Giombi and X.~Yin,
  ``The Higher Spin/Vector Model Duality,''
J.\ Phys.\ A {\bf 46}, 214003 (2013).
[arXiv:1208.4036 [hep-th]].
}

\lref\JainNZA{
  S.~Jain, M.~Mandlik, S.~Minwalla, T.~Takimi, S.~R.~Wadia and S.~Yokoyama,
  ``Unitarity, Crossing Symmetry and Duality of the S-matrix in large N Chern-Simons theories with fundamental matter,''
JHEP {\bf 1504}, 129 (2015).
[arXiv:1404.6373 [hep-th]].
}
\lref\JainGZA{
  S.~Jain, S.~Minwalla and S.~Yokoyama,
  ``Chern Simons duality with a fundamental boson and fermion,''
JHEP {\bf 1311}, 037 (2013).
[arXiv:1305.7235 [hep-th]].
}

\lref\JainPY{
  S.~Jain, S.~Minwalla, T.~Sharma, T.~Takimi, S.~R.~Wadia and S.~Yokoyama,
  ``Phases of large $N$ vector Chern-Simons theories on $S^2 \times S^1$,''
JHEP {\bf 1309}, 009 (2013).
[arXiv:1301.6169 [hep-th]].
}
\lref\GiombiYA{
  S.~Giombi and X.~Yin,
  ``On Higher Spin Gauge Theory and the Critical O(N) Model,''
Phys.\ Rev.\ D {\bf 85}, 086005 (2012).
[arXiv:1105.4011 [hep-th]].
}

\lref\SezginRT{
  E.~Sezgin and P.~Sundell,
  ``Massless higher spins and holography,''
Nucl.\ Phys.\ B {\bf 644}, 303 (2002), Erratum: [Nucl.\ Phys.\ B {\bf 660}, 403 (2003)].
[hep-th/0205131].
}
\lref\KlebanovJA{
  I.~R.~Klebanov and A.~M.~Polyakov,
  ``AdS dual of the critical O(N) vector model,''
Phys.\ Lett.\ B {\bf 550}, 213 (2002).
[hep-th/0210114].
}

Another source of information about these theories arises by taking $N$ and $k$ to infinity with fixed ratio.  Then, one can use large $N$ techniques to compute many observables \refs{\GiombiYA\AharonyJZ\GiombiKC\MaldacenaJN\AharonyNH\GiombiMS\AharonyNS\JainPY\JainGZA
\JainNZA\InbasekarTSA\MinwallaSCA-\GurAriXFF} and they turn out to confirm this duality.  Also, these dual large $N$ theories have the same gravitational dual description \refs{\SezginRT,\KlebanovJA,\GiombiYA,\AharonyJZ,\GiombiKC}, which involves an unusual type of theory with almost massless
fields of very high spin (for one reference out of many on this topic, see \VasilievVF).  This leads to additional evidence for the duality between the different field theories.  Understanding
the field theory dual of this unusual sort of gravitational theory was one motivation for study of these Chern-Simons-matter theories and their dualities.

Going back to finite $N$ and $k$, the theories with unitary gauge groups have monopole operators whose properties can be analyzed and  matched with  operators on the other side of the duality \RadicevicYLA.

Ofer Aharony combined all these elements and spelled out three conjectured Chern-Simons-matter dualities \AharonyMJS
\eqn\Oferdg{\eqalign{
N_f\ {\rm  fermions\ coupled\ to }\ U(k)_{-N+{N_f\over 2}, -N+{N_f\over 2}}  \qquad &\longleftrightarrow \qquad N_f\ {\rm scalars\ coupled\ to\ } SU(N)_k \cr
N_f\ {\rm  fermions\ coupled\ to\ } SU(k)_{-N+{N_f\over 2}}   \qquad &\longleftrightarrow \qquad N_f\ {\rm  scalars\ coupled\ to\ } U(N)_{k,k}\cr
N_f\ {\rm  fermions\ coupled\ to\ } U(k)_{-N+{N_f\over 2}, -N-k+{N_f\over 2}}   \qquad &\longleftrightarrow \qquad N_f\ {\rm  scalars\ coupled\ to\ }U(N)_{k,k+N} }}
where the subscript denotes the Chern-Simons level and $U(L)_{M,K}\equiv( SU(L)_M \times U(1)_{LK})/\Z_L$.  If there are no matter fields we substitute $N_f=0$ and find the rigorously established well known level-rank dualities.

Here we specialize to $N=k=N_f=1$ and interpret $SU(1)$ as trivial.  This turns \Oferdg\ to
\eqn\Oferd{\eqalign{
{\rm A\  fermion\ coupled\ to\ }U(1)_{-{1\over 2}}  \qquad &\longleftrightarrow \qquad {\rm A\ scalar} \cr
{\rm A\ fermion}\qquad &\longleftrightarrow \qquad {\rm A\  scalar\ coupled\ to\ }U(1)_{1} \qquad \cr
 {\rm\ A\ fermion\ coupled\ to\ } U(1)_{-{3\over 2}}  \qquad &\longleftrightarrow \qquad {\rm A\ scalar\ coupled\ to\ } U(1)_{2} } }
We interpret these dualities to mean that the fermions are coupled to gauge fields without additional interactions.  In particular, in the second duality the fermion is free.  On the other hand the scalar in the first duality should be interpreted to be in a Wilson-Fisher fixed point and in the other dualities this Wilson-Fisher theory is gauged.

Even before we get into the details, the dualities \Oferdg\ and the special cases \Oferd\ pose a puzzle.  Some of these theories are purely bosonic and can be formulated on non-spin manifold.\foot{At least classically, these theories and most theories considered in the present paper  can be defined only on  orientable manifolds, as their Chern-Simons couplings  require an orientation.  Every oriented three-manifold $M$ admits
a spin structure, but generically $M$ may admit multiple inequivalent spin structures. In this context, when we speak of a ``non-spin'' three-manifold $M$, we mean a three-manifold
without a chosen spin structure.  For a detailed explanation of how a Chern-Simons action can depend on the choice of spin structure, even though naively it is purely bosonic,
see Appendix A.}   Their suggested dual theories involve fermions and seemingly can only be formulated on a spin manifold with a choice of spin structure.  For example, the right hand side of the first and the third dualities in \Oferd\ involve fundamental bosons and their Chern-Simons couplings are consistent without a choice of spin structure.  (The right hand side of the second duality in \Oferd\ involves bosons but with a Chern-Simons coupling that makes sense only on a spin manifold.)  So if the dualities are correct, it must be possible to formulate
the theories on the left hand side without choosing a spin structure, even though those theories contain fermions.

Below we will extend the dualities \Oferd\ in several ways.   First, we will couple the global symmetries to background gauge fields and will constrain their Chern-Simons contact terms \ClossetVP.  Second, we will change some of the dynamical or background gauge fields to spin$_c$ connections (see section 1.4).  This will allow us to place the theories on non-spin manifolds.  Third, we will show that these dualities follow from each other; assuming any one of them we can derive the other two dualities.
Finally, we will derive a number of other dualities including a fermion/fermion duality that will be discussed in the next section,
and we  will explore in detail many of their properties.

 \subsec{Condensed Matter Perspective}

The particle-vortex duality of bosons reviewed  in section 1.1 is tremendously useful in condensed matter physics. It gives a powerful conceptual way to access novel phases and phase transitions of systems of interacting bosons. It is thus natural to ask if there are similar dualities for fermionic systems with a conserved global $U(1)$ current. For Dirac fermions in $2+1$ dimensions a few such  dualities have been proposed over the years. To set the stage for the results of the present paper we briefly review these proposals. We will be somewhat telegraphic - more precise statements will be made later in the paper.

We begin with a recently proposed \refs{\WangQMT,\MetlitskiEKA} duality between two different descriptions, both involving a single two-component massless Dirac fermion, of the spatial boundary of a $3+1$-dimensional topological insulator.  We will refer to this below as a fermion-fermion duality.  It is well known that  the free massless Dirac theory
\eqn\freeD{{\cal L}_0 = i \bar \Psi \slashchar{D}_A \Psi  }
describes a possible phase for the surface.
At the topological insulator boundary, this theory preserves time reversal symmetry (the parity anomaly of this theory is cancelled by a bulk contribution). $D_A$ is a covariant derivative with $A$ an external background gauge field.
The topological insulator surface also admits a number of other phases which are stabilized by strong interactions between the underlying electrons. A dual description of this surface
theory capable of describing the possible phase diagram was described in \refs{\WangQMT,\MetlitskiEKA}  and takes the form
\eqn\dualD{{\cal L}_{dual} = i \bar \chi \slashchar{D}_a \chi  + {1\over 4\pi} Ada}
This description has had considerable success, but  as  stated it cannot be precisely correct, since the coupling $A da/4\pi$ is not gauge-invariant mod $2\pi$, assuming that the gauge fields obey standard Dirac quantization.  Moreover, obvious fixes such as a  nonstandard Dirac quantization law for $a$ run into one problem or another.  One
outcome of the present paper will be to resolve
this situation by slightly correcting the duality statement, roughly by adding a topological field theory in \dualD.   As for whether this refinement  is important for condensed matter physics, this depends on the application
that one has in mind.  In general, the refinement is likely to be important for applications
to gapped phases\foot{In particular it allows for a smooth derivation of the topological field theory of gapped phases where global aspects are automatically correctly captured.}, but less important for applications to gapless phases.

 A rough analogy can be made between the dual Dirac theory in its original version \dualD\ and the  effective field theories of flux attachment
 (such as the HLR theory \HalperinMH) popular in  the literature on the Fractional Quantum Hall Effect.  Some simple versions of these theories make use of emergent gauge fields with improperly quantized Chern-Simons couplings, and in that case   it is known
 that a more precise and completely gauge-invariant description can be given by adding additional fields  (see Appendix C).  Our refinement of eqn.\ \dualD\ is somewhat similar.

Refs. \refs{\WangQMT,\MetlitskiEKA} also raised the possibility that \freeD\  and \dualD\ in fact flow to the same IR fixed point, and provided some suggestive supporting arguments (see also \RoscherWOX). A derivation of the equivalence of these two theories when one spatial direction is discretized (leading to what is known as a ``wire construction") has also appeared \MrossIDY.

Given that \dualD\ is ill defined, how can these facts be true?  We will present a modified version of \dualD\ with properly normalized Chern-Simons terms and will explore its dynamics.  We will recover the above claims in a clearer setting and will relate the duality between \freeD\  and the modified version of \dualD\ to similar statements about other dualities in the web that we discuss.  Also, we will embed this statement as
part of a larger web of dualities including supersymmetric and large $N$ dualities that are
treated in the high energy literature.

The dual Dirac theory \dualD\ has been related \refs{\WangCTO,\MetlitskiEKA,\MetlitskiYQA}
to the bulk electric-magnetic duality  of  $3+1$-dimensional $U(1)$ gauge theory coupled to the global $U(1)$ symmetry of a topological insulator.   Again, in section 6 of the present paper, we will provide a more precise framework for arguments of this nature.

The existence of such a dual Dirac description leads to a number of fundamental results in the theory of strongly interacting electronic systems.
We sketch these briefly. A synthesis with more details is in \WangFQL.
Introducing a uniform magnetic field at the topological insulator surface\foot{This requires a topological insulator  of charged fermions with a time-reversal symmetry under which the electric charge is odd. (In relativistic terminology, this corresponds to a $\CC\CT$ rather than $\CT$ symmetry.) A material of this type is called a class AIII topological insulator in the condensed matter literature. }   leads to a mapping to the famous problem of the half-filled Landau level of two dimensional electrons in the quantum hall regime. The existence of the dual Dirac theory gives a theoretical basis for a recently proposed description by Son \SonXQA\  of the metallic state found in experiments in the half-filled Landau level.  The classic theory \HalperinMH\ for this state -- due to Halperin, Lee, and Read (HLR) -- has long been known to not incorporate a symmetry present when the Hamiltonian of the electron gas is restricted to the lowest Landau level.  The proposed new description -- dubbed the Dirac composite fermion theory -- provides an elegant alternate to the standard HLR theory that includes this symmetry.  The Dirac composite fermion theory makes some specific predictions for numerical calculations which seem to be verified in recent work \GeraedtsPVA.  Further the theory makes predictions \refs{\SonXQA,\PotterCDN,\WangGQJ} for experiments that may distinguish it from what is expected within the HLR theory.

In a different direction, the dual Dirac theory resolves a number of puzzling conceptual questions about possible surface states of topological insulators in $3+1$-D.
In the presence of strong interactions between the electrons, the surface of the topological insulator may be gapped while preserving all physical global symmetries. Such a symmetric gapped boundary state supports anyon excitations and is described by a Topological Quantum Field Theory. However  the symmetries are realized anomalously, {\it i.e}, in a manner not possible in a strictly two-dimensional system. Such gapped boundaries were originally constructed in \refs{\WangUKY\MetlitskiBPA\ChenJHA-\BondersonPLA} and were shown to have non-abelian anyons.  One such symmetric gapped state --  known as the T-Pfaffian -- was obtained through soluble lattice models in \ChenJHA.  A different gapped boundary state (with twice as many distinct quasiparticles as the T-Pfaffian) was constructed in \refs{\WangUKY,\MetlitskiBPA}   through a procedure known as vortex condensation by starting with the surface superconductor (which spontaneously breaks the $U(1)$ but preserves time reversal).   A recent field theoretic description   of the topological insulator surface  based entirely on clearcut considerations of weak coupling   \SeibergRSG\ also finds this more complex state and not the simpler T-Pfaffian state.  (It was also possible to find T-Pfaffian$\times U(1)_2$, but not T-Pfaffian by itself.)  Despite its appearance in the solvable lattice model, the T-Pfaffian was previously hard to fit into the understanding of the possible phase diagram at the topological insulator surface. In particular its relation with several conventional boundary states (such as the free Dirac fermion, or the superconductor) was obscure.  Further \ChenJHA\ found two versions of the T-Pfaffian distinguished by the action of time reversal of which only one corresponds to a possible  surface state of the conventional topological insulator but it was not clear which one. The dual Dirac
liquid constructs \refs{\WangQMT,\MetlitskiEKA} the T-Pfaffian as a simple paired state of the dual fermions, thereby  ``explaining" its existence in the lattice constructions. Further the ambiguity on which of the two T-Pfaffians corresponds to the conventional topological insulator could be resolved \MetlitskiYQA.  Again after refining the dual Dirac theory in the way that we propose here, the relation of this theory to T-Pfaffian can be described via clear-cut arguments at weak coupling.  (We will prefer to replace the ``paired state'' treatment of these theories by an explicit and concrete description in terms of fundamental scalars.)

Many years ago, it was shown \refs{\WilczekDU,\PolyakovMD,\JainTX}  that coupling to a $U(1)$ gauge field at level 1 can shift the spin of a quasiparticle by $1/2$, converting bosons to fermions. We schematically write
\eqn\Polyakovbf
{{\rm A\ fermion}\qquad \longleftrightarrow \qquad {\rm A\  scalar\ coupled\ to\ }U(1)_{1} \qquad }
This was originally formulated for coupling of  gapped quasiparticles to $U(1)_1$.   In the present  paper, we will, in a sense, extend this
relationship of massive particles to a critical point
where the particles are gapless and the theory with the scalar coupled to $U(1)_1$ becomes equivalent to a free fermion theory.

There is another version of this in which the fermion is coupled to a $U(1)$ gauge field and the scalar is at its Wilson-Fisher critical fixed point.
In fact, the literature contains proposals for such a duality  \refs{\ChenCD,\BarkeshliIDA}.  We will state it
in a more precise way and we will loosely write it as
\eqn\xydualf
{{\rm A\  fermion\ coupled\ to\ }U(1)_{-1/2}  \qquad  \longleftrightarrow \qquad {\rm A\ scalar}}
(We note, however, that the meaning of $U(1)_{-1/2}$ is often expressed in an oversimplified way.  See the discussion in the next subsection.)

We refer to dualities such as \Polyakovbf\ or \xydualf\  as boson-fermion dualities. They may be thought of as
relativistic versions of the flux attachment transformation familiar from the theory of the quantum Hall effect.

There are a number of fundamental questions that these various dualities raise.
In both \Polyakovbf\ and \xydualf, time-reversal symmetry is manifest on one side and not on the other.  What implications does time-reversal have for the understanding of these dualities?
How  are  these  boson-fermion dualities related to each other, and to particle-vortex duality for bosons and to the corresponding duality for fermions that was discussed above?
Finally how are these dualities related to electric-magnetic duality in $3+1$ dimensions?

 In this paper we will address all of these questions.  We will find a web of dualities that contains all of these dualities and additional ones.  Assuming one of the boson-fermion dualities  leads to a derivation of all of the other dualities in the web. As a bonus, understanding the duality web leads to the refined version mentioned above of the dual Dirac theory.   We will also learn that on one side of the boson-fermion dualities,  time-reversal itself acts as a  duality transformation that interchanges particles and vortices of the same statistics.  The relation to electric-magnetic duality gives an appealing way to understand this rather subtle realization of time-reversal.

\subsec{Review of Some Background Material}

In preparation, we review here some general properties of gauge fields, Chern-Simons terms, fermion path integrals, topological insulators, and monopole operators.

Our conventions are that upper case letters denote classical gauge fields and lower case letters denote dynamical fields.  $A$ and $a$ are spin$_c$ connections, while other letters denote ordinary $U(1)$ gauge fields.

In  condensed matter systems made from fermions of odd charge, there is a spin/charge relation which states that all local operators of integer spin carry even charge and all operators with half-integer spin carry odd charge. The consequences of this relation can be subtle.  An elegant although formal way to capture these relations is to introduce
 a spin$_c$ connection\foot{Condensed matter physicists might find the discussion of this concept in \MetlitskiYQA\ accessible.} $A$ that couples to the conserved charge carried by the fermions \refs{\MetlitskiYQA,\SeibergRSG}. (We introduce $A$ as a background classical field, or if the fermions are naturally coupled
 to a $U(1)$ gauge field, we reinterpret this as a $\spinc$ connection.)  A $\spinc$ connection is locally the same as a $U(1)$ gauge field, but its Dirac quantization is different. Its fluxes satisfy
\eqn\spincc{\int_C {dA\over 2\pi} = \half \int_C w_2\ \mod\ \Z ,}
where $C\subset X$ is an oriented two-cycle in our spacetime $X$ and $w_2$ is the second Stieffel-Whitney class \NakaharaNW\ of $X$.

Let us consider a single spin$_c$ connection $A$ and a number of $U(1)$ gauge fields $B^i$. Then, the properly normalized Chern-Simons terms are
\eqn\CSno{\eqalign{
&{k_{ij}\over 4\pi} B^idB^j + {q_{i}\over 2\pi} B^idA + {\hat k\over 4\pi} AdA +(2\hat k +16n) \CSg\cr
&k_{ij}, q_{i}, \hat k , n \in \Z \cr
&k_{ii} = q_{i}\ \mod\ 2~,}}
where $\CSg$ is a gravitational Chern-Simons term.
See Appendix B and \SeibergRSG\ for more details.\foot{{ Though not necessarily deduced from well-definedness of Chern-Simons couplings, these constraints will be familiar to condensed matter physicists well-versed in  the $K$-matrix description of topological ordered states of fermionic matter. Local (``transparent'') operators  (i.e, ones creating particles that braid trivially with other particles) couple to $B^i$ with charges $l_i = \sum_j k_{ij} m_j$ where $m_j$ are integers. These have self statistics $\theta_m = \pi m^T k m$ where we have used an obvious matrix notation.  Choosing the $i$th entry of $m$ to be $1$, and the rest $0$, the self-statistics will be $\pi k_{ii}$. This is a boson if $k_{ii}$ is even and a fermion if $k_{ii}$ is odd. Requiring that bosons (fermions) carry even (odd) charge under $A$ gives $k_{ii} = q_i\ \mod 2$. The $AdA$ and $\CSg$ terms may be interpreted as combining the TQFT described by the first two terms with another $2+1$-D gapped system (made out
of the same microscopic fermions)  which has no non-trivial quasiparticles.  This added system will then have an integer electrical Hall conductivity $\sigma_{xy} = \hat k$, and  a chiral central charge (equal to the thermal Hall conductivity $\kappa_{xy}$ in units of $\kappa_o = {\pi^2 T \over 3}$ where $T$ is the temperature) of edge modes $= \hat k + 8n$. The mismatch between the thermal and electrical  Hall conductivities  by a multiple of $8$ in such a system can be understood through the arguments in the Appendix of \WangLCA, and references therein.}  }  Below we will use such expressions where some of these fields will be dynamical (and will be denoted by lower case letters).

Consider a $2+1d$ Dirac fermion coupled to a background gauge field or spin$_c$ connection $\CA$ via the Dirac operator $\slashchar{D} _\CA$
\eqn\freef{i\bar \Psi \slashchar{D} _{\CA}\Psi ~.}
This theory needs regularization.  One choice of regularization leads to the partition function \refs{\AlvarezGaumeNF,\WittenABA}
\eqn\Ztd{Z_{2+1d}=|\det \slashchar{D} _\CA|e^{- {i\pi\over 2}\eta(\CA)}~.}
It is common instead of $\exp(-i \pi \eta/2)$ to write here $\exp(-i \CSi(\CA)/2)$ (or a similar and more complete expression with a gravitational correction)
where $\CSi(\CA)=(1/4\pi)\int \CA d\CA$ is a properly normalized Chern-Simons coupling.
This is adequate for many purposes but is not entirely correct as $\CSi(\CA)/2$ (even with the gravitational correction) is only gauge-invariant
mod $\pi$, so that $\exp(-i \CSi(\CA)/2)$ is not gauge-invariant.  A careful treatment of the path integral of $\Psi$
leads instead, with one regularization,  to eqn.\ \Ztd.
The relation between this description with $\pi\eta/2$ and the more familiar but slightly less precise one with $\CSi(\CA)/2$
is as follows.  Although $\CSi(\CA)/2$ is not gauge-invariant mod $ 2 \pi$ and so is not a satisfactory term in an effective action,
its variation $\delta \CSi(\CA)/2$ is perfectly well-defined and gauge-invariant.  Moreover, the Atiyah-Patodi-Singer index
theorem implies that as long as the Dirac operator has no zero-modes, $\delta (\pi \eta/2) = \delta \CSi(\CA)/2$.  Because of this,
unless one asks certain delicate questions (the theory of a topological insulator is the most obvious place where such questions arise) one
can proceed informally with $\CSi(\CA)/2$ rather than $\pi \eta/2$.  In effect this is done in most of the literature and the theory regularized as in eqn.\ \Ztd\ is commonly
called $U(1)_{-1/2}$.  We will follow that terminology here.

Eqn.\ \Ztd\ is only one possible regularization of the fermion path integral.  Other
 regularizations can be parameterized by adding well-defined local  counterterms.  Specifically, we can add properly normalized Chern-Simons terms \CSno.
 For example, for $\CA$ a spin$_c$ connection, we can add ${k\over 4\pi} \CA d\CA+2k\CSg$ with integer $k$. (We then call the resulting theory $U(1)_{k-1/2}$.) The parity anomaly is the statement that the partition function \Ztd\ cannot be made real by adding such a counterterm.

When we add a mass term for the fermion and integrate it out, the low energy interaction includes the phase from \Ztd\ and an additional factor $\exp(i\sign(m){\pi\over 2}\eta(\CA))$.  So for $m$ positive the phase is canceled and for $m$ negative it is $\exp(-i\pi\eta(\CA)) = \exp(-{i \over 4\pi} \int \CA d\CA -2i\int\CSg)$.  Part of the subtlety of the subject is in the last
statement; the Atiyah-Patodi-Singer theorem can be used to replace $\exp(-i\pi\eta(\CA)) $ with  $\exp(-{i \over 4\pi} \int \CA d\CA -2i\int\CSg)$, but there is no such replacement for
$\exp(-i\pi\eta(\CA)/2) $.

Next, we would like to review various ways to cancel the phase in \Ztd\ and make the theory $\CT$-invariant.
First, if $\CA=2B$ for some $U(1)$ gauge field $B$, then we can add the counter term ${2\over 4\pi}BdB$ to make the answer real and $\CT$-invariant.  Clearly, this violates the spin/charge relation.

The topological insulator instead restores $\CT$-invariance by making $\CA$ a $3+1d$ field.  Then we can add a bulk term $\pi \int_{bulk} (\hat A(R)+{1\over 8 \pi^2}d\CA d\CA )$ (see Appendix B) and write the Lagrangian
\eqn\freefb{i \bar \Psi \slashchar{D}_A \Psi +{1\over 8\pi} AdA+\CSg~,}
where the improperly quantized Chern-Simons terms of $A$ and the metric are shorthand notation for the bulk term.
Then the partition function
\eqn\Zfd{Z_{3+1d}=|\det \slashchar{D} _\CA|e^{- {i\pi\over 2}\eta(\CA) +i \pi \int_{bulk} \left(\hat A(R)+{1\over 8 \pi^2}d\CA d\CA \right) }~}
is real and $\CT$-invariant.\foot{In fact, by the Atiyah-Patodi-Singer index theorem,
it equals $|\det\slashchar{D}_\CA|(-1)^{\cal I}$, where $\cal I$ is the index of the four-dimensional Dirac operator, computed with APS boundary conditions \WittenABA.}

The main examples in \SeibergRSG\ combined these two mechanisms.  There $\CA= 4nb+A$ with $A$ a classical $3+1d$ spin$_c$ bulk field and $b$ a $2+1d$ gauge field on the boundary.  In this case the ``bulk'' contribution in \Zfd\ can be written as
\eqn\Zfdp{e^{  i \pi \int_{bulk} (\hat A(R)+{1\over 8 \pi^2}d\CA d\CA )} = e^{ {i\over 4\pi}\int_{boundary} (8n^2 bdb + 4nbdA)} e^{ i\pi \int_{bulk} (\hat A(R)+{1\over 8 \pi^2}dA dA )}~.}
The second factor is from the bulk.  The first factor involves properly normalized Chern-Simons terms that depend only on the fields at the boundary. This is consistent with $b$ being a boundary field.  The genuine bulk term in \Zfdp, which depends on $A$, is a classical term.  Therefore, the Abelian sector in the examples in \SeibergRSG\ can be interpreted as a counterterm that must be added to the classical theory in order to make it $\CT$-invariant.

Returning to \freef, an interesting background field $\CA$ can be constructed by removing a point $\CP$ from our $2+1d$ manifold and specifying ``monopole boundary conditions'' \refs{\BorokhovIB,\BorokhovCG} on an $S^2$ surrounding $\CP$, $\int _{S^2} d\CA = 2\pi$.  Both $\Psi$ and $\bar \Psi$ have zero-modes leading after
quantization to two different states differing by a factor of $\Psi$ (or $\bar\Psi$).  These two states have spin zero and their electric charges differ by $1$.  One way to determine the charges is to add, as in \freefb, a bulk term ${1\over 8\pi} \CA d\CA$ .  Then the theory is $\CT$ and $\CC\CT$ invariant.
This determines the charges to be $\pm \half$.  The charges without this bulk term can be determined by noticing that in a monopole background the bulk term shifts the charges
of all monopole states by $+\half$.  Therefore, in the theory \freef\ without that term the charges are $0$ and $-1$.\foot{It is common in the literature to view the fermion determinant as real and to ``add by hand'' $-{1\over 8\pi } \int AdA$ as an approximation to $-{i\pi\over 2}\eta(A)$.  In this presentation the charges of the monopole receive a contribution of $-\half$ from this Chern-Simons term and $\pm \half$ from the fermion zero-modes.  This leads to the charges $0$ and $-1$.  Here we do not add such a term by hand and we do not approximate $\eta$ by a Chern-Simons term.}

We will denote such an insertion by $\CM_\CA$.  But strictly, this is not a simple local operator in the theory \freef.  One way to understand this assertion is that this insertion changes the background far from the point $\CP$.  This can also be seen by noting that the operator $\Psi$ is not single valued in that background and correspondingly its angular modes have integer spin (rather than half-integer spin).  Below we will slightly abuse the notation and will denote the two spin zero monopoles with charges $0$ and $-1$ as $\CM_A$ and $\bar\Psi\CM_A$, as if these were local operators.

\subsec{Outline of the Paper}

In section 2 we present our web of dualities.  Assuming one boson/fermion duality we derive many other dualities including the known boson/boson particle/vortex duality and a fermion/fermion duality.  We discuss some of the properties of these dualities, emphasizing the action of time-reversal, which is often subtle.

In section 3 we give a plausible argument (which falls short of a proof) for the dualities.  We present it first for the particle/vortex duality and then for one of the boson/fermion dualities.

Section 4 is devoted to a detailed analysis of a special $\CT$-invariant fermion/fermion duality.  We explain its subtle $\CT$ reversal symmetry and how it can be used in a topological insulator.  As a check, we match the global symmetries and the operators between the two sides of the duality.

In section 5 we deform that fermion/fermion duality and gap it.  This way we derive the known T-Pfaffian state of a topological insulator.

Section 6 clarifies the relation between these dualities and $S$-duality in $3+1$ dimensions.  Viewing the $2+1$-dimensional theory as living on the boundary of a $3+1$-dimensional space and coupled to a $U(1)$ gauge field in bulk provides a nice context for the $2+1$-dimensional dualities.

In section 7 we describe some applications of our work to condensed matter physics.  In particular, we discuss the role of $\CC\CT$ symmetry in a half filled Landau level.

In Appendix A we discuss some subtleties of spin Chern-Simons theory that are important in our work.  Appendix B describes the almost trivial $U(1)_1$ Chern-Simons theory.  And Appendix C shows how some of the techniques we use clarify some issues in flux attachment.

After completing this work, we became aware of two forthcoming papers \refs{\KarchTong,\MuruganNastase},  which partially overlap with our work.

\newsec{A Web of Dualities}

\subsec{A Free Fermion is Dual to Gauged Wilson-Fisher}

We assume the duality in \Polyakovbf\ (i.e, the second duality in \Oferd) between a free fermion theory coupled to a classical spin$_c$ connection $A$ and a gauged version of the $O(2)$ Wilson-Fisher fixed point also coupled to $A$
\eqn\basicduality{
i \bar \Psi \slashchar{D}_A \Psi \qquad \longleftrightarrow  \qquad |D_b\phi|^2 -|\phi|^4 + {1 \over 4\pi} bdb +{1\over 2\pi} bdA ~.}
We would like to make several comments about this assumed duality.
\item{1.}  The theory in the left hand side is free, while the theory in the right hand side appears interacting.  The assumed duality states that the theory in the right hand side is in fact free and describes a free fermion.
\item{2.} The free fermion theory is $\CT$-invariant with an anomaly, so that if we attach it to a bulk and add to the Lagrangian $+{1\over 8\pi}AdA +\CSg$, it is $\CT$-invariant.  Then it describes the boundary of a topological insulator.  The assumption of the duality \basicduality\ states that the same is true for the right hand side.

\item{3.} The monopole operator $\CM_b$ in the right hand side carries $U(1)_b$ charge $1$ and $U(1)_A$ charge 1 because of the classical Chern-Simons couplings.  This means that $\phi^\dagger\CM_b$ is $U(1)_b$ gauge-invariant.  It has spin $\half $ because of the relative angular momentum between the electrically charged $\phi^\dagger$ and the magnetically charged $\CM_b$,
and we can identify it with the free fermion of the free theory \eqn\fermionid{\Psi=\phi^\dagger\CM_b ~.}

\item{4.} The free fermion theory can be gapped be perturbing it by a $\CT$-violating mass term $m\bar\Psi\Psi$.  Depending on the sign of $m$, the gapped theory has a Chern-Simons contact term for $A$ with coefficient $0$ or $-1$.  In the latter case, integrating out $\Psi$ also generates a gravitational Chern-Simons couplings $-2 \CSg$, where $\CSg$ is described in Appendix B.  So the effective
action for this sign of $m$ is $-{1\over 4\pi}AdA-2\CSg$.   The corresponding operator in the bosonic theory is $m|\phi|^2$ and it leads to a gapped spectrum. For one sign of $m$ the field $\phi$ condenses and Higgses the $U(1)_b$ gauge symmetry making the IR theory completely trivial.  For the other sign, $\phi$ becomes massive but the $U(1)_b$ symmetry remains.  Then the low energy theory is ${1 \over 4\pi} bdb +{1\over 2\pi} bdA={1\over 4\pi}(b+A)d(b+A)-{1\over 4\pi}Ad A$.  As explained in Appendix B, the $U(1)_1$ theory ${1\over 4\pi}(b+A)d(b+A)$ is nearly trivial and can be replaced with $-2 \CSg$.  (We will call
this process ``integrating out $b$.'')
So we reduce to  $-{1\over 4\pi}AdA -2\CSg$ , as in the free fermion theory.\foot{{ This may be interpreted physically by saying that one phase has Hall conductivities $\sigma_{xy} = \kappa_{xy} = 0$ while the other has $\sigma_{xy} = -1, {\kappa_{xy} \over \kappa_0} = -1$.}}  Although the $U(1)_1$ factor is trivial, its couplings change the massive $\phi$ particles from being bosons to fermions.  This is known as flux attachment \refs{\WilczekDU,\JainTX}.  Here we extend this phenomenon to the massless theory.
\item{5.}  Consider the two theories in a background of $A$ corresponding to a monopole.  We discussed it in the free theory above and we saw that there are two such spin zero objects $\CM_A$ and $\bar\Psi\CM_A$ with $U(1)_A$ charges $0$ and $-1$. In terms of the a monopole background in the interacting bosonic theory $\tilde \CM_A$ they are identified as
    \eqn\monoAid{\CM_A =\phi^\dagger\tilde \CM_A\qquad, \qquad \bar \Psi\CM_A = \CM_b^\dagger\tilde\CM_A~.}

\subsec{Time-Reversal Symmetry and Particle/Vortex Duality}

Assuming the duality \basicduality, we can easily derive additional dualities.  First, applying $\CT$ (with $\CT(A)=-A$, $\CT(b)=-\hat b$ and $\CT(\phi)=\hat \phi$) to both sides we find
\eqn\Tbasicduality{
i \bar \Psi \slashchar{D}_A \Psi +{1\over 4\pi} AdA+2\CSg \qquad \longleftrightarrow  \qquad |D_b \hat \phi|^2 -|\hat \phi|^4 -{1 \over 4\pi}\hat b d \hat b -{1\over 2\pi}\hat b dA.}
On the left hand side, we have included $c$-number couplings $AdA/4\pi+2 \CSg$ to the background gauge and gravitational fields ($\CSg$ is a gravitational Chern-Simons coupling,
described in Appendix B).  These couplings are generated by the usual time-reversal or ``parity'' anomaly of the fermion $\Psi$.  Equivalently, we may move these couplings to the right hand side and write
\eqn\Tbasicdualityp{
i \bar \Psi \slashchar{D}_A \Psi\qquad \longleftrightarrow  \qquad |D_{\hat b} \hat \phi|^2 -|\hat \phi|^4 -{1 \over 4\pi} \hat bd\hat b -{1\over 2\pi} \hat bdA -{1\over 4\pi} AdA-2\CSg~.}
 We have added hats to the transformed fields to highlight that they are not the same as the original fields; the relation between the original fields and the transformed fields could be  complicated and nonlocal.

The two dualities \basicduality\ and \Tbasicdualityp\ imply a purely bosonic duality
\eqn\bosonicduality{
|D_b\phi|^2 -|\phi|^4 + {1 \over 4\pi} bdb +{1\over 2\pi} bdA \qquad \longleftrightarrow  \qquad |D_{\hat b} \hat\phi|^2 -|\hat\phi|^4-{1 \over 4\pi} \hat b d\hat b -{1\over 2\pi} \hat bdA -{1\over 4\pi} AdA-2\CSg~.}
This purely bosonic duality can be derived by starting with the particle/vortex duality \particlevortex, adding ${1\over 4\pi} BdB + {1\over 2\pi} BdA$ to both sides ($A$ is a spin$_c$ background field), and turning $B$ into a dynamical field $b$.  This leads to
\eqn\particlevortexb{
|D_b\phi|^2 -|\phi|^4  +{1\over 4\pi} bdb + {1\over 2\pi} bdA \qquad \longleftrightarrow  \qquad |D_{\hat b} \hat\phi|^2 -|\hat\phi|^4+{1\over 2\pi} \hat b db +{1\over 4\pi} bdb + {1\over 2\pi} bdA~.}
The Chern-Simons terms on the right hand side can be written as $ {1\over 4\pi} (b+\hat b + A) d (b+\hat b + A) - {1\over 4\pi} (\hat b+A)d(\hat b+A)$.  The first term is a decoupled trivial $U(1)_1$ sector, which can be replaced by $-2\CSg$.  Then \particlevortexb\ becomes \bosonicduality.  This proves that $\CT$-invariance of \basicduality\ follows from particle/vortex duality \particlevortex.  In particular, the scalar field $\phi$ transforms to its vortex field $\hat \phi$. Since the mass of the boson and the vortex are opposite in sign according to the charge-vortex duality, the boson mass term $m_b|\phi|^2$ breaks time-reversal symmetry. This is consistent since the boson mass term is dual to the Dirac mass term $m_f\bar{\Psi}\Psi$, which also breaks time-reversal symmetry.

Alternatively, this shows that the assumed boson-fermion duality \basicduality\ implies the known particle/vortex duality \particlevortex.  At any rate, landing on our feet in this way gives us more confidence in \basicduality.

\subsec{More Dualities}

Another duality is obtained from \basicduality\ by turning $A$ into a dynamical field.  We write $A= a$ and add ${1\over 2\pi}  a dB-{1\over 4\pi} BdB$, where $B$ is a classical $U(1)$ background gauge field. The duality \basicduality\ becomes\foot{An important fact needs to be clarified here.  The gauge theory of the fermions can be deformed by a fermion mass term $m$.  In many situations the point $m=0$ is preferred by symmetries and then it is natural to set $m$ to zero.  Because of the parity anomaly, this is not the case here.  More precisely, if we give the gauge field $a$ a kinetic term and flow to the IR a fermion mass term can be generated.  We know that the theory is in different phases for $m$ positive and $m$ negative (see also below).  And we will assume that as we vary $m$ these two phases are separated by a second order phase transition.  We define $m=0$ to be that critical point and the Lagrangians that we will write will be shorthand notation for that critical point.}
\eqn\gaugefirst{i \bar \Psi \slashchar{D}_{ a} \Psi +{1\over 2\pi}  a dB -{1\over 4\pi} BdB\qquad \longleftrightarrow  \qquad |D_b \phi|^2 -|\phi|^4 + {1 \over 4\pi} bdb +{1\over 2\pi} bd a + {1\over 2\pi} a dB-{1\over 4\pi} BdB ~. }
We can integrate out $ a$ on the right hand side
using the path integral identity
\eqn\zelfo{\int D a \exp\left({i\over 2\pi}\int a d c\right)=\delta(c),}
where $\delta(c)$ is a properly normalized delta function that sets $c=0$ up to a gauge transformation (here ``properly normalized'' means that with appropriate gauge-fixing, $\int Dc\,\delta(c)=1$).
This often-useful identity, which depends on the fact that $a$ couples precisely via $a d c/2\pi$ for some $U(1)$ gauge field $c$, can be described by saying that $a$ behaves as a Lagrange multiplier
setting $c$ to 0.  It leads in the present instance to a constraint setting $b=-B$, and thus to
\eqn\secondduality{
i \bar \chi \slashchar{D}_a\chi +{1\over 2\pi} adB-{1\over 4\pi} BdB \qquad \longleftrightarrow \qquad |D_{-B}\phi|^2 -|\phi|^4  ~.}
On the left hand side we see a theory which is usually referred to as a single fermion coupled to QED with Chern-Simons level $-1/2$; on the right hand side we find the $O(2)$ Wilson-Fisher  fixed point theory.  Both of them are coupled to a classical $U(1)$ gauge field $B$.  Here we can identify the scalar $\phi$ of the bosonic theory (which has charge $-1$ under $B$) as the operator $\CM_a^\dagger$ in the left hand side.

It is interesting that the right hand side of \secondduality\ is purely bosonic and can be formulated on a non-spin manifold.  If $a$ had been a $U(1)$ gauge field, the left hand side of \secondduality\ could be placed only on a spin manifold with a given spin structure.  This would have led to a contradiction, showing that the duality \secondduality\ could not be true.  However, with $a$ a spin$_c$ connection, as we have assumed, this contradiction does not exist and both sides of the duality can be formulated without a choice of spin structure.

The duality \secondduality, which was the first one in \Oferd, has antecedents in \refs{\ChenCD,\BarkeshliIDA}, as was described in the introduction.  We have extended previous statements to make $a$ a spin$_c$ connection and to add the classical gauge field $B$ with its Chern-Simons contact term.  Below we will see how the time-reversal symmetry that is obvious on the right hand side acts on the left hand side. This has been  mysterious.

Starting with \secondduality, we can add a Chern-Simons contact term ${k\over 4\pi} BdB$ to the two sides and turn $B$ into a dynamical gauge field $b$.  For $k$ even, we couple the new conserved current $db/2\pi$ to a background $U(1)$ gauge field $C$, and for $k$ odd we couple it to a spin$_c$ connection $A$. This way we can find many dualities, most of them  new.

For example, with $k=2$ we have
\eqn\manse{ i \bar \chi \slashchar{D}_a\chi +{1\over 2\pi} adb+{1\over 4\pi} bdb+ {1\over 2\pi } bdC \qquad \longleftrightarrow \qquad |D_{-b} \phi|^2 -|\phi|^4 +{2\over 4\pi} bdb + {1\over 2\pi } bdC ~.}
The left hand side can be written as $ i \bar \chi \slashchar{D}_a\chi +{1\over 4\pi} (b+a+C)d(b+a+C)- {1\over 4\pi } (a+C)d(a+C) $.  The second term is a decoupled $U(1)_{1}$ sector, which can be replaced with $-2\CSg$, and we end up with
\eqn\thirdduality{
i \bar \chi \slashchar{D}_a\chi -{1\over 4\pi} ada- {1\over 2\pi} adC- {1\over 4\pi} CdC -2\CSg \qquad \longleftrightarrow \qquad |D_{-b} \phi|^2 -|\phi|^4  + {2\over 4\pi} bdb +{1\over 2\pi} bdC ~.}
This duality is the third one in \Oferd.  We extended $a$ to be a spin$_c$ connection and we have added the background $U(1)$ gauge field $C$ with its Chern-Simons contact term.

\subsec{A Fermion/Fermion Duality}

Another interesting case is obtained with $k=-1$.  Here \secondduality\ becomes
\eqn\mansek{i \bar \chi \slashchar{D}_a\chi +{1\over 2\pi} adb-{2\over 4\pi} bdb+ {1\over 2\pi } bdA \qquad \longleftrightarrow \qquad |D_{-b} \phi|^2 -|\phi|^4 -{1\over 4\pi} bdb +{1\over 2\pi } bdA ~.}
Here $A$ is a spin$_c$ connection.
Using \Tbasicduality\ in the right hand side we find a duality between two fermionic theories
\eqn\mansekf{
i \bar \chi \slashchar{D}_a\chi +{1\over 2\pi} adb-{2\over 4\pi} bdb+ {1\over 2\pi } bdA \qquad \longleftrightarrow \qquad i \bar \Psi \slashchar{D}_{A} \Psi +{1\over 4\pi} AdA+2\CSg~.}

We can couple \mansekf\ to a bulk and add $-{1\over 8\pi}AdA-\CSg$ to find a $\CT$-invariant theory
\eqn\mansekfb{
i \bar \chi \slashchar{D}_a\chi +{1\over 2\pi} adb-{2\over 4\pi} bdb+{1\over 2\pi } bdA -{1\over 8\pi} AdA -\CSg\qquad \longleftrightarrow \qquad i \bar \Psi \slashchar{D}_A \Psi +{1\over 8\pi} AdA+\CSg~.}
Below we will analyze this duality in detail.  This is a more precise version of the basic fermion-fermion duality proposed in \refs{\WangQMT,\MetlitskiEKA}. It retains many of the same physical features, as we elaborate later.

We have derived all these dualities with the assumption of \basicduality.  Most of the other boson-fermion dualities above could be taken as an equally good starting points.  However, although the duality \basicduality\ implies all these dualities including the known bosonic duality \particlevortex\ and the new fermionic duality \mansekfb, purely boson or fermionic dualities do not imply boss-fermi dualities.
So the assumption \basicduality\ is stronger than \particlevortex\ or \mansekfb.  The latter dualities could be true and the former false.

\subsec{More on Time-Reversal Symmetry}

All the dualities discussed so far can be derived starting from the basic duality \basicduality. Since we showed in the discussion following \Tbasicdualityp\ that time-reversal symmetry is compatible with the basic duality \basicduality, all the other dualities should also be compatible with time-reversal operation. In particular, when one side of a duality is manifestly time-reversal invariant, the other side should also be time-reversal invariant. There are two such examples: the boson/fermion duality \secondduality, and the fermion/fermion duality \mansekfb. Below we illustrate how time-reversal symmetry works out for the boson/fermion duality \secondduality. The basic idea is the same as in \Tbasicdualityp, namely time-reversal transforms the theories to their duals. We discuss time-reversal properties for the fermion/fermion duality separately in later sections.

We first look at the left hand side of \secondduality\ , using the fermion/fermion duality \mansekf:
\eqn\vortexfermiondual{\eqalign{
\CL&= i \bar \chi \slashchar{D}_a\chi +{1\over 2\pi} adB-{1\over 4\pi} BdB  \cr
&\longleftrightarrow\qquad  i\bar{\tilde{\chi}}\slashchar{D}_{\t a}\tilde{\chi}+{1 \over 2\pi}{\t a}dc -{2 \over 4\pi}c dc -{1 \over 4\pi}ada+{1 \over 2\pi}ad(c +B)-{1 \over 4\pi}BdB-2 \CSg  \cr
&\longleftrightarrow\qquad i\bar{\tilde{\chi}}\slashchar{D}_{\t a}\tilde{\chi}+{1 \over 2\pi}{\t a}dc -{1 \over 4\pi}c dc +{1 \over 2\pi}c dB \cr
&\longleftrightarrow\qquad i\bar{\tilde{\chi}}\slashchar{D}_{\t a}\tilde{\chi}+{1 \over 4\pi}{\t a}d{\t a}+{1 \over 2\pi}{\t a}dB+{1 \over 4\pi}BdB+2\CSg,
}}
where the second line comes from applying the fermion/fermion duality \mansekf\ to $ i \bar \chi \slashchar{D}_a\chi $,  the third line comes from eliminating $a$, and the last line comes from eliminating $c$. Each step of elimination uses the equivalence
of $U(1)_{\pm 1}$ to a multiple of the $c$-number coupling $\CSg$ (see Appendix B).

Time-reversal can now be implemented as
\eqn\Tfermionvortex{
\CT: \qquad \chi\leftrightarrow\tilde{\chi}, \qquad a\leftrightarrow \t a, \qquad B\leftrightarrow -B.
}
The action on $\CT$ generates extra terms ${1 \over 4\pi}{\t a}d{\t a}+2\CSg$ associated to the usual ``parity'' anomaly.
A precise formulation is as follows.  Instead of \Tfermionvortex, what we should really say is $\CT: \chi\to \tilde{\chi}'$, where $\chi$ and $\tilde{\chi}$ are regularized through \Ztd\ with phase factor $e^{-{i\pi \over 2}\eta}$ in the partition function, while $\tilde{\chi}'$ is regularized with the conjugate phase factor $e^{{i\pi \over 2}\eta}$. The theory  $i\bar{\tilde{\chi}'}\slashchar{D}_{\t a}\tilde{\chi}'$ with the natural regulator for $\t\chi'$ is
equivalent to the theory
$i\bar{\tilde{\chi}}\slashchar{D}_{\t a}\tilde{\chi}+{1 \over 4\pi}{\t a}d{\t a}+2\CSg$ with the conventional regulator for $\t\chi$.

We conclude that the theory $U(1)_{-\half}$ has a quantum $\CT$ symmetry, which is not visible classically.  This fact trivially follows, if the duality \secondduality\ is true, simply because the right hand side has such a classical symmetry.  The conclusion \Tfermionvortex\ shows that this symmetry does not act simply on the fundamental fields $\chi$ and $a$ but maps them to some dual fields $\tilde \chi$ and $\tilde a$.

Notice that the Dirac mass term is time-reversal invariant here: the mass of the dual fermion $\tilde{\chi}$, according to the duality, is opposite to the original mass of $\chi$. Therefore the mass term, which naively breaks $\CT$, is actually $\CT$-invariant. This is fully consistent with the Dirac mass being dual to the boson mass in Wilson-Fisher theory.

\newsec{A ``Derivation'' of the Dualities}

In this section we will present ``derivations'' of the the boson/boson and boson/fermion dualities.  The reason for the quotation marks is that the derivation is not rigorous.  We will have to make some plausible qualitative assumptions.  Using these assumptions the duality will follow.

\subsec{Deriving the Boson/Boson Particle/Vortex Duality}

We begin with a discussion of the standard bosonic particle-vortex duality from a standpoint that enables easy passage to the boson-fermion duality in the next subsection.

We start with the Lagrangian
\eqn\startL{|D_{b}\phi|^2 +|D_{\hat b}\hat\phi|^2 -V(|\phi|,|\hat \phi|) +{1\over 2\pi} bd\hat b +{1\over 2\pi }b dB}
where $B$ is a classical field.

This model has two global $U(1)$ symmetries associated with conservation of $db$ and $d\hat b$. However we will soon break the conservation of $d\hat b$ explicitly by adding a monopole operator to the Lagrangian.

\midinsert\bigskip{\vbox{{\epsfxsize=6in
       \nobreak
    \centerline{\epsfbox{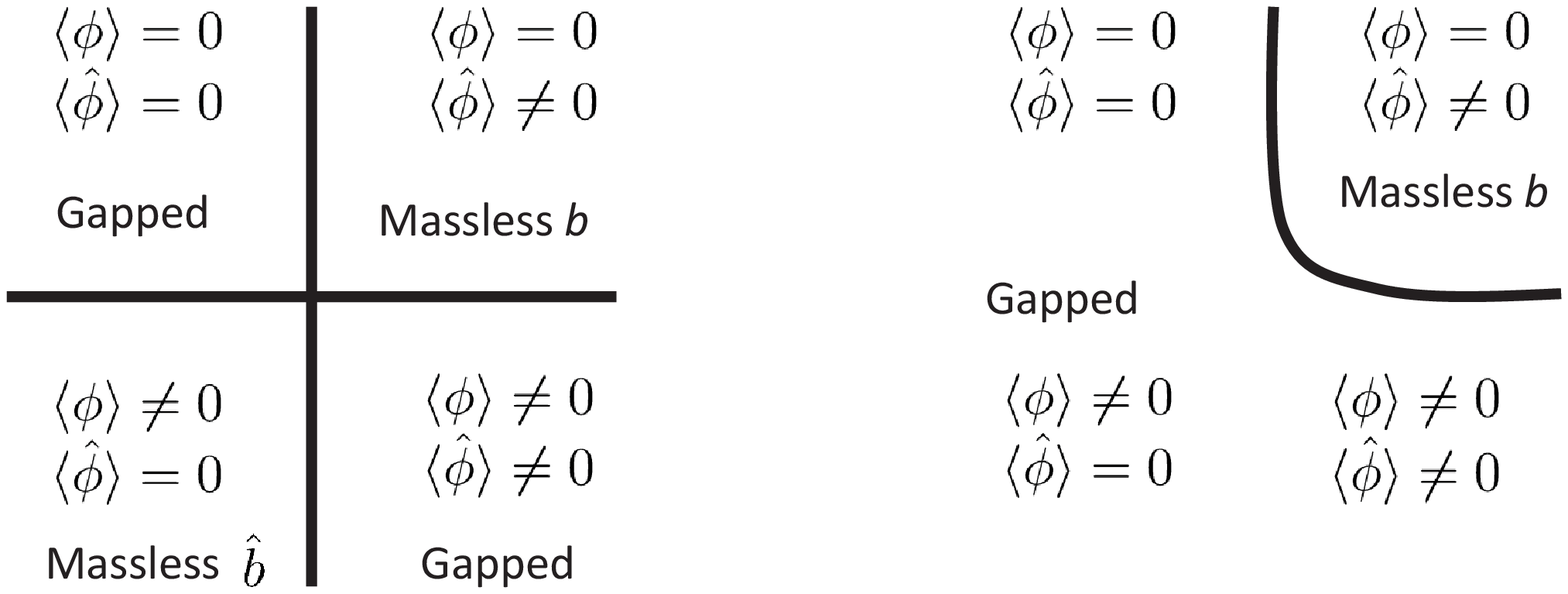}}
       \nobreak\bigskip
    {\raggedright\it \vbox{
{\bf Figure 1.}The left panel represents the four phases of \startL.  The right panel is a suggestion about the behavior when a monopole operator of $\hat b$ is added to the Lagrangian. }}}}}
\bigskip\endinsert

This theory has four phases depending on whether $b$ or $\hat b$ are Higgsed (see Figure 1). These are as follows.

In the phase with $\langle\phi\rangle , \langle\hat \phi\rangle  \not =0$, the two $U(1)$ gauge symmetries are Higgsed and the spectrum is gapped.

In the phase with $\langle\phi\rangle = \langle\hat \phi\rangle =0$, the two $U(1)$ gauge symmetries are not Higgsed.  $\phi$ and $\hat \phi$ are massive and can be integrated out.  The two gauge fields couple through $ {1\over 2\pi} bd\hat b +{1\over 2\pi }db B$ making the spectrum gapped and the low energy TQFT is trivial.

In the phase with $\langle\phi\rangle =0$ with $ \langle\hat \phi\rangle  \not =0$, the gauge symmetry $U(1)_{\hat b}$ is Higgsed and $U(1)_b$ is not Higgsed.  Then the low energy spectrum includes a massless boson, which is the dual of $b$.  It couples to $B$.

The fourth phase is obtained from the previous one by exchanging the hatted and the un-hatted fields.

Next we explicitly break the global symmetry whose current is $d\hat b$.  (Since we planned to do that we did not include in \startL\ a coupling $ d\hat b \hat B $.)  The monopole operator $\CM_{\hat b}$ carries charge $1$ under the gauged $U(1)_b$ and therefore we add to the Lagrangian the operator
\eqn\addedmon{\CO=\phi^\dagger \CM_{\hat b}~.}

How does this affect the four phases?  The two gapped phases are not changed.  The phase with massless $b$ is also not affected.  In that phase the massless $b$ is the Nambu-Goldstone boson of the spontaneously broken global symmetry whose current is $db$.  The only difference is in the phase with $\langle\phi\rangle \not=0$ with $ \langle\hat \phi\rangle  =0$.  Here the massless $\hat b$ boson is lifted by the operator \addedmon.
We see that we are left with two phases.  One of them has a massless boson and the other is gapped.

We suggest that the phase diagram is simply the right panel in Figure 1 with a single smooth transition line between two phases and with no additional structure.

Let us describe the phase transition between the two phases.  For $\langle \hat \phi\rangle \to \infty$, we can integrate out $\hat \phi$ and $\hat b$. After tuning the potential $V$, we describe the critical point by
\eqn\Lone{|D_b\phi|^2 -|\phi|^4+ {1\over 2\pi} b dB.}  We recognize here a gauged version of the Wilson-Fisher fixed point coupled to a classical gauge field $B$. The other limit has $\langle \phi \rangle =0$ and $\phi$ is very massive.  Integrateng  out $\phi$  and again tuning the potential, we find
\eqn\Ltwot{  |D_{\hat b}\hat \phi|^2-|\hat\phi|^4+ {1\over 2\pi} b d(\hat b+B)~.}
Integrating out $b$ via \zelfo, we end up with
\eqn\Ltwo{ |D_{-B}\hat \phi|^2 -|\hat\phi|^4~.}
We recognize this as as the Wilson-Fisher fixed point coupled to a classical gauge field $B$.

We conclude that \Lone\ and \Ltwo\ are two limits of the same transition, thus establishing the particle-vortex duality between them.

\subsec{Deriving the Boson/Fermion Duality}

Next we similarly study the duality between the free Dirac theory, and the massless boson theory coupled to a level $1$ Chern-Simons gauge field \basicduality\ or its $\CT$-image \Tbasicdualityp.

Consider a field theory\foot{ For condensed matter physicists, this may be motivated as follows.
Consider electrons in a 2d system undergoing an ``integer quantum Hall" phase transition from a trivial insulator ($\sigma_{xy} = 0)$ to an integer quantum Hall state (${\sigma_{xy} = -1})$. It may be useful to keep in mind lattice models of free fermions where we go from a filled band with Chern number $0$ to one with Chern number $-1$. This phase transition is captured by a continuum field theory model of a single massive Dirac fermion $\psi$. The transition occurs when this Dirac mass passes through zero.
 Note that weak short ranged interactions added to this system do not change the universal low energy physics of the transition or the two phases.

 We introduce a ``parton" description of the electron operator by writing $\psi = \hat{\phi}^\dagger \chi $. This representation has a $U(1)$ gauge redundancy, and a reformulation in terms of $\chi$ and $\hat{\phi}$ comes with a $U(1)$ gauge field $\hat{b}$. Both $\chi$ and $\hat{\phi}$ carry charge $1$ under $U_{\hat{b}}(1)$ (so that the electron $\psi$ is gauge-invariant). On a lattice
 we assume that the $\chi$ has the same structure of hopping matrix elements as the original electrons. The $\hat{\phi}$ will be described by a model of bosons hopping on the lattice with some short ranged repulsion (a bose Hubbard model).  Note that in this construction we must include monopoles of $\hat{b}$.
  In a suitable continuum limit we will then get the field theory we study. }  with
 \eqn\Lprtn{{\cal L} = i\overline{\chi}\slashchar{D} _{\hat{b}+A} \chi + m \overline{\chi} \chi + |D_{\hat{b}} \hat{\phi}|^2 -  V( |\hat \phi|^2 )}
As written, this theory has two global $U(1)$ symmetries: one is simply $U_A(1)$, and the other is associated with the conservation of $d\hat b$.  As in the previous subsection, soon we will explicitly break the latter symmetry by including a monopole operator in the Lagrangian.

\midinsert\bigskip{\vbox{{\epsfxsize=6in
       \nobreak
    \centerline{\epsfbox{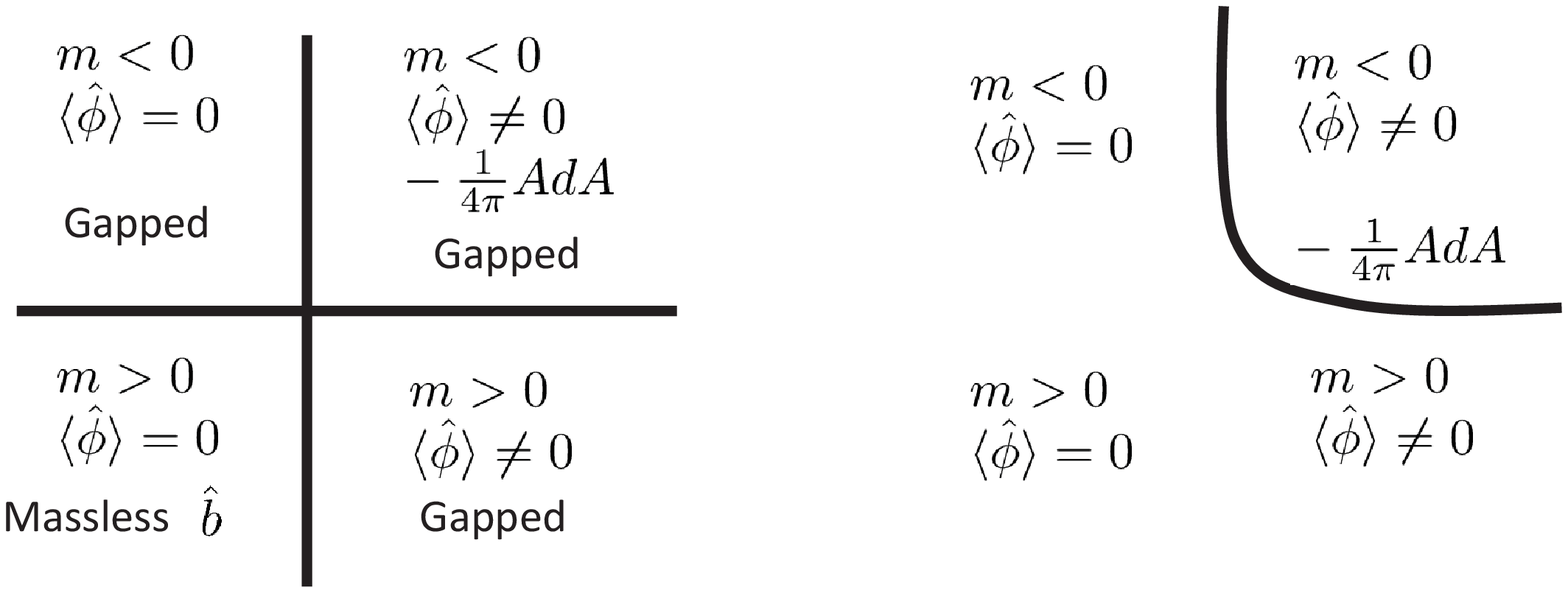}}
       \nobreak\bigskip
    {\raggedright\it \vbox{
{\bf Figure 2.} The left panel represents the four phases of \Lprtn.  The right panel is a suggestion about the behavior when a monopole operator of $\hat b$ is added to the Lagrangian.  In the phases with a Chern-Simons term we suppressed the gravitational Chern-Simons term.}}}}}
\bigskip\endinsert

The fermionic operator $\psi = \hat{\phi}^\dagger \chi $ is invariant under the gauge $U_{\hat{b}}(1)$ symmetry and carries charge $1$ under the global $U_A(1)$ symmetry.  We will refer to this operator as the electron.

Let us study the phases and phase transitions of this continuum theory and start without the added monopole operator.  Clearly, the system has four phases corresponding to the two values of $\sign(m)$ and to whether $\langle \hat{\phi} \rangle = 0$ or $\langle \hat{\phi} \rangle\neq 0$.  The phases are depicted in the left panel of Figure 2.

When $\langle \hat{\phi} \rangle \neq 0$, the gauge field $\hat{b}$ is Higgsed and can be integrated out.  Then we can identify $\chi$ with the electron $\psi$. (Note that $\psi \sim \langle \hat{\phi}^\dagger \rangle \chi$.)  As $m$ changes sign, we go from a trivial phase with no induced Chern-Simons terms to a different phase,  where there is an induced term $-{1 \over 4\pi} AdA - 2\CSg$.  In condensed matter physics, this is an integer quantum Hall phase transition of non-interacting electrons.  The critical point is described by the free massless Dirac fermion theory:
 \eqn\LfreeD{{\cal L} = i\overline{\psi} \slashchar{D}_{A} \psi}

What about the phases with $\langle \hat{\phi} \rangle = 0$?  Here $\hat b$ is not Higgsed. Since $\hat \phi$ is massive we can integrate it out.
When $m >0$,  we integrate out $\chi$ and there is no Chern-Simons term induced.  We are left with a theory with a massless gauge field $\hat b$, which can be dualized to a massless scalar.  It can be interpreted as the Nambu-Goldstone boson of the spontaneously broken global symmetry, whose current is $d\hat b$.

When the mass of $\chi$ changes sign to $m < 0$, the integral of $\chi$ induces the  Chern-Simons terms
 \eqn\cb{   - {1 \over 4\pi} (A + \hat b) d(A + \hat b)  - 2\CSg~,}
giving a mass to $\hat b$.  According to the analysis in Appendix B, the  $U(1)_{-1}$ path integral cancels the $-2\CSg$, and we end up with a trivial gapped phase.\foot{Since we are planning to add to the Lagrangian a monopole operator of $\hat b$, we did not include in \Lprtn\ a term of the form ${1\over 2\pi} \hat B d\hat b$.  But since so far we have not yet added this monopole operator, we could ask how such a term would have changed our conclusions.  In that case \cb\ would have been replaced by
\eqn\cbr{ - {1 \over 4\pi} (A + \hat b-\hat B) d(A + \hat b-\hat B) +{1\over 4\pi} \hat B d\hat B  -{1\over 2\pi} \hat B dA - 2\CSg~.}
This leads effectively to
\eqn\cbrr{ {1\over 4\pi} \hat B d\hat B  -{1\over 2\pi} \hat B dA ~}
and therefore in the presence of $\hat B$ this phase is not completely trivial.}

These four phases are depicted in the left panel of Figure 2.

Next, we add a monopole operator of $\hat b$ to the Lagrangian.  As in the previous subsection, the three gapped phases are not modified.  But since this operator explicitly breaks the global symmetry whose current is $d\hat b$, the massless $\hat b$ Nambu-Goldstone boson acquires a mass.  It is reasonable that the system has only two phases, as depicted in the right panel of Figure 2.
In  condensed matter terminology, these two phases are the trivial insulator and the quantum Hall insulator.

When $m$ is small and $\langle \hat\phi\rangle $ is large the transition is described by a massless fermion and the Lagrangian \LfreeD.  The other limiting case is $m$ large and negative and $\langle \hat \phi \rangle $ small.  There the transition can be described by
\eqn\cbt{|D_{\hat b}\hat\phi|^2 -|\phi|^4   - {1 \over 4\pi} (A + \hat b) d(A + \hat b)  - 2\CSg~.}
Assuming there is a unique fixed point controlling this transition, we conclude that the critical point of \cbt\ is described by the free massless Dirac theory.  This establishes the duality \Tbasicdualityp.  As in section 2, this also leads to its $\CT$-image duality \basicduality.

\newsec{Properties of the Fermion/Fermion Duality}

In this section we discuss in detail the fermion/fermion duality \mansekfb\
\eqn\mansekfbf{
i \bar \chi \slashchar{D}_a\chi +{1\over 2\pi} adb-{2\over 4\pi} bdb+ {1\over 2\pi } bdA -{1\over 8\pi} AdA -\CSg \qquad \longleftrightarrow \qquad i \bar \Psi \slashchar{D}_A \Psi +{1\over 8\pi} AdA +\CSg~,}
where we include the bulk term ${1\over 8\pi} AdA+\CSg$ to make the theory $\CT$-invariant.  This invariance is manifest in the free fermion side, but it is not manifest in the interacting fermion side.  Below we will demonstrate that the left hand side is indeed $\CT$-invariant, we will present more checks of this duality, and  we will show how to use the theory on the left hand side of the duality to construct the T-Pfaffian state.

\subsec{Using the $U(1)_2\times U(1)_{-1}$ theory}

Around \Zfd, we discussed how to turn the $2+1d$ free fermion theory to a $\CT$-invariant theory.  Here we will present another way to do it. The known semion/fermion theory \FidkowskiJUA\ can be made explicitly $\CT$-invariant with an anomaly by writing it as a $U(1)_2\times U(1)_{-1}$ theory \SeibergRSG
\eqn\twoone{{2\over 4\pi} bdb - {1\over4\pi} b'db' -{1\over 2\pi} A db' - {1\over 2\pi}(A+A') db ~, }
where $b$ and $b'$ are dynamical $U(1)$ gauge fields, and $A$ and $A'$ are classical spin$_c$ connections.  (We could have used $B=A+A'$, which is a classical $U(1)$ gauge field.)

Consider the $\CT$ transformation
\eqn\twooneTg{\eqalign{
&\CT(A)=-A \cr
&\CT(A' ) = A' \cr
&\CT(b)=b-b' -2 A \cr
&\CT(b')=-b'+2b - A' - A
~,
}}
which is consistent with the spin$_c$ property of $A$ and $A'$.  On-shell this transformation squares to one.  Under this transformation the theory is invariant up to a shift by the ``anomaly''
\eqn\anomaly{ {1\over 4\pi}A' d A' - {1\over 4\pi} A d A  ~.}
As always, the anomaly is well defined as a $2+1d$ term, but it cannot be described as the transformation of a $2+1d$ term.

Let us go back to a theory of fermions $\chi$ coupled to some $A'$ with the usual action $i\bar\chi\slashchar{D}_{A'}\chi$.  Our conventions are that the time-reversal
anomaly in this theory is $+{1\over 4\pi} A' dA'+2\CSg$ (see \Zfd).    So the anomaly can be canceled by subtracting \twoone\ from the fermion theory and adding the standard bulk term $+{1\over 8\pi}AdA +\CSg$.
Next, we make $A'$ dynamical and denote it by $a$.  This leads to the $\CT$-invariant Lagrangian
\eqn\dualtheoryt{
i\bar \chi \slashchar{D} _{a}\chi - {2\over 4\pi} bdb + {1\over 4\pi} (b'+A)d(b'+A) + {1\over 2\pi}a db + {1\over 2\pi}A db-{1\over 8\pi} AdA+\CSg
~.}
The term ${1\over 4\pi} (b'+A)d(b'+A)$ can be replaced with $-2\CSg$ according to Appendix B, and we end up with
\eqn\dualtheory{
\CL=i\bar \chi \slashchar{D} _{a}\chi - {2\over 4\pi} bdb + {1\over 2\pi}a db + {1\over 2\pi}A db-{1\over 8\pi} AdA-\CSg
~.}
We recognize it as the left hand side of the fermion/fermion duality \mansekfb.  The discussion above shows that this theory is $\CT$-invariant,  just like the free fermion theory in the right hand side of \mansekfb.
The equation of motion of $b$ in the theory based on $\CL$ of \dualtheory\ is
\eqn\ateom{2db=da +dA~.}
If we naively integrate out $b$ by using this equation of motion to set $b=(a+A)/2$ and suppress the gravitational term, we find
\eqn\dualtheoryen{\CL_{Naive}= i\bar \chi \slashchar{D} _{a}\chi+{1\over  4\pi}a dA +{1\over 8\pi} a da ~.}
 Taking into account the meaning of the fermion path integral used in this paper, this is essentially  the dual Dirac theory appearing in
 \refs{\SonXQA,\WangQMT,\MetlitskiEKA\WangFQL,\GeraedtsPVA\MulliganGLM-\WangGQJ}. However this process of eliminating $b$ is not strictly valid
as it does not take into account the fact that the gauge fields can have nontrivial fluxes satisfying Dirac quantization.  In  particular, in general $(a+A)/2$ does not satisfy Dirac quantization
and therefore we cannot set $b=(a+A)/2$.
 Indeed, as detailed in Appendix C, exactly the same issue arises in the relationship of the standard flux attachment theories (such as HLR)  of quantum Hall  systems with a more precise version that does not involve fractional coefficients of Chern-Simons terms.  For the present theory likewise the attempt to integrate out $b$ leads to  the troubling fractional coefficients of the two Chern-Simons terms in \dualtheoryen.   So here we will not integrate $b$ and will work with \dualtheory. Nevertheless we will see that this more precise formulation retains many of the physical features that follow formally from \dualtheoryen.

\subsec{Matching the Symmetries and the Operators in the Fermion/Fermion Duality}

Here we will study the symmetries and operators in the theory \dualtheory\ and will compare them with those of the free theory \freefb.

The theory \dualtheory\ has three conserved currents, the fermion bilinear $\chi^\dagger\chi$ and the two monopole currents $da$ and $db$.  Two linear combinations of these are gauged because they couple to $a$ or $b$.  This leaves us with a single global $U(1)$ symmetry, which couples to $A$. We will denote it as $U(1)_A$.
This conclusion about the global symmetries matches with the free theory \freefb, which also a single global $U(1)_A$ symmetry coupled to $A$.

We see that the theory \dualtheory\ does not have well defined monopole number symmetry. In order to explore this issue in more detail, let us attempt to construct monopole operators for $a$ and $b$ and denote them $\CM_a$ and $\CM_{b}$.  These objects are not gauge-invariant.  The Chern-Simons terms in \dualtheory\ lead to the gauge $U(1)_a\times U(1)_{b}$ charges and to global $U(1)_A$ charges
\eqn\CMgauge{\matrix{ & U(1)_a &U(1)_{b} & U(1)_A\cr
\CM_a &0&1 &0\cr
\CM_{b} & 1 &-2 &1\cr
\chi &1&0&0}}
where for completeness we added also $\chi$.  From these we can form gauge-invariant local operators
\eqn\gaum{\CM_a^{2n} \CM_{b}^n\bar\chi^n ,}
where in $\bar\chi^n$ we mean an operator with $n$ fermions and some derivatives, which are necessary because of Fermi statistics.  Note that the equation of motion \ateom\ implies that the $a$ flux is twice the $b$ flux, which is consistent with \gaum.  This is not surprising because the $b $ equation of motion guarantees invariance under $U(1)_{b}$.  We see that the local operators must have an even  monopole charge
 of $a$.  The simplest such operator has $n=1$
\eqn\simplesm{\Psi=\CM_a^2 \CM_{b}\bar\chi ~.}
Its $U(1)_A$ charge is $+1$.  We identify it with the free fermion $\Psi$ of \freefb.

We can also make operators gauge-invariant by attaching them to open Wilson lines.  Using the charges \CMgauge, our building blocks are
\eqn\opewl{\eqalign{
\CM_a e^{i\int b} \cr
\CM_{b} e^{i \int (a -2 b)} \cr
\chi e^{i\int a} ~.}}
And we can construct other objects using polynomials in these and their conjugates.

Let us understand these objects in more detail, starting with $\CM_a e^{i\int b} $.  An $S^2$ that surrounds $\CM_a$  is pierced by the open line at a point $p$.  The equations of motion of $a$ and $b$ show that $db=0$ and that
the total flux $\int_{S^2} da =2\pi $ must be localized as a delta function at $p$.  This monopole has a single $\chi$ and a single $\bar \chi$ zero-mode both localized around $p$, because that
is where the flux is. Their quantization leads to to two states with $U(1)_a$ charges $0$ and $-1$.  They correspond to
\eqn\monoone{\CM_a e^{i\int b}   \qquad, \qquad \bar \chi \CM_a e^{i\int (b-a)} ~.}
We could also multiply these by powers of $\CM_{b}e^{i\int (a -2 b)}$, or add fermions by multiplying by $\chi e^{i\int a}$, or their conjugates.

Next we move to monopole number two and construct such operators using $\CM_a^2$.  Here there are two $\chi$ zero-modes and two $\bar \chi$ zero-modes leading to states with $U(1)_a$ charges $q$, $q-1$, $q-2$ for some $q$.  The value of $q$ is determined by the background $b$ flux.  First, as with the single monopole \monoone, we can simply attach a Wilson line, leaving the $b$ flux to vanish. This leads to $q=0$ and the operators
\eqn\monotwo{\CM_a^2 e^{2i\int b}   \qquad, \qquad \bar\chi \CM_a^2 e^{i\int (2b -  a)} \qquad, \qquad \bar\chi^2 \CM_a^2 e^{2i\int (b -  a)} ~.}
These operators are products of the ones in  \monoone.
We could also change the background $b$ flux by multiplying the operator by powers of $\CM_{b}e^{i\int( a -2 b)}$ or add fermions by multiplying by $\chi e^{i\int a}$, or their conjugates.  A special case is the local operator \simplesm.

Comparing with the free fermion \freefb, both theories have only a single global symmetry $U(1)_A$ and the gauge-invariant local operators match \simplesm.  As we have just discussed, the interacting theory also has operators that are ends of lines.  But since they cannot be characterized by quantum numbers, their identification in a dual description is difficult.  In the dual free theory \freefb, they could correspond to complicated line operators constructed out of the free fermion.

Just as in the free theory, we can also consider a monopole of $A$.  We denote such an insertion by $\tilde \CM_A$ to distinguish it from the monopole insertion $\CM_A$ of the free fermion theory, which will be identified below.  Because of the term ${1 \over 2\pi} Adb$ in \dualtheory, $\tilde\CM_A$ is not gauge-invariant and carries $U(1)_{b}$ charge.  This can be canceled by attaching a line.  Another, more interesting possibility is to make $\tilde \CM_A$ gauge-invariant by multiplying it with $\CM_a^\dagger$ or $ \bar \chi  \CM_a \CM_{b}$.  This gives objects that we can identify with the monopole insertions of the free theory:
\eqn\singlmon{\CM_A=\bar \chi\tilde\CM_A \CM_a  \CM_{b}(\CP) \qquad, \qquad  \bar\Psi\CM_A=\tilde \CM_A\CM_a^\dagger(\CP).}

\subsec{Time-Reversal Properties of Monopole Operators}

We now study how monopole operators transform under time-reversal. We focus on two operators: the operator $\CM_a$ and the local fermion $\CM_a^2\CM_b\bar{\chi}$.

Let us first consider $\CM_a$. In the previous version of the Dirac duality \dualtheoryen\ discussed in the literature, a manifestation of the  fact that the path integral is not really gauge-invariant (because
the Chern-Simons couplings are not properly quantized) is that the monopole operator $\CM_a$ cannot be made gauge-invariant.
Since time-reversal acts as $\CT: \CM_a\to \bar{\chi}\CM_a$, we have no choice but to assign $U(1)_a$ gauge charge $\pm1/2$ to the two operators obtained by quantizing zero-modes.

It is instructive to see how this issue is cured in the more properly defined theory \dualtheoryt. Here the gauge-invariant operator is $\CM_ae^{i\int b}$. The crucial point is that the flux configuration of $\CM_a$ is not time-reversal invariant. Rather, according to \twooneTg\ (setting $A=0$ for simplicity), the fluxes should transform as
\eqn\Tflux{\eqalign{
\int d a=2\pi \qquad &\to \qquad \int d a=2\pi \cr
\int db'=0 \qquad &\to \qquad \int d (-b'+2b-a)=0 \cr
\int db=0 \qquad &\to \qquad \int d (b-b')=0,
}}
for which the solution is simply $\int d a=\int d b=\int d b'=2\pi$. Therefore the monopole operator $\CM_ae^{i\int b}$ should transform to
\eqn\Tmonopole{
\chi\CM_a\CM_{b'}\CM_{b}e^{i\int(-b+b') }=\left(\CM_a e^{i\int b}\right)\left(\CM_{b'} e^{i\int b'}\right)\left(\chi\CM_{b} e^{-i\int 2b}\right).}

All the three factors in the above line are separately gauge-invariant (notice that $\CM_{b}$ carries $U(1)_a$ charge, which is canceled by $\chi$). The second and third factor both carry spin-$1/2$, so the total spin is still $0$ which is consistent. Therefore unlike for the old theory in \dualtheoryen, gauge-invariance is fully compatible with time-reversal symmetry on the monopole operators in the properly defined theory.

We now consider the local fermion operator $\CM_a^2\CM_b\bar{\chi}$. Using the time-reversal transformation \twooneTg, it is straightforward to see that the flux configuration associated with $\CM_a^2\CM_b$ is time-reversal invariant. In fact, for local operators such as the physical fermion, it suffices to use a simplified version of the time-reversal transformation: we can take the on-shell condition for $b'$ (namely $b'=-A$) in \twooneTg, to get
\eqn\Tsimplified{\eqalign{
\CT(A)&=-A \cr
\CT(a)&=a \cr
\CT(b)&=b-A.
}}
The flux configuration of $\CM_a^2\CM_b\bar{\chi}$ is obviously invariant under this $\CT$-transformation, since the $A$-flux vanishes.

There are two fermion zero-modes $\chi_1,\chi_2$ associated with this flux configuration. Since time-reversal $\CT$ acts like $\CC\CT$ on $\chi$, we have
\eqn\TdoubleM{
\CT: \CM_a^2\CM_b \to \bar{\chi}_1\bar{\chi}_2\CM_a^2\CM_b.
}

Therefore $\CT$ acts on the local fermion as
\eqn\Tlocalf{
\CT: \bar{\chi}_1\CM_a^2\CM_b \to \bar{\chi}_2\CM_a^2\CM_b \to -\bar{\chi}_1\CM_a^2\CM_b,
}
which gives $\CT^2=(-1)^F$, fully consistent with the free fermion theory.

\newsec{Gapping the Interacting Fermion Theory}

In this subsection we will add scalars to the interacting fermion theory \dualtheory\ and consider a gapped phase of the system. (In the condensed matter literature this is discussed as the condensation of a Cooper pair formed out of the dual fermions. As we said above, we prefer to describe it by adding an elementary scalar field.)  It was suggested in \refs{\WangQMT,\MetlitskiEKA}   in the original version of the dual theory \dualD, that this leads to the T-Pfaffian state.  We have emphasized above that especially when discussing topological aspects of gapped phases, one has to be careful with global considerations and in particular, it is crucial to use a description of the theory with properly normalized Chern-Simons terms.  Here we will work this out with the refined version \dualtheory\ and we will show that, indeed, the T-Pfaffian state can be obtained this way.

Although the theory \dualtheory\ is dual to a free fermion theory, the gapped theory that will emerge from our manipulations is not dual to something that we could
have constructed starting with the free fermion theory.  By adding scalars in the UV of the interacting fermion theory and by changing parameters in the UV theory to a phase with expectation values of these scalars, we go beyond the range of validity of the IR duality implied by \mansekfbf. No obvious manipulation on the free fermion
theory will match what we will do in the dual theory.   However, as in \SeibergRSG, if we add the scalars to the dual theory in a $\CT$-invariant way and their expectation value does not break that symmetry, we are assured that the gapped theory that we derive is $\CT$-invariant with exactly the same bulk interactions as the free fermion theory.

There are various constraints on scalars that might be included.
Using $\CT(b)=b-A$ (eqn.\ \Tsimplified), it is easy to see that if we want a scalar field to transform homogeneously under $\CT$ as defined in \twooneTg\ or \Tsimplified, and to preserve the spin/charge relation, it cannot couple to $b$.  Whether a scalar can couple to $a$ or to $A$ depends on how it transforms under $\CT$.  Even without $\CT$,
the sum of the $a$ and $A$ charges of a scalar must be even, as $a$ and $A$ are $\spinc$ connections.  For our purposes,  the
 simplest possibility is to introduce a scalar $\Phi$ with $U(1)_a$ charge $-2$, which can couple as $\Phi \chi\chi$.  In order to preserve $\CT$-invariance, $\Phi$
  should transform as $\CT(\Phi)=-\bar \Phi$ (in
terms of real components $\Phi=(\phi_1+i\phi_2)/\sqrt 2$, this means that  $\CT\phi_k=-\phi_k,~k=1,2$). The expectation value of $\Phi$
 gives mass to the fermions and spontaneously breaks $U(1)_a \to {\Z}_2$.  More precisely, the nonzero expectation value of $\Phi$ breaks $U(1)_a$ and breaks $\CT$, but it preserves a $\Z_2 \subset U(1)_a$ gauge symmetry and another time-reversal transformation $\CT'$, which is a combination of the original $\CT$ and a broken gauge transformation.  Then, the low-energy theory is gapped and topologically non-trivial.

As in \SeibergRSG, we describe first the Abelian sector of the theory.
To describe the symmetry-breaking from $U(1)_a$ to $\Z_2$, we introduce a new $U(1)$ gauge field $c$, dual to the phase of $\Phi$, with a coupling $2c da/2\pi$.
Adding this to
 the Chern-Simons couplings that were present already in \dualtheory, we have
\eqn\dualtheorye{
\CL_{Abelian}= {2\over 2\pi} cda- {2\over 4\pi} bdb + {1\over 2\pi}a db+ {1\over 2\pi}A db -{1\over 8\pi} AdA-\CSg .}
Note that we do not include a term of the form ${1\over 8\pi} ada$ from integrating out the fermions.
(The $a$ dependence of the fermion path integral goes into the construction of the Ising sector
that we mention momentarily.)   In eqn.\ \dualtheorye, we recognize $a$ as a Lagrange multiplier setting $b  \sim -2c$ (up to a gauge transformation).  Eliminating $a$
and imposing the constraint, we get
\eqn\dualtheoryep{\CL_{Abelian\ eff.}=-{8\over 4\pi} c d c-{2\over 2\pi} cd A  -{1\over 8\pi} AdA -\CSg~.}
This is a $U(1)_{-8}$ theory of $c$ coupled to $A$.  As in \SeibergRSG, we add to this the Ising sector with its three lines $1$, $W_\psi$, $W_\sigma$. Performing the ${\Z}_2$ quotient, we find the T-Pfaffian theory.

We can go back to the UV theory \dualtheory\ and, as in \SeibergRSG, identify there the quasi-particles in the gapped phase.  The expectation value of $\Phi$ breaks $U(1)_a\to {\Z}_2$ and leads to vortices as follows.
\item{$v=\pm1$} The basic vortex has a single fermion zero-mode.  It has spin $0$ and electric charge $\pm{1\over 4}$.  It is represented in the topological theory by $W_\sigma e^{\pm i\oint c}$.
\item{$v=\pm 2$} Two vortices are a monopole operator of $a$.  As we discussed above, this cannot be a local gauge-invariant operator.  Instead, it leads to a quasi-particle.  The $\chi$ zero-modes lead to states with spins $\pm{1\over 4}{\rm mod}  1$ and  electric charges $\pm \half$.  They are represented in the topological theory by the lines $e^{\pm 2i\oint c} $ and $W_\psi e^{\pm 2i\oint c} $.
\item{$v=\pm 3$} Vorticity $3$ leads to  quasiparticles with spin $\half\ \mod\  1$ and electric charge $\pm{3\over 4}$.  They are represented by $W_\sigma e^{\pm 3i\oint c}$.
\item{$v=4$} Finally, the vorticity $4$ states are related to $\CM_a^2$ we discussed above. They are $e^{ 4i\oint c} $ and $W_\psi e^{4i\oint c} $.  The first has spin $0~{\rm mod} ~ 1$ and charge $1$ and its correlation functions are nontrivial.  The second is a transparent line, which can be
interpreted as the world line of the free fermion $\Psi$ of the T-Pfaffian theory, which in turn is the fermion of the dual free fermion theory.  In the context of a topological insulator, this is the underlying electron.

\newsec{Relation to $S$-Duality in the Bulk}

The goal of this section is to present a four-dimensional interpretation of the dualities that have been described in this paper.  For every three-dimensional
duality that we have discussed, we will describe a four-dimensional duality from which it follows.  We should stress that this will not be a proof of the dualities.

Concretely, a $3+1$ dimensional theory of a free $U(1)$ gauge field exhibits $S$-duality.  It can be derived using straightforward manipulations in the functional integral, which can be performed even when the four-dimensional manifold has a boundary.  Below we will review this subject.  Then we will show that these known $S$-duality transformations can be naturally combined with our assumed $2+1$ dimensional dualities to lead to a consistent picture.  This will allow us to derive the $2+1$ dimensional dualities from a new assumption about the $3+1$ dimensional theory.

\subsec{Review Of Electric-Magnetic Duality}

First we recall basic facts about electric-magnetic duality in bulk for free $U(1)$ gauge theory. We consider a $U(1)$ gauge field $B$ with field strength
$F_{\mu\nu}=\partial_\mu B_\nu-\partial_\nu B_\mu$.   The electromagnetic action in Lorentzian signature is
\eqn\thaction{I_\em=\int d^4x\sqrt{-g}\left(-{1\over 4e^2}F_{\mu\nu}F^{\mu\nu}+{\theta\over 32\pi^2}\epsilon^{\mu\nu\lambda\rho}F_{\mu\nu}F_{\lambda\rho} \right).}
It is convenient to combine the gauge coupling $e$ and theta-angle $\theta$ to a complex coupling parameter
\eqn\deftau{\tau={\theta\over 2\pi}+{2\pi i\over  e^2}}
that takes values in the upper half plane.  Then, in terms of $F^{\pm}_{mn}={1\over 2}\left(F_{mn}\pm {i\over 2} \epsilon_{mnpq}F^{pq}\right)$, the action is
\eqn\naction{I_\em(B;\tau)=-{i\over 8\pi}\int d^4x\sqrt{-g}\left(\bar\tau \,F^+_{mn}F^{+mn}   -\tau F^-_{mn}F^{-mn}        \right). }

The basic electric-magnetic duality transformation starts with the action \naction\ with $F$ expressed in terms of the gauge field $B$.  Using differential form notation, we add to the action
\eqn\addedterm{{1\over 2\pi} \int   F' \wedge(F-dB)~,}
where $F$ and $F'$ are arbitrary two-forms.  Clearly, this does not affect the theory, since a path integral over $F'$ will just give a constraint setting $F = dB $. Because of
this fact, once
\addedterm\ is added, we can treat $F$ in the original action \naction\
 as an independent variable, forgetting its original definition as $dB$. To get something non-trivial, we instead integrate first over $F$. Since the action is quadratic in $F$, the integral over $F$ is  Gaussian.  It leads to
\eqn\nactiontt{-{i\over 8\pi}\int \left(\bar\tau' \,F^{'+}  F^{'+}   -\tau' F^{'-} F^{'-}        \right) -{1\over 2\pi} \int   F'dB~, }
with
\eqn\sd{\tau'=-{1\over \tau}. }
Now the path integral of the $U(1)$ gauge field $B$ gives a delta function constraint saying that
 $F'$ is closed and its periods  are integer multiples of $2\pi$, so we can interpret $F'$ as the field strength of a  dual gauge field $B'$. Restoring indices, we find that the dual action is
\eqn\nactiont{I_\em(B';\tau')=-{i\over 8\pi}\int d^4x\sqrt{-g}\left(\bar\tau' \,F^{'+}_{\mu\nu}F^{'+\mu\nu}   -\tau' F^{'-}_{\mu\nu}F^{'-\mu\nu}        \right). }

In sum, the theory with action $I_\em(B;\tau)$ is equivalent to a theory with another $U(1)$ gauge field $B'$, field
strength $F'_{mn}=\partial_m B'_n-\partial_n B'_m$,
and action \nactiont\ with coupling constant $\tau'$.
This derivation essentially holds on any oriented four-manifold $X$, though a little more care shows that there is  a $c$-number anomaly involving the Riemannian curvature
of $X$  \WittenGF.\foot{The definition of $\tau$ used in that paper is $\tau=\theta/2\pi+4\pi i/g^2$, where $g^2=2e^2$.
This choice was made in order to agree with conventions often used in $U(N)$ gauge theory for $N>1$.}

What is often called $S$-duality is the group generated by $S(\tau)=-1/\tau$ along with
\eqn\vaction{T(\tau)=\tau+1. }
Thus $T$ generates the transformation $\theta\to\theta+2\pi$.  $S$ and $T$ together generate an infinite discrete group called $SL(2,\Z)$.
An element of $SL(2,\Z)$ is a $2\times 2$ integer-valued matrix of determinant 1:
\eqn\matrixp{M=\pmatrix{ a& b \cr c& d}, ~~ad-bc=1, }
acting on $\tau$ as
\eqn\atrixp{M(\tau)={a\tau+b\over c\tau+d}. }
Note that the element $-1\in SL(2,\Z)$ acts trivially on $\tau$ (so the group that acts faithfully on $\tau$ is
actually the quotient group $SL(2,\Z)/\{\pm 1\}=PSL(2,\Z)$).  Accordingly, $-1$ is a symmetry for any $\tau$
and (taking $\tau\to i\infty$) must correspond to a classical symmetry.  It can be shown that this
symmetry is charge conjugation, which acts by $B\to -B$.\foot{If instead of adding \addedterm\ we had subtracted it, this would have changed the sign of $F'$ and $B'$.  We
would get a second duality transformation acting as $\tau\to -1/\tau$ but differing from the first by the element $-1\in SL(2,\Z)$, that is by $B'\to-B'$.}

Actually, as noted in \WittenGF, with $\theta$ defined as in eqn.\ \thaction, $I_\em$ is invariant mod $2\pi$ under $T:\theta\to\theta+2\pi$
only if $X$ is a spin manifold.  (One needs to know that the quantity $J$ defined in Appendix A  is always integer-valued.)
Without a spin structure, one only has invariance under $T^2:\theta\to\theta+4\pi$.   So on a four-manifold $X$ that is not necessarily a spin manifold,
$U(1)$ gauge theory is invariant only under the subgroup of $SL(2,\Z)$ that is generated by $S$ and $T^2$.   This subgroup is of index 3.
The significance of this was clarified only recently \MetlitskiYQA.  In $U(1)$ gauge theory, one can define
Wilson-'t Hooft operators with integer electric and magnetic charges $(n_e,n_m)$.
If we consider the charges mod 2, then we are left with three basic Wilson-'t Hooft operators, of charges $(1,0)$, $(0,1)$, and $(1,1)$.
One can define three variants of $U(1)$ gauge theory in which two of the three basic line operators are taken to be bosonic and the third to be fermionic.\foot{It is not
possible for all of the basic line operators to be bosonic, since if two are bosonic, the angular momentum in the electromagnetic field forces the third to be fermionic.
Likewise it is not possible for two of the basic line operators to be fermionic and only one bosonic.  It has been argued that in a purely $3+1$-dimensional
theory, it is not possible for all three basic line operators to be fermionic \refs{\MPS,\McG}.  It is also possible to have a version of $U(1)$ gauge theory in which all
of the line operators are allowed to be either bosonic or fermionic.  This implies that one can consider a neutral fermion, so the theory with
a full set of line operators in this sense only makes sense on a spin manifold.
This leads to  full $SL(2,\Z)$ symmetry.  In effect, this option was assumed in \WittenGF.}
These three choices are permuted by $SL(2,\Z)$.  In either of the two versions in which the $(1,0)$ line operator is bosonic, the gauge field $B$ is an ordinary
$U(1)$ gauge field.   These two versions differ by whether the $(0,1)$ line operator is bosonic, and they are exchanged by $T$.  The third option is that the $(1,0)$
line operator is fermionic.  For this to make sense on a manifold $X$ that is not a spin manifold, the gauge field that appears in the action and
couples to the $(1,0)$ line operator
must be a $\spinc$ connection rather than an ordinary $U(1)$ gauge field.

In our previous analysis in eqns.  \addedterm-\nactiont, we transformed one $U(1)$ gauge field $B$ to another $U(1)$ gauge field $B'$.
One might wonder how one can modify the derivation to instead transform $B$ to a $\spinc$ connection\foot{A nontrivial theorem says that every orientable four-manifold
admits a $\spinc$ structure. (In five or more dimensions, this is not true.)  Clearly we will have to invoke this theorem at some point, since on a four-manifold $X$ that does
not admit any $\spinc$ structure, a $U(1)$ gauge field could not be dual to a $\spinc$ connection.  The theorem will be invoked in the next footnote.}
 $A$.
For this, we return to the action $I_\em(B,\tau)$ of eqn.\ \naction, but now we write it as $I_\em(B,\tau+1)+I'$, where
 $I'=-{1\over 4\pi}\int dB dB$. Again we add the term \addedterm\ with $F$ and $F'$ arbitrary two-forms.  But  now we replace $dB$ by $F$ only in $I_\em(B,\tau+1)$,
 not in $I'$.   Integrating out $F$ and including $I'$ now leads to
\eqn\nactiontts{-{i\over 8\pi}\int \left(\bar\tau' \,F^{'+}F^{'+}   -\tau' F^{'-}F^{'-}        \right) -{1\over 2\pi} \int   F'dB -{1\over 4\pi}\int dB  dB~, }
with
\eqn\sds{\tau'=ST(\tau)=-{1\over \tau+1}. }   Now we want to integrate out $B$.  On a spin manifold, $I'=-{1\over 4\pi}\int d B d B$ is an integer multiple of $2\pi$
and has no influence on the path integral.  In general, it is congruent mod $2\pi$ to  $\pi\int w_2 \wedge d B/2\pi $ where\foot{We are being slightly cavalier, since $w_2$ is
a class in mod 2 cohomology rather than a differential form.  However, using the theorem mentioned in the last footnote, we can replace $w_2$ in this derivation by $G/2\pi$, where $G$ is the curvature  or field strength
of any
$\spinc$ connection.  The argument then proceeds as in the text, using the fact that with $F=dB$,  $\int_X F\wedge F/(2\pi)^2$ is congruent mod 2 to $\int_X F\wedge G/(2\pi)^2$ and the fact that
periods of $G/2\pi$ are congruent mod 2 to periods of $w_2$.} $w_2$ is the second Stieffel-Whitney class of $X$,
which is the obstruction to a spin structure on $X$.  The path integral over the abelian gauge field $B$ still gives a constraint saying that $F'$ is closed,
But now instead of learning that $F'$ has periods that are integer multiples of $2\pi$, we learn that $F'+\pi w_2$ has periods that have that property.  As in eqn.\ \spincc,
this amounts to saying that $F'$ is the curvature of a $\spinc$ connection $A'$.  Thus we have shown that the duality transformation $ST$ maps
a $U(1)$ gauge field $B$ to a $\spinc$ connection $A'$.

\subsec{Duality On A Manifold With Boundary}

Now we consider the case of a $U(1)$ gauge field $B$ that propagates only in ``half'' of  spacetime, say the half-space $x^3\geq 0$ in Minkowski spacetime,
where $x^3$ is one
of the three spatial coordinates.  We call the half-space $X_+$ and denotes its boundary as $M$. ($X$ will denote all of spacetime and $X_-$ will be the opposite
half-space.) The boundary of $X_+$ is a $2+1$-dimensional spacetime $M$.
We assume that the gauge field $B$ on $X_+$ is coupled to some
degrees of freedom that propagate on $M$.   The part of the action that involves
$B$ is generically
\eqn\baction{{1\over 2\pi}\int_M d^3x\,J^\mu B_\mu -{i\over 8\pi}\int_{x^3\geq 0} d^4x\left(\bar\tau \,F^+_{\mu\nu}F^{+\mu\nu}   -\tau F^-_{\mu\nu}F^{-\mu\nu}        \right), }
where $J_\mu$ is a conserved current defined only on $M$ and the $1/2\pi$ is  a  convenient normalization.

The duality transformation of this action was anticipated in \WittenYA\ and was worked out in \GaiottoAK, section 4.4.  Here we will give a derivation by adapting the steps above.  We express the action in terms of the gauge field $B$, add the term \addedterm, and then express the bulk term in the action \baction\ in terms of the arbitrary two-form $F$.  We cannot do that in the first term in \baction\ because it depends explicitly on $B$.  Next, we integrate out $F$ to convert the bulk part of the action to \nactiontt\ and we integrate out the bulk components of $B$
(that is, we integrate over $B$ keeping its boundary value fixed) to learn that locally $F'=d B'$ is the field strength of a $U(1)$ gauge field.  This turns the second term in \nactiontt\ to $-{1\over 2\pi}\int dB'dB$, which can be expressed as a Chern-Simons term on the boundary, namely $-{1\over 2\pi}\int BdB'$.  At this point all that remains of $B$ is its boundary value, which is a purely $2+1$ dimensional gauge field.  We denote the boundary value by $b$ and we end up with the dual action
\eqn\dualaction{{1\over 2\pi}\int_M  d^3x\left( J^\mu b_\mu- \epsilon^{\mu\nu\rho} b_\mu \partial_\nu B'_\rho\right)-{i\over 8\pi}\int_{x^3\geq 0} d^4x\left(\bar\tau' \,F^{'+}_{\mu\nu}F^{'+\mu\nu}   -\tau' F^{'-}_{\mu\nu}F^{'-\mu\nu}        \right)~,}
with $\tau'=-1/\tau$.

Let us take stock of this answer.  We started with a theory in which a bulk gauge field $B$ couples to the current $J/2\pi$ on the boundary.  After
duality, the original bulk gauge field $B$ has been replaced by a new gauge field $B'$ with coupling parameter $\tau'=-1/\tau$.  And the restriction of $B$ to the boundary, which we denote by $b$,  now behaves as a purely $2+1$-dimensional gauge field.  It retains its original coupling to $J/2\pi$ and  couples to $B'$ by $\int_M b d B'/2\pi$.

In general, we may consider on $M$ any theory with degrees of freedom that we generically call $\Phi$, coupled to  a background $U(1)$ gauge field $B$ by
some action $I(\Phi,B)$.   Now let us suppose that $B$ is actually a dynamical gauge field, propagating on the half-space $X_+$, with coupling parameter $\tau$.  To apply a bulk
duality transformation $\tau\to -1/\tau$, we apply the procedure that was summarized in the last paragraph.  The dual theory is obtained by replacing $B$ in bulk by a dual gauge field $B'$, treating the restriction of $B$ to the boundary as a purely $2+1$ dimensional gauge field $b$, and including the $b d B'/2\pi$ coupling:
\eqn\faction{I(\Phi,b)-{1\over 2\pi}\int_M b d B' -{i\over 8\pi}\int_{x^3\geq 0} d^4x\sqrt{-g}\left(\bar\tau' \,F^{'+}_{\mu\nu}F^{'+\mu\nu}   -\tau' F^{'-}_{\mu\nu}F^{'-\mu\nu}        \right).}

As in \WittenYA, this explains how to implement the duality transformation
\eqn\howto{S=\pmatrix{0&-1\cr 1&0},~~\tau\to -1/\tau,}
in the context of a gauge field on a manifold with boundary coupled to an arbitrary system on the boundary.  The full group $SL(2,\Z)$ is generated by $S$ and
\eqn\owto{T=\pmatrix{1&1\cr 0&1 }}
with the relation $(ST)^3=1$.  The action of $T$ is actually more elementary.  In bulk, $T$ shifts the theta-angle by $2\pi$.  On a spin manifold without boundary, that is a symmetry
(in the absence of a spin structure, we can instead consider in a similar way the operation $T^2$).
In the presence of a boundary, shifting the bulk theta-angle by $2\pi$ is not a symmetry, but we can compensate for this
by adding
a Chern-Simons coupling $-(1/4\pi)\int_M B d B$ to the boundary theory.  So this gives the action of $T$ on the boundary couplings:
\eqn\nowto{I(\Phi,B)\qquad \longrightarrow \qquad  I(\Phi,B)-{1\over 4\pi}\int_M B d B.}
The fact that these relations do satisfy $(ST)^3=1$ can be demonstrated by a short, formal calculation.\foot{This calculation can be found
in \WittenYA.  However, note that what we call $S$ and $T$ were called $S^{-1}$ and $T^{-1}$ in that paper.  (Exchanging $S$ with $S^{-1}=-S$
can be understood as combining $S$ with charge conjugation, and replacing $T$ with $T^{-1}$ is equivalent to changing the sign of the theta-angle.) The relation $(ST)^3=1$ is unchanged.}

Below we will often find it convenient to use $-S$ instead of $S$.  Its action on $\tau$ is the same as the action of $S$, but it involves also a factor of charge conjugation.  Therefore, the action of $-S$ is obtained by changing $ B'\to -B'$ in  \faction.

The reader may be surprised that in this derivation, we insist that the bulk gauge field $B$ propagates only in the half-space $X_+$.  What happens
if $B$ propagates throughout all of a four-manifold $X$, in which $M$ is embedded?  This question can be answered using the facts explained above, but the answer may be slightly more
complicated than one anticipates.  First, let us interpret a gauge field $B$ that propagates throughout $X$ as a pair consisting of a gauge field $B_+$ on $X_+$
and another gauge field $B_-$ on $X_-$, with a constraint setting $B_+=B_-$ (up to a gauge transformation) along $M$.  As above, we implement the constraint
by including a $U(1)$ gauge field $c$ that only propagates on $M$ with a coupling $(1/2\pi)\int_M c d(B_+-B_-)$.  It is also a useful abbreviation to write, for
example, $\int_{X_+}\L(B_+,\tau)$ for the bulk action of a gauge field $B_+$ on $X_+$ with coupling parameter $\tau$.    Then the system consisting of a gauge field
$B$ on $X$ that is free in bulk but has a coupling $I(\phi,B)$ along $M$ can be described by the action:
\eqn\lowto{I(\Phi,B_+)-{1\over 2\pi}\int cd(B_+-B_-) +\int_{X_+}\L(B_+,\tau)+\int_{X_-}\L(B_-,\tau).  }
(We could equally well replace $I(\Phi,B_+)$ by $I(\Phi,B_-)$, because of the constraint that effectively sets $B_+=B_-$ along $M$.)
Thus we have a theory with two half-space gauge fields $B_+$ and $B_-$ and boundary couplings $\hat I(\Phi,B_+,B_-)=I(\Phi,B_+)+{1\over 2\pi}\int c d(B_+-B_-)$.  Now, as above, we implement the duality transformation $\tau\to -1/\tau$ for this system by replacing $B_\pm$ in the bulk by new gauge fields $B'_\pm$, denoting the boundary values of $B_\pm$ by $b_\pm$, and  adding new couplings $-(1/2\pi)\int_M\left(
b_+ d B'_++b_- d B'_-\right)$.  Thus we arrive at
\eqn\doubled{I(\Phi,b_+) -{1\over 2\pi}\int \biggl(cd(b_+-b_-) +b_+ d B'_+ +b_- d B'_-\biggr) +\int_{X_+}\L(B'_+,-1/\tau)+\int_{X_-}\L(B'_-,-1/\tau)~. }
The path integral over $b_-$ can be performed, and gives a constraint setting\foot{By $B_-'|$, we mean the restriction of $B_-'$ to $M$.} $c=B'_-|$.  Denoting $b_+$ simply as $b$, we are left with
\eqn\oubled{I(\Phi,b)-{1\over 2\pi}\int b d(B'_+-B'_-)  +\int_{X_+}\L(B'_+,-1/\tau)+\int_{X_-}\L(B'_-,-1/\tau)~. }

Note that here there is {\it not} any constraint setting $B'_+|=B'_-|$ (we cannot get such a constraint by integrating over $b$, since $b$ also appears in $I(\Phi,b)$).
Thus in the dual description there are really separate gauge fields propagating in $X_+$ and in $X_-$. This may come as a surprise, though there is an example
involving the D3-D5 system that is relatively well-known to string theorists.\foot{Let $X$ be the $3+1$-dimensional worldvolume of a D3-brane and $M$ a codimension one subspace
on which $X$ intersects a D5-brane.  On $X$, there propagates a (supersymmetric) $U(1)$ gauge theory.  The $U(1)$ gauge multiplet is coupled on $M$
to a charged hypermultiplet, which one can think of as the supersymmetric extension of a charged boson.  After a $\tau\to -1/\tau$ duality, one has a D3-NS5 system.
Now there are separate $U(1)$ gauge fields $B'_+$ and $B'_-$ on the two sides of the NS5-brane.  The standard description of the system is that $B'_+$ and $B'_-$
 are coupled on $M$ to a ``bifundamental hypermultiplet'' (this is simply
a hypermultiplet that couples with charges $(1,-1)$ to $B'_+$ and $B'_-$).  Interestingly, this standard description differs from the one we get in eqn.\ \doubled\
by a supersymmetric version of charge-vortex duality (the relevant
duality was originally developed in \refs{\IntriligatorEX,\KapustinHA}).
Comparing the two descriptions gives a derivation of this supersymmetric charge-vortex duality from
the underlying string theory dualities.}

Going back to the one-sided case, one may ask what is the physical interpretation of a $U(1)$ gauge field that only propagates in a half-space.  The originally
envisaged application was to the AdS/CFT correspondence.  Since free $U(1)$ gauge theory is conformally invariant and anti de Sitter space is conformally
equivalent to a half-space in Minkowski spacetime, the question of how duality acts on a $U(1)$ gauge field in a half-space has applications to AdS/CFT duality.  In condensed matter
physics, one may interpret a half-space gauge field $B$ as an emergent gauge field that only propagates in the world-volume $X_+$ of some material,
and may be coupled to boundary degrees of freedom on the surface of the material.

Finally, we would like to point out that, by adapting the analysis of
\nactiontts,  we can similarly describe dualities involving $\spinc$ connections rather than ordinary $U(1)$ gauge fields on a manifold with boundary.

\subsec{A Four-Dimensional Interpretation Of Particle-Vortex Dualities}

We will use these tools to describe a four-dimensional interpretation of the dualities studied in the present paper.  We start with the
standard bosonic particle-vortex duality.

We consider a  $U(1)$-invariant theory of a complex boson $\phi$ on $M$ at the Wilson-Fisher fixed point.  We couple $\phi$ to a half-space gauge field
$B$ via the action \eqn\coupac{
\int_M d^3 x \sqrt{-g}\left(|D_B\phi|^2-|\phi|^4\right)+\int_{X_+}\L(B,\tau).}
As usual, $|\phi|^4$ is shorthand for a quartic coupling that is tuned to the Wilson-Fisher fixed point.  For the present purposes, we may set $\theta=0$ so that $\tau=2\pi i/e^2$.
For $e\to 0$, $B$ effectively decouples and behaves as a background gauge field coupled to the Wilson-Fisher theory.

Now instead, let us go to $\tau=i$, that is $e^2=2\pi$.  This is a fixed point of the transformation $S:\tau\to -1/\tau$.  We introduce now the
{\it assumption} that the combined system \coupac\ is $S$-invariant at $\tau=i$.  We know that this is true in bulk, so the assumption is that the particular boundary
coupling in \coupac\ is $S$-invariant.  If so, the relevant operator $|\phi|^2$ of the
Wilson-Fisher theory must be odd under $S$, as will follow from the analysis below.

$S$ will exchange the deformation away from $e^2=2\pi$ to $e\ll1$ to a deformation to $e\gg 1$.  Thus under our assumption, the Wilson-Fisher theory coupled to a background
field, which we get at $e\ll1$, has a dual description that is obtained by taking $e\gg 1$ in eqn.\ \coupac.   The dual theory has coupling parameter $-1/\tau$.

The trouble with this is that a description that involves $e\gg 1$ is not very useful.  To get a more useful description, we apply an $S$-duality transformation to the dual
theory found in the last paragraph.  This will introduce a new gauge field $B'$ in $X_+$, now once again with coupling parameter $\tau$.  As we know by now,
the duality is implemented on the boundary by replacing $B$ by a  purely $2+1$-dimensional gauge field $b$ in the original boundary couplings, and adding a $b d B'/2\pi$
coupling.  So we arrive at
 \eqn\noupac{
\int_M d^3 x \sqrt{-g}\left(|D_b\phi|^2-|\phi|^4\right)-{1\over 2\pi}\int_M b d B' +\int_{X_+}\L(B',\tau).}
Actually, here we can drop the prime from $B'$.  Since the bulk action for $B'$ has the same coupling parameter $\tau$ as the bulk action for $B$ in eqn.\ \coupac,
$B'$ must coincide with $B$ up to a classical symmetry of the bulk theory.  The only relevant classical symmetry would be charge conjugation, $B\to -B$.  Given
that \coupac\ and \noupac\ are equivalent, by combining such an equivalence with charge conjugation, if necessary, we can assume that the relation is $B'=B$
rather than $B'=-B$.

The conclusion is that {\it assuming} the theory \coupac\ is selfdual at $\tau=i$, it follows that the two theories \coupac\ and \noupac\ are equivalent for any $\tau$.
Taking $\tau\to i\infty$, we recover the standard bosonic particle-vortex duality.  (To match with our earlier equations, set $b\to -b$.)

Every $2+1$-dimensional duality proposal in the present paper can similarly be deduced from an analogous duality conjecture in $3+1$ dimensions.
Moreover, the $2+1$-dimensional duality web, wherein one duality conjecture can be deduced from another, has a counterpart in $3+1$ dimensions.

To explain this, we start with the basic bose-fermi duality, from which follow all of the purely $2+1$-dimensional dualities considered in this paper.   We will
explain how to interpret the bose-fermi duality in $2+1$ dimensions as a consequence of a conjecture in $3+1$ dimensions.

Let us start with the basic duality of eqn.\ \basicduality, between a Dirac fermion $\Psi$ coupled to a background $\spinc$ connection $A$,
and a Wilson-Fisher boson that is coupled to a $U(1)$ gauge field $b$ as well as to $A$:
\eqn\upac{i\bar\Psi\slashchar{D}_A\Psi \qquad \longleftrightarrow \qquad |D_b\phi|^2-|\phi|^4 +{1\over 4\pi}b d b +{1\over 2\pi}b d A. }
We observe that the theory on the right hand side can be obtained by applying the operation $-ST^{-1}$ to a Wilson-Fisher boson coupled to a background gauge field $B$:
\eqn\zupac{ |D_B\phi|^2-|\phi|^4. }
This suggests the following interpretation.  First of all, the inverse of $-ST^{-1}$ is $TS$, which transforms $\tau$ to $\tau'=1-1/\tau$.
Now we postulate an equivalence between two half-space theories.  In theory I, the Dirac fermion $\Psi$ on $M$ is coupled to the boundary values of a half-space
$\spinc$ connection $A$; the bulk action for $A$ has coupling parameter $\tau$.  The action is thus
\eqn\lupac{\int_Md^3x \sqrt{- g} \,i \bar\Psi \slashchar{D}_A\Psi+\int_{X_+}\L(A,\tau).  }
In theory II, a Wilson-Fisher boson on $M$ is coupled to the boundary values of a half-space $U(1)$ gauge field  $B$, but now the coupling parameter
is $\tau'$:
\eqn\tupac{\int_M d^3x\sqrt {-g}\,\left(|D_B\phi|^2-|\phi|^4\right)+\int_{X_+}\L(B,\tau'). }
We postulate that these theories are equivalent for all $\tau$.  To try to extract a purely $2+1$-dimensional duality from this statement, we take $\tau\to i\infty$,
whereupon theory I reduces to a Dirac fermion coupled to a background $\spinc$ connection $A$.  Unfortunately, when $\tau\to i\infty$, we have $\tau'\to 1$.
The description of theory II via the action \tupac\ is not useful for $\tau'\to 1$.  To get something useful, we apply a duality transformation $-ST^{-1}$ to theory II.
This transforms the gauge coupling parameter from $\tau'$ back to $\tau$.  Following the reasoning of \MetlitskiYQA, the $-ST^{-1}$ transformation in bulk
maps the $U(1)$ gauge field $B$ to a $\spinc$ connection rather than an ordinary $U(1)$ gauge field, so we write $A'$ (rather than $B'$) for the new bulk
connection.  Apart from this, the action of $-ST^{-1}$ on theory II is obtained in the usual way: we act with $T^{-1}$ by adding a new boundary coupling $B d B/4\pi$,
and then we act with $-S$ by replacing $B$ in the boundary couplings with a purely $2+1$-dimensional gauge field $b$, with an additional coupling $b d A'/2\pi$ to the new
bulk field $A'$.
As in our discussion of the bosonic case, since the half-space theories of $A$ and $A'$ have the same coupling parameter $\tau$, we can assume that these fields
are simply related by $A=A'$.
Thus we end up with
\eqn\rupac{\int_M d^3s \sqrt {-g}\,\left(|D_b\phi|^2-|\phi|^4\right) +{1\over 4\pi}\int_M b d b +{1\over 2\pi}\int_M b d A+\int_{X_+}\L(A,\tau)~, }
which should be equivalent to \lupac\ for any $\tau$.  In the limit $\tau\to i\infty$, the equivalence of \lupac\ and \rupac\ reduces to the purely $2+1$-dimensional duality
of eqn.\ \upac.

In section 2,  this one duality was used as a starting point to build up a web of dualities.
This was done by successive application of $S$ and/or $T$ transformations and/or time-reversal
$\cal T$ to the left
and right hand sides of eqn.\ \upac.  All these operations have bulk counterparts, so every statement in section 2 can be ``promoted'' to a $3+1$-dimensional statement.
Since it is particularly elegant, we will give as an example the fermion-fermion duality of eqn.\ \mansekf.  This is a duality between a Dirac fermion $\Psi$
coupled to a background $\spinc$ connection $A$, and a Dirac fermion $\chi$ coupled to purely $2+1$-dimensional fields $a$ and $b$ as well as $A$:
\eqn\ipac{i\bar\chi\slashchar{D}_a\chi+{1\over 2\pi}adb -{2\over 4\pi}bdb +{1\over 2\pi}bdA\qquad \longleftrightarrow\qquad
i\bar\Psi\slashchar{D}_A\Psi+{1\over 4\pi}AdA+2\CSg. }
A bulk theory related to the right hand side of eqn.\ \ipac\ is simply
\eqn\zipac{\int_M d^3x \sqrt{-g}\, i\bar\Psi \slashchar{D}_A\psi +{1\over 4\pi}\int_M A d A+\int_{X_+}\L(A,\tau). }
This theory is $\cal T$-invariant if $\rm{Re}\,\tau=-1/2$, that is if
\eqn\yipac{\tau=-{1\over 2}+i x}
for some real $x$.  (Taking $x\to\infty$, this statement reduces to the statement that eqn.\ \mansekfb\ is $\cal T$-invariant.)  On the left hand side of eqn.\ \ipac, we see
the ingredients that are required to act with $ST^{2}S$ on the theory of a Dirac fermion $\chi$, successively introducing new fields $a$ and $b$ with suitable
couplings.  Imitating the procedure that we used to give a $3+1$-dimensional interpretation
to eqn.\ \upac, this suggests that we should start with a bulk theory whose coupling parameter is obtained by transforming $\tau$ by the inverse of $ST^{2}S$.
That inverse is $ST^{-2} S$, and maps $\tau$ to $\tau'=\tau/(2\tau+1)$.  We note that if $\tau=-1/2+ix$, then
\eqn\mipac{\tau'={1\over 2}+{i\over 4x}.}
At such a value of $\tau'$, the theory
\eqn\ripac{\int_M d^3x\sqrt{g}\,i\bar\chi\slashchar{D}_{A'}\chi + \int_{X_+}\L(A',\tau') }
is $\cal T$-invariant.

 We postulate that theory \zipac, with any coupling parameter $\tau$, is equivalent to theory \ripac\ with the corresponding coupling parameter $\tau'$.
 As usual, a naive attempt to deduce from this a purely $2+1$-dimensional duality runs into trouble.  If we take $\tau\to i\infty$, then theory \zipac\ reduces
 to the theory of a $2+1$-dimensional Dirac fermion $\Psi$ coupled to a background $\spinc$ connection $A$.  But for $\tau\to i\infty$, we have $\tau'\to 1/2$, a limit in which
 the description \ripac\ is not useful.   To compensate for this, we apply $ST^{2}S$ to \ripac, giving a new description in which the bulk coupling
 parameter is the original $\tau$ and hence the bulk $\spinc$ connection is the original $A$:
 \eqn\zzipac{\int_M d^3x \sqrt{g}\,i\bar\chi\slashchar{D}_a\chi+\int_M\left(-{1\over 2\pi}a db -{2\over 4\pi}b db-{1\over 2\pi}b dA\right)
 +\int_{X_+}\L(A,\tau). }
 Now, after a sign change of $b$, in the limit $\tau\to i\infty$ the equivalence of \zipac\ and \zzipac\ leads to the purely $2+1$-dimensional duality \ipac, as promised.

\subsec{Time-Reversal Symmetry}

We now show that the bulk duality between \lupac\ and \tupac\ is fully compatible with time-reversal symmetry. This provides further evidence for the bulk duality. It also gives a simple interpretation of the unconventional time-reversal symmetry in the three-dimensional dualities \basicduality\ and \secondduality, discussed in \Tbasicduality\ and \Tfermionvortex.

Let us start with the equivalence between \lupac\ and \tupac\ on a closed manifold without boundary. Consider the case with $\tau={1\over 2}+{i\over 2}\rm{tan}(\alpha/2)$. The naive time-reversal transform $\CT_0$, which takes $\theta\to-\theta$ and keeps $e^2$ unchanged, acts on $\tau$ as $\CT_0: \tau\to-\bar{\tau}$. For $\tau={1\over 2}+{i\over 2}\rm{tan}(\alpha/2)$ this means $\CT_0: \tau\to\tau-1$. This shift can be restored by a $T$ transform (which is legitimate since $A$ is a spin$_c$ connection). Therefore the real time-reversal symmetry for \lupac\ should be $T\CT_0$. We recognize this as the usual time-reversal symmetry of
the gauged topological insulator.

Now what about the bosonic side \tupac? For $\tau={1\over 2}+{i\over 2}\rm{tan}(\alpha/2)$ we have $\tau'=TS\tau=1-1/\tau=-e^{-i\alpha}$. The naive $\CT_0$ changes it to $\tau''=-\bar{\tau'}=e^{i\alpha}$. But one can make another $S$ transform to bring it back to $S\tau''=-1/\tau''=-e^{-i\alpha}=\tau'$. So the real time-reversal on \tupac\  should be $S\CT_0$.

The above statement can be derived more systematically as follows. We want to find a time-reversal operation $g\CT_0$, where $g\in SL(2,\Z)$ and
\eqn\TITb{
g\CT_0 (TS\tau)=TS\tau,
}
for any $\tau$ that satisfies
\eqn\TITa{
T\CT_0\tau=\tau.
}
A direct calculation shows that $\CT_0 TS=T^{-1}S\CT_0$. After some simple manipulation we get
\eqn\TITc{
g=-TSTST=S,
}
which is what we expected.

The above result has a very simple physical picture. For $\tau={1\over 2}+{i\over 2}\rm{tan}(\alpha/2)$, the $TS$ transform maps the charge $(1/2,1)$ and $(1/2, -1)$ dyons into the basic charge and monopole particles. Since time-reversal exchanges the two original dyons, after the transform time-reversal should act by exchanging the charge and monopole particles, namely as an $S$ operation.

We now turn to manifolds with boundary. The surface Dirac fermion in \lupac\ is not invariant under the naive $\CT_0$, rather under $T\CT_0$ -- this is simply the physics of the topological insulator surface, as we explained in discussing eqn.\ \freefb. What about the bosonic side? The bosonic surface state in \tupac\ is invariant under $\CT_0$. It is also invariant under the surface $S$ operation, as demonstrated by the charge/vortex duality. Therefore the bosonic side \tupac\ with a boundary is also $S\CT_0$-invariant.

Even though the bosonic surface theory in \tupac\ is invariant under $S\CT_0$, the surface field $\phi$ does not transform classically. Rather, the symmetry transforms it to its vortex dual $\hat{\phi}$. Physically we can interpret $\phi$ and $\hat{\phi}$ as the surface correspondents of the electric and magnetic particles in the bulk, which correspond to the $(1/2,\pm1)$ dyons in the dual fermion theory in \lupac. This offers a simple interpretation of the time-reversal property of the boson/fermion duality in \basicduality, discussed under \Tbasicduality.

There is a parallel story when $\tau'=ix$.  The bosonic side \tupac\ is now manifestly $\CT_0$-invariant. Here $\tau=(-S)T^{-1}\tau'=-1/(\tau'-1)=-{1\over -1+ix}$. $\CT_0$ takes $\tau$ to $\tau''=-\bar{\tau}=-{1\over 1+ix}$, which can be restored to $\tau$ by $-ST^{-2}S$, namely $\tau=-1/(-2-1/\tau'')$. Again formally this is obtained by asking for a time-reversal implementation $h\CT_0$ with $h\in SL(2,\Z)$, such that
 \eqn\TRa{
 h\CT_0 (-ST^{-1})\tau'=-ST^{-1}\tau',
 }
 for any $\tau'$ that satisfies $\CT_0\tau'=\tau'$. This can similarly be solved and we get
 \eqn\TRb{
 h=-ST^{-2}S.
 }

There is also a simple physical picture for this result: for the bosonic QED in \tupac\ with $\tau=ix$, the $(1,-1)$ and $(1,1)$ dyons are fermions and are exchanged by time-reversal symmetry. After the $-ST^{-1}$ transform, the $(1,-1)$ dyon becomes the $(1,0)$ charge, and the $(1,1)$ dyon becomes a $(q,2)$ dyon ($0<q<1$). The exchange operation of the two fermions becomes the $-ST^{-2}S$ operation.

Now we ask how the surface transforms under such time-reversal. The bosonic side is trivially $\CT$-invariant. For the fermionic side, after the naive $\CT_0$, the Dirac fermion action
becomes $\sqrt{- g} \,i \bar\Psi \slashchar{D}_A\Psi+{1\over 4\pi}AdA+2\CSg$. The $-ST^{-2}S$ operation further transforms it to
\eqn\surfaceSdual{
i\bar\chi\slashchar{D}_a\chi+{1\over 4\pi}ada-{1\over 2\pi}adb +{2\over 4\pi}bdb +{1\over 2\pi}bdA+2\CSg.
}

By acting a simple $\CT$-transform on the fermion-fermion duality in \mansekf, we see that the above line is exactly dual to the free Dirac fermion $\sqrt{- g} \,i \bar\Psi \slashchar{D}_A\Psi$. Therefore the fermion theory \lupac\ with a boundary is also invariant under $ST^{-2}S\CT_0$.

The surface fermions $\Psi$ in \lupac\ transforms to their dual fermions $\chi$ under time-reversal. Physically they can be interpreted as corresponding to the $(1,\pm1)$ dyons in the bulk. This also gives a simple interpretation of \Tfermionvortex, in which time-reversal also exchanges the $\chi$ fermion with its dual.

\newsec{Relation to Problems in Condensed Matter Physics}

As explained in the introduction, the fermion-fermion duality has many important applications in condensed matter physics. We have already described how the gapped T-Pfaffian state appears naturally from \dualtheory.
Thus the fermion-fermion duality  allows for a field theoretic viewpoint on how this state arises at the surface of a topological insulator in agreement with lattice constructions.

The other important application is to the theory of the half-filled Landau level which we now turn to.  Let us briefly recap the relationship to the topological insulator surface. More detail may be found, eg, in \WangFQL. We are interested in the physics of electrons in two space dimensions in a strong magnetic field at Landau level filling $\nu = {1 \over 2}$. Upon projecting to the Lowest Landau Level, and restricting to a two body interaction (e.g.\ just Coulomb repulsion), the Hamiltonian has an extra discrete anti-unitary symmetry not present in a UV system of non-relativistic electrons. This symmetry is known as a particle-hole symmetry. It has the effect of exchanging the empty Landau level with the filled one. The half-filled Landau level can be obtained by either starting with the empty Landau level and adding electrons or by starting with a filled Landau level and removing electrons. The particle-hole transformation relates these two ways of reaching the half-filled state.

This same physical situation can be realized  by starting with a relativistic Dirac fermion. Consider a single two-component massless Dirac fermion $\psi$. This can be rendered $\CT$ and $\CC\CT$ invariant by placing at the spatial boundary of a $3+1$-D topological insulator. Note the both of these are anti-unitary symmetries.  The electron is a Kramers doublet under $\CT$. Under  $\CC\CT$, the electrical charge density is odd, and the electrical current is even.  An external non-zero uniform magnetic magnetic field breaks $\CT$ but preserves $\CC\CT$.  Now it is well known that Dirac electrons in a uniform field form Landau levels which famously includes one at zero energy. The $\CC\CT$ symmetry guarantees that this is half-filled. In the presence of interactions at the UV scale this then maps the low energy physics to that of the half-filled Landau level with the $\CC\CT$ operation playing the exact same role as the particle-hole transformation mentioned in the previous paragraph.

Thus the particle-hole symmetric half-filled Landau level can be UV completed while preserving charge conservation and $\CC\CT$ symmetries by placing it at the surface of a $3+1$-D topological insulator.
Now let us describe this system using \dualtheory . We note that the duality interchanges the role of $\CT$ and $\CC\CT$. In particular $\chi$ is now a Kramers doublet under $\CC\CT$. We need to study \dualtheory\ in the presence of a uniform background magnetic field associated with the $A$ gauge field.  Before doing so we note that in  relating the results to the standard half-filled Landau level obtained starting with non-relativistic fermions, we need to add a background Chern-Simons  term for $A$. Specifically, at the TI surface, the empty $0$th Landau Level is assigned a Hall conductivity of $-{1 \over 2}$ while the filled one is assigned ${1 \over 2}$.
In the standard Landau level problem of non-relativistic fermions, the Hall conductivity assignments are shifted by ${1 \over 2}$ (so that the empty Landau level has zero Hall conductivity).  Similar statements apply to the thermal Hall conductivity. This amounts to adding a term
${1 \over 8\pi}A dA + \CSg$ to both sides of \mansekfbf .  Further in order to connect smoothly with standard results on composite fermions, we will do a charge conjugation transformation $\chi \rightarrow \chi^\dagger, a \rightarrow -a, b \rightarrow -b$ on \dualtheory. This has the effect of changing the sign of the $bdA$ term while leaving the rest of the Lagrangian unchanged.  Thus our proposed theory of the half-filled Landau level is
\eqn\phcfl{
\CL=i\bar \chi \slashchar{D} _{a}\chi - {2\over 4\pi} bdb + {1\over 2\pi}a db - {1\over 2\pi}A db ~.}

It is understood that there is a non-zero uniform magnetic field $B$ associated with $A$.  Let us consider the equations of motion.  First we note that the physical electric $U_A(1)$ current is
\eqn\physJ{J = - {1 \over 2\pi} db}
We denote the average value of the time component as $\rho$ (the physical electron density).
The equation of motion of $b$ (essentially \ateom\ after accounting for the sign change of the $bdA$ term) gives the average effective magnetic field (usually denoted $B^*$) seen by the composite fermions
\eqn\Bstar{ B^* = B - 4\pi \rho}

Varying with respect to $a_0$, we get the condition
\eqn\agauss{\rho_\chi - {1 \over 4\pi} \epsilon_{ij} \partial_i (a_j  -  2b_j) = 0}
Here $\rho_\chi$ is the average density of composite fermions.  The second term is the contribution from the variation of $\eta[a]$ that is present in our definition of the fermion path integral. Though $\eta[a]$ is not identical to the level-$1/2$ CS term, its variation is identical to the variation of the level-$1/2$ CS term. We thus find that
\eqn\cfrho{\rho_\chi = {1 \over 4\pi} B}

The equation for $B^*$ is identical  to the HLR theory (and is such that $B^*$ vanishes at $\nu = {1 \over 2}$). However  $\rho_\chi$ (the average composite fermion density) is different. Here it is given by the external magnetic field rather than by the physical electron density.
The finite average density of composite fermions means that (if we ignore the dynamics of the gauge fields $a$ and $b$) they will form a Fermi surface. Further under $\CC\CT$ the composite fermions are Kramers doublets.  There will thus be a Berry phase of $\pi$ when the composite fermion goes around the Fermi surface.

These are very much the same features as those postulated by Son \SonXQA; thus, not surprisingly (given that the equations of motion of the refined version \dualtheory\ is the same as the original version in \dualD),  at this level the present theory retains these features.
Including the dynamics of $a$ and $b$ we obtain a description of the particle-hole symmetric composite fermi liquid that, while better defined than the previous version, will still lead to essentially the same predictions in numerics and experiments. For instance the suppression of $2K_f$ backscattering of particle-hole symmetric operators \GeraedtsPVA\ will also obtain in the present theory (assuming that the dynamical gauge fields do not dramatically alter the essential physics, as is reasonable in the presence of the long range Coulomb interaction \HalperinMH). The electrical and thermal Hall conductivities will take exactly the values required by the particle-hole symmetry\foot{The electrical Hall conductivity $\sigma_{xy} = {e^2 \over 2h}$ and the thermal Hall conductivity $\kappa_{xy} = L_0 T \sigma_{xy}$ with $L_0 = {\pi^2 k_B^2 \over 3e^2}$ (the free electron Lorenz number). $e$ is the electron charge and $T$ the temperature.}.

The improvement offered by the present theory is very useful if we wish to correctly obtain the topological field theory of a Jain state\JainTX\ proximate to $\nu = {1 \over 2}$ (just like with the HLR action and its improved parton version discussed in Appendix C).  Indeed if we break particle hole symmetry explicitly then we can add a Dirac mass to \phcfl . In the limit of large mass, it is readily seen that \phcfl\ goes over to the parton version of HLR  while the original proposal by Son goes over to the original version of HLR.

\bigskip
\noindent {\bf Acknowledgments}

We are grateful to S.~Minwalla and S.~Sachdev for helpful discussions.  We thank A.~Karch and D.~Tong for communicating their results to us prior to their publication.
The work of NS was supported in part by DOE grant DE-SC0009988.  The work of TS was supported by NSF DMR-1305741, and  partially supported by a Simons Investigator award from the Simons Foundation. The work of CW was supported by the Harvard Society of Fellows. The work of EW was supported in part by NSF Grant PHY-1314311.
Opinions and conclusions expressed here are those of the authors and do not
necessarily reflect the views of funding agencies.

\appendix{A}{Spin Structure Dependence of Chern-Simons Couplings}

If $M$ is a three-dimensional spin manifold and $b$ is a $U(1)$ gauge field, then there is a level 1 Chern-Simons interaction that can be written
informally as
\eqn\informal{\ICS={1\over 4\pi}\int_M b d b. }
It is well-defined mod $2\pi$.   However, this function really does depend on the spin structure of $M$; in general, if the spin structure is changed, $\ICS$
is shifted by an integer multiple of $\pi$.

Since a spin structure is not visible in eqn.\  \informal, the claim that $\ICS$ depends on the spin structure may come as a slight surprise.  Matters become clearer,
however, if we give a precise definition of $\ICS$.  The informal definition in eqn.\  \informal\ makes sense in a topologically trivial situation, but not if $b$ has Dirac string
singularities.  For a more general definition, let $X$ be an oriented four-manifold with boundary $M$, such that the spin structure of $M$ extends to a spin structure on $X$,
and the $U(1)$ gauge field $b$ can be extended over $X$.  Then set
\eqn\normal{\ICS={1\over 4\pi}\int_X f \wedge f,}
where $f=db $ is the curvature of $b$.  An $X$ with the properties that we have assumed always exists, and the Chern-Simons function $\ICS$ defined as in eqn.\  \normal\
is independent mod $2\pi$ of the choice of $X$ and of the extension over $X$ of the spin structure and gauge field of $M$.  Independence of the choices is proved by first
showing that if $X$ is a four-dimensional spin manifold without boundary and $b$ is a $U(1)$ gauge field on $X$,
then
\eqn\ormal{J={1\over 2}\int_X {f\over 2\pi}\wedge {f\over 2\pi} }
is an integer.  Here it is essential that $X$ is a spin manifold; without this hypothesis, $J$ would in general be a half-integer and $\ICS$ in eqn.\  \normal\ would be independent
of the choices only modulo $\pi$.  Once $J$ is known to be an integer when $X$ has no boundary, the fact that $\ICS$ is independent of the choices mod $2\pi$ follows
in a standard way.\foot{If $X_1$ and $X_2$ are two oriented manifolds of boundary $M$ over which the spin structure and gauge field of $M$ have been
extended, then by gluing them together along their boundary after reversing the orientation of $X_2$, one makes a spin manifold $X$ without boundary.
The difference between $\ICS$ defined using $X_1$ or using $X_2$ is $\pi\int_X f\wedge f/(2\pi)^2$, and this is a multiple of $2\pi$ because of integrality of $J$.}

Now that we have given a general definition of $\ICS$, it is clear why this function may depend on the spin structure of $M$: this spin structure affects
which choices of $X$ are allowed.
For a concrete example, we take $M$ to be a three-torus $T^3$, which we factorize
as $T^2 \times S^1$.  We take a gauge field with one unit of magnetic flux along $T^2$
\eqn\rmal{\int_{T^2}{db\over 2\pi}=1,}
and trivial
in the $S^1$ direction. (Thus $b$ is a pullback from $T^2$ to $T^2\times S^1$.)
 Spin structures on $T^3$ are classified according to whether fermions are periodic or antiperiodic in going around the three
directions in $T^3$; we write $+++$ and $++-$ for spin structures that are periodic in the $T^2$ directions, but periodic or antiperiodic in the $S^1$ direction.

We claim that in this situation, $\ICS$ is equal to 0 for the $+++$ spin structure, and to $\pi$ for the $++-$ spin structure.  Before beginning a technical explanation,
we explain the physical interpretation.  We quantize $U(1)_1$, with action $\ICS$, on $T^2$ and view $S^1$ as the ``time'' direction.  Having put a unit of
flux on $T^2$, to satisfy the Gauss law constraint we also need a charge $-1$ quasiparticle on $T^2$.  On $T^2\times S^1$, we place a single Wilson
line, running in the $S^1$ direction, representing this quasiparticle.   In this situation, the space $\CH$ of physical states is 1-dimensional, or equivalently
$\Tr_{\CH}\,1=1$.  Here in a path integral approach, $\Tr_\CH\,1$ is computed by a path integral with $++-$ spin structure.  To decide  if the one state in $\CH$ is bosonic or fermionic,
we compute $\Tr_\CH \,(-1)^F$, which we do via a path integral with $+++$ spin structure.  The fact that shifting the spin structure from $++-$ to $+++$ shifts
$\ICS$ by $\pi$ means that it changes the sign of the $U(1)_1$ path integral, so that $\Tr\,(-1)^F=-1$ and the one state in $\CH$ is fermionic.

Technically, it is straightforward to compute $\ICS$ in this example for the case of $++-$ spin structure.  We can take $X=T^2 \times D$, where $D$ is a disc
with boundary $S^1$.  The $++-$ spin structure on $T^2\times S^1$ extends over $T^2\times D$, and we can extend the gauge field $b$ over $T^2 \times D$
by simply taking it to be trivial in the $D$ direction (that is, we take $b$ to be a pullback from $T^2$ to $T^2\times D$).  With this choice, $f\wedge f$ vanishes
pointwise on $X$ so the integral in \normal\ trivially vanishes and $\ICS=0$.

By contrast, the computation of $\ICS$ with $+++$ spin structure is not straightforward, because there is no elementary choice of $X$.  It can be shown that any $X$ with
boundary $T^2\times S^1$
over which the $+++$ spin structure extends has signature 8 mod 16, which shows that any obvious $X$ will not do.  For a relatively simple choice of $X$,  let
$X_0$ be a rational elliptic surface.  From a topological point of view, $X_0$ is a four-manifold of signature 8 that has a surjective map $\varphi:X_0\to S^2$ with the generic fiber being a two-torus.  Let $p\in S^2$ be a point such that $\varphi^{-1}(p)$ is a copy of $T^2$.  Then this is also so for every point in a small open ball $U_0$ containing $p$, and
$\varphi^{-1}(U_0)=U_0\times T^2$.  Let $U\subset U_0$ be an even smaller open ball containing $p$, and let   $D$ be a disc obtained by omitting $U$ from  $S^2$. Set $X=\varphi^{-1}(D)$.  The map $\varphi$ when restricted to $S^1=\partial D$
is a trivial fiber bundle with fiber $T^2$ (because $S^1\subset U_0$) so the boundary of $X$ is $\varphi^{-1}(S^1)=T^2\times S^1$.  Moreover, $X$ is simply-connected, with a unique spin structure
up to isomorphism, and this spin structure restricts to the $+++$ spin structure on $X$.  It can be shown by fairly standard topological arguments\foot{Because  $\varphi^{-1}(U_0)$
is a product $U_0\times T^2$, one can pick an extension of the gauge field over $\varphi^{-1}(U_0)$ as a pullback from $T^2$.  After doing this, one can further extend
the gauge field over $X$ and so over all of $X_0$.  With this choice $f\wedge f=0$ in $\varphi^{-1}(U_0)$ and in particular that is true in $X_0\backslash X$ (i.e. the complement of $X$ in $X_0$).  Therefore $\int_{X}f\wedge f/(2\pi)^2$,
which we wish to evaluate, is the same as $\int_{X_0}f\wedge f/(2\pi)^2$.  Since $X_0$ is compact, the latter integral can be interpreted as an intersection number.
One can choose the extension of the gauge field so that $f/2\pi$ is Poincar\'e dual to a section $s$ of $\varphi:X_0\to S^2$ and since (by standard properties of the rational elliptic
surface) $s\cap s=-1$, we get with this choice of extension of the gauge field $\int_X f\wedge f/(2\pi)^2=-1$.  With any choice, the integral would be an odd integer.}
 that the $U(1)$ gauge
field $b$ on $M$ extends over $X$  with $\int_X f\wedge f/(2\pi)^2$ equal to an odd integer.  Using this result in eqn.\  \normal, one finds that with the $+++$ spin structure,
$\ICS=\pi$ mod $2\pi$.

With a little more work, one can show that a similar result holds if one changes the spin structure on $T^2$:  with a $\pm \pm -$ spin structure on $T^2\times S^1$
and the same $b$ as before, $\ICS=0$, but with $\pm \pm +$, one has $\ICS=\pi$.

In contrast to what we have just explained for $U(1)$ gauge fields, for a $\spinc$ connection  there is a level one Chern-Simons coupling that is well-defined
mod $2\pi$ with no choice of spin structure.  It is studied in Appendix B below.

\appendix{B}{An Almost Trivial Theory}

At numerous points in this paper, we encounter an almost trivial theory $U(1)_1$.
In some checks of the consistency of our statements, it is important to know that this theory
is actually equivalent to a purely classical theory in which the dynamics of the $U(1)_1$ gauge
field are replaced by a $c$-number function of the gravitational background.

There are two slightly different versions of this statement, depending on whether we are on a spin manifold
or a $\spinc$ manifold.  The spin case is slightly easier to explain, so we begin there.  The $U(1)_1$ action
is
\eqn\action{ I_1={{1}\over{4\pi}}\int_M b d b. }
It is well-defined mod $2\pi$ on a spin manifold, that is on a three-manifold $M$ with a chosen spin structure.

Like most Chern-Simons theories, this one is not quite a topological field theory, since it has a framing anomaly (which is related to the
central charge of the corresponding conformal field theory in two dimensions).  In a sense,
the theory is completely determined by its framing anomaly.  In a general $U(1)_k$ theory, on a  Riemann surface $\Sigma$ of genus $g$ (and
for any spin structure if $k$ is odd),
the number of physical states is $k^g$. So $U(1)_1$ has precisely one physical state $\Psi$ for any $\Sigma$.  The mapping
class group of $\Sigma$ (its group of topologically nontrivial diffeomorphisms, modulo trivial ones) can only act on $\Psi$ by phases,
that is by complex numbers of modulus 1, and this action is completely determined by the framing anomaly.   But a knowledge of the physical
states and mapping class group action on any $\Sigma$, together with general axioms of topological field theory, completely determine
the path integral on any three-manifold.  So the $U(1)_1$ path integral on any three-manifold is determined by its framing anomaly together with
general axioms.

The most relevant axiom is as follows.   Suppose a manifold $M$ without boundary is obtained by gluing together two manifolds $M_1$ and $M_2$
that have the same boundary $\Sigma$ (and are oriented so that their orientations match after the gluing).  Let $\Psi_\alpha$, $\alpha=1,\dots,n$
be an orthonormal basis of physical states on $\Sigma$ in some topological field theory.   Let $Z_M$ be the path integral on $M$ in this theory.
Similarly let
 $Z_{M_i,\Psi_\alpha}$ be a path integral on $M_i$ with initial or final state $\Psi_\alpha$ inserted on $\Sigma=\partial M_i$.
The basic gluing law of topological field theory says that $Z_M=\sum_\alpha Z_{M_1,\Psi_\alpha}Z_{M_2,\Psi_\alpha}$.  In the  case of $U(1)_1$,
there is only one physical state $\Psi$ on $\Sigma$ and the formula collapses to
\eqn\collapsed{
Z_M=Z_{M_1,\Psi}Z_{M_2,\Psi}.}

To find a purely $c$-number theory that reproduces the $U(1)_1$ path integral, we just need to find a function of the metric of $M$ that has the same
framing anomaly and gluing law as $U(1)_1$.  Such a function can be constructed from the gravitational Chern-Simons function.
This can be defined as follows.  Let $M$ be the boundary of the four-dimensional spin manifold $X$.  Then we define
\eqn\defOm{\Omega_0 = \pi \int_X \hat A(R) , }
where $\hat A(R)$ is a quadratic function\foot{To be precise, $\int_X\hat A(R)={1\over 48(2\pi)^2}\int_X{ \rm{tr}}\,R\wedge R$.} of the Riemann tensor of $X$, known as the $\hat A$ genus.  For a four-dimensional spin manifold $X$
without boundary, $\int_X\hat A(R)$ is an even integer.  This can be used to prove that modulo $2\pi$, $\Omega_0$ depends only on the metric
on $M$ and not on the choice of $X$ or of the metric on $X$.   Accordingly, the function $\exp(i\Omega_0)$ is a well-defined function of the metric
of $M$.  Moreover, it satisfies a gluing law just analogous\foot{This follows formally from the fact that $\exp(i\Omega_0)=\exp(i\int_M\CSg)$,
where $\CSg$ is a gravitational Chern-Simons three-form.  The factorization analogous to \collapsed\ is formally
 $\exp(i\int_M\CSg)=\exp(i\int_{M_1}\CSg)\cdot \exp(i\int_{M_2}\CSg)$.
The subtlety here is that as $\CSg$ picks up a total derivative term under a diffeomorphism, to define the phases of $\exp(i\int_{M_1}\CSg)$ and
of $ \exp(i\int_{M_2}\CSg)$ requires some choices.  This is precisely analogous to the fact that the factorization in \collapsed\ depends
on the choice of phase of the state $\Psi$.} to eqn.\ \collapsed.

The partition function of $U(1)_1$ is, however, not\foot{Actually, $\exp(-i\Omega_0)$ is the partition function of the Ising spin topological
field theory (the three-dimensional theory that is related to the Ising model in two dimensions).  This is explained in section 3.6 of \SeibergRSG.  In that explanation, $\exp(-i\Omega_0)$ is written as $\exp(-i\pi\eta/2)$, where here $\eta$ is the eta-invariant of the Dirac
operator coupled to gravity only.  The two can be related via the
Atiyah-Patodi-Singer index theorem.  More generally, $\exp(-in\Omega_0)$ is the partition function of the spin topological field theory $SO(n)_1$.  This
family of theories is discussed in Appendix C.5 of \SeibergRSG.  If we sum $\exp(-in\Omega_0)$ over spin structures, we get the partition
function of the non-spin theory ${\rm Spin}(n)_1$, also discussed in the same Appendix.  The sign in the exponent of $\exp(-i\Omega)$ depends on
an orientation convention.  For our purposes, we can most easily fix this sign by considering the $\spinc$ case, which is treated below.  With the $\spinc$
action $I'$ as defined later
and the argument of the path integral being $e^{iI'}$, the $A$-dependence of the path integral is $\exp\left(-{i\over 4\pi}\int_MA dA\right)$, so the
partition function of the $\spinc$ theory equals $e^{-i\Omega'}$ and that of the spin theory is $e^{-i\Omega}$.}  $\exp(i\Omega_0)$ but $\exp(-2i\Omega_0)=\exp(-i\Omega)$,
where
\eqn\zolo{\Omega=2\Omega_0.}
The factor of 2 can be verified by comparing framing anomalies.   Another route is as follows.    Let us remember first that
the action $I_1$ defined in eqn.\ \action\ is not gauge-invariant if $M$ has nonempty boundary $\Sigma$.  Gauge-invariance can be restored by
adding boundary degrees of freedom, and there is a simple choice of boundary state of $U(1)_1$ such that a chiral Dirac fermion propagates on $\Sigma$.
Likewise, $\exp(-i\Omega_0)$ is not diffeomorphism-invariant on a manifold with boundary.  But diffeomorphism invariance can be restored by
incorporating on $\Sigma$ a single chiral Majorana-Weyl fermion, whose gravitational anomaly cancels the failure of diffeomorphism invariance
of $\exp(-i\Omega_0)$.  A chiral Dirac fermion  is the same as two chiral Majorana-Weyl fermions, so the partition function of $U(1)_1$ is
actually $\exp(-2i\Omega_0)=\exp(-i\Omega)$.

In addition to reproducing the partition function of $U(1)_1$, one may also wish to reproduce the expectation values of observables.
The only observable of $U(1)_1$ is the line operator $W_\ell=\exp \left(i\oint_\ell b\right)$ that represents the propagation of a trivial neutral fermion around a loop
$\ell\subset M$.   As is typically the case for line operators in 3d topological field theory, the expectation value of $W_\ell$ is not a topological
invariant, but has a framing anomaly, in this case corresponding to spin 1/2. $W_\ell$ has a purely gravitational interpretation because
it is a ``transparent line'' that has trivial braiding with itself and all other line operators. To explain the purely gravitational interpretation of $W_\ell$, we
simply make use of the Riemannian connection of $M$.  Let $\omega$ be the Riemannian connection on the tangent bundle of $M$.  It has structure
group $SO(3)$.  However, after restricting it to $\ell$, we can project this connection to the Lie algebra of the $SO(2)$ subgroup of $SO(3)$ that
rotates the normal plane to $\ell$.  Let us write $\omega'$ for the projected connection.  Even without a spin structure on $M$, we can define
the object $\exp\left(in\oint_\ell \omega'\right)$ for any integer $n$.  It describes propagation around $\ell$ of a particle of spin $n$.   If $M$ is endowed
with a spin structure, then the corresponding parallel transport  $\exp\left(in\oint_\ell \omega'\right)$ is defined for half-integral $n$.   The
operator $W_\ell$ of $U(1)_1$ simply corresponds to $\exp\left({i\over 2}\oint_\ell\omega'\right)$.  Note that $\exp\left({i\over 2}\oint_\ell\omega'\right)$ is
of modulus 1 but is not a topological invariant; its phase changes continuously when one varies the metric of $M$, or the embedding of $\ell$ in $M$.
This dependence reproduces the framing anomaly\foot{This interpretation of the framing anomaly may be unfamiliar, but actually can be seen in an
early calculation \refs{\PolyakovMD,\ShajiIS} of bose-fermi
transmutation via coupling to Chern-Simons gauge fields.
The study of the self-linking integral (eqn.\ (3) of \PolyakovMD) amounts to showing that in $U(1)_1$,
the variation of $\langle W_\ell\rangle$ when $\ell$ is varied is the same as the variation of $\exp\left({i\over 2}\oint \omega'\right)$. The calculation
is done on a flat manifold, but because the considerations involved are local (and on dimensional grounds, the framing anomaly does not
depend on the curvature tensor of $M$), this is not material.} of $W_\ell$.  Apart from the framing anomaly, the main property of $W_\ell$ is that it is multiplied
by $-1$  if one changes the spin structure of $M$ in a way that is nontrivial when restricted to $\ell$.  This property is shared by
$\exp\left({i\over 2}\oint_\ell\omega'\right)$.

Now let us discuss the analog of this on a $\spinc$ manifold.  First of all, on a $\spinc$ manifold, there is no generalization of $\Omega_0$, but there is
a generalization of $\Omega=2\Omega_0$, namely
\eqn\lastone{\Omega'={1\over 4\pi}\int_M AdA +\Omega , }
where $A$ is the $\spinc$ connection. This expression is well-defined mod $2\pi$ (see Appendix A.3 of \SeibergRSG),  so on a $\spinc$ manifold, there is a $c$-number theory with partition function $\exp(-i\Omega')$.
There is also a $\spinc$ version of $U(1)_1$ gauge theory, with action
\eqn\spincaction{I'={1\over 4\pi}\int_M\left(b d b+2 b d A\right) ={1\over 4\pi}\int_M(b+A)d(b+A)-{1\over 4\pi}\int_M A d A. }
Here $b$ is a $U(1)$ gauge field, $A$ is a background $\spinc$ connection, and therefore $b+A$ is also a $\spinc$ connection.

Just like the spin version of $U(1)_1$, this theory has only one quantum state on any two-manifold, so its path integral is completely
determined by the framing anomaly and the dependence on $A$.  The obvious guess is that this path integral is just $\exp(-i\Omega')$.
Since this certainly has the correct framing anomaly, we really only need to verify that the dependence on $A$ is correct.  To show this,
we simply pick a spin structure on $M$, after which we can make a change of variables from $b$ to $b'=b+A$.  The $b'$ path integral
gives the familiar $\exp(-i\Omega)$, and the last term on the right hand side of eqn.\ \spincaction\ gives a factor $\exp\left(-{i\over 4\pi}\int_M AdA\right)$.
These combine together to $\exp(-i\Omega')$.  The subtlety in this derivation is that although $\exp(-i\Omega')$ does not depend on the
choice of spin structure that was made in defining $b'$, when we factor it as $\exp(-i\Omega)\cdot \exp\left(-{i\over 4\pi}\int_MA d A\right)$,
each factor does depend on that choice.

In the $\spinc$ context, the possible line operators that can be defined using only the $\spinc$ connection $A$ and the Riemannian connection
$\omega$  are $\exp\left(i\oint_\ell(qA+j\omega')\right)$, where $q$ is an integer, $j$ is
a half-integer, and $q$ is congruent to $2j$ mod 2.  In particular, the line operator $\exp(i\oint_\ell b)$, which represents a spin $1/2$ particle of charge $-1$,
corresponds to $\exp\left(-i\oint_\ell(A+{1\over 2}\omega')\right)$.

\appendix{C}{Effective Field Theories With Flux Attachment}

Here we briefly describe the effective field theories that arise through flux attachment in describing quantum Hall phenomena of electrons in two space dimensions.  This will serve as an insightful and familiar venue to discuss the kind of issues (and their cures) with quantization of Chern-Simons terms that arise in the less familiar field theory that arises in the fermion-fermion duality.

As is well known from the theory of anyons \WilczekDU, flux attachment is a formal procedure to trade particle statistics in two space dimensions. In the quantum Hall context, the essential idea is to attach some integer number $m$ of fictitious flux quanta to the microscopic electrons (at some Landau level filling $\nu$) to convert them to either bosons \ZhangWY\ (by choosing $m$  odd) or to fermions \refs{\FradkinWY,\HalperinMH}\ (by choosing $m$ even). The charge-flux composites thus obtained are known either as composite bosons or composite fermions depending on their statistics. A useful approximation is obtained at special filling factors $\nu = {1 \over m}$ in a ``flux smearing'' mean field treatment where the attached fictitious flux exactly cancels the flux of the externally imposed magnetic field. Fluctuations about this mean field are described by an effective field theory of composite particles moving in zero net field but coupled to a dynamical $U(1)$ gauge field with a Chern-Simons term.

For concreteness we focus on the composite fermion case \JainTX, and specialize to $m = 2$, though everything below carries over straightforwardly to other integer $m$. The structure of the resulting composite fermion effective field theory takes the form
\eqn\Lcf{L_{HLR} = L[\psi_{CF}, c+ A] + {1 \over 8\pi} cdc~.}
Here $c$ is the internal $U(1)$ gauge field, and $A$ is a background ``probe'' $U(1)$ gauge field.  $L[\psi_{CF}, c+ A]$ is a non-relativistic Lagrangian describing a finite density of  the composite fermions $\psi_{CF}$. For electrons at filling factor $\nu = {1 \over 2}$, this is precisely the famous Halperin-Lee-Read Lagrangian used to described the observed metallic state.

Note however the wrong quantization of the Chern-Simons term: it is $1/2$ the proper allowed value. However we remember that the flux attachment condition is that the physical electron density $\rho$  (= composite fermion density) $\rho_{CF}$ is given by
\eqn\rhoflux{\rho = - {1 \over 4\pi} (\partial_x c_y - \partial_y c_x)}
On a compact space integrating this we get
\eqn\Nflux{N = - {N^\phi_c \over 2}}
where $N$ is the total electron number and $N^\phi_c$ is the total number of flux quanta of the CS gauge field $c$.

Thus $N^\phi_c$ is necessarily even and the right flux quantization for $c$ is $4\pi$ and not $2\pi$.  To emphasize this let us write
$c = 2c'$  with the usual flux quantization for $c'$.  This transforms \Lcf\ to
\eqn\Lcfb{L_{HLR} = L[\psi_{CF}, 2c'+ A] + {2 \over 4\pi} c'dc'}

This seems to be a properly defined quantum field theory but there is now a physical problem. To appreciate  this consider using this Lagrangian away from $\nu = 1/2$. An illustrative extreme is to consider the empty Landau level $\nu = 0$ when there are no electrons (and hence no composite fermions).  The composite fermion field may then be safely integrated out but we are left with a $U(1)_2$ Chern-Simons theory which is a non-trivial TQFT.  Clearly we have a physically incorrect description of the empty vacuum! Thus though \Lcfb\ is mathematically well-defined it is physically incorrect.

A resolution of this problem may be attempted  by noticing that much of the non-triviality of the $U(1)_2$ theory (degenerate ground states on the torus, semion excitations, etc) arise from the existence of fields that carry charge $1$ under $c'$. But in the original Lagrangian \Lcf\ these will correspond to fields with $1/2$ charge under $c$ which are usually understood to not exist in such flux attachment effective theories. So we might suppose that \Lcfb\ with the additional restriction that we disallow fields with charge $1$ under $U_{c'}(1)$
is the sensible physical theory.  But does such a restriction spoil the mathematical consistency of the theory? Clearly a cleaner resolution of the issues with \Lcf\ or \Lcfb\ is called for.

A cleaner formulation of the composite fermion theory is known in the condensed matter literature and is obtained through the following construction (known as slave particle or parton construction). We represent the microscopic electron operator $\psi$ as a product
\eqn\qhprtn{\psi = \chi \phi}
where $\chi$ is a fermion field and $\phi$ is a boson field. This representation comes with an internal $U(1)$ gauge redundancy associated with opposite phase rotations of $\chi$ and $\phi$.  An effective field theory in terms of the $\chi$ and $\phi$ fields will faithfully represent the electron system so long as we include a $U(1)$ gauge field $a$ under which $\chi$ has charge $1$ and $\phi$ has charge $-1$.  We assign  physical $U_A(1)$ charge $1$ to $\phi$ and $0$ to $\chi$. As the bosons carry the physical electric charge their density will match the density of electrons, and they will see the externally imposed magnetic field. This construction leads to an effective field theory
\eqn\prtnbL{L_{prtn} = L[\chi, a]  + L[\phi, - a + A_{tot}]}
Here $A_{tot}$ includes both the vector potential of the background magnetic field, and a probe gauge field $A$.

At filling fraction $\nu = {1 \over 2}$ we can envisage a state of matter where the average magnetic flux of the internal gauge field $a$ is zero. Then the bosons see all of the external magnetic field and hence are themselves at filling ${1 \over 2}$ while the fermions $\chi$ move in an average net zero magnetic field. Then $L[\chi, a]$ is the standard Lagrangian of
non-relativistic fermions at finite density coupled to a $U(1)$ gauge field $a$.  The $\phi$ will themselves form a quantum Hall state of bosons at $\nu = 1/2$. An effective field theory for such a bosonic quantum Hall state is well known, and takes the form
\eqn\Lbhalf{ - {2 \over 4\pi} b db + {1 \over 2\pi} (A - a) db}

Using this in \prtnbL, we obtain the  effective field theory
\eqn\hlrprtn{L_{HLR-prtn} = L[\chi,a] - {2 \over 4\pi} b db + {1 \over 2\pi} (A - a) db}

We note that all Chern-Simons terms are properly quantized. Further if we proceed naively and integrate out $b$ using its equation of motion, we obtain precisely the standard HLR Lagrangian in \Lcf .  However this integration is not strictly valid, and it is more precise to keep $b$ as a dynamical field. Unlike \Lcfb, it is readily seen that \hlrprtn\ gives physically sensible answers. For instance if we take the empty Landau level, then we can again integrate out the $\chi$-fields, and simply be left with \Lbhalf .  The integration over $a$ sets $b = 0$ and we get a trivial theory exactly as expected.

The Lagrangian \hlrprtn\ represents a clean formulation of the flux attachment idea that is both physically correct and mathematically consistent.

It is instructive to understand the relationship between \Lcfb\ and \hlrprtn.  We show below that the two theories differ precisely by a decoupled $U(1)_2$ factor. To expose this we replace \Lcfb\ by
\eqn\Lcfbp{L_{HLR}' = L[\psi_{CF}, 2c'+ A] + {2 \over 4\pi} c'dc' - {1 \over 2\pi} b d(c - 2c')}
where $b$ is a Lagrange multiplier that simply sets $c = 2c'$ (up to a gauge transformation). Now we are free to replace $2c'$ by $c$ in other terms in \Lcfbp. We do this in the first term but not the second and we substitute $c=a-A$ and $c'=\tilde c-b$
to find
\eqn\Lcfbp{L_{HLR}' = L[\psi_{CF}, a] - {2 \over 4\pi} b db +{1 \over 2\pi} b d(A-a) + {2 \over 4\pi} \tilde {c} d\tilde {c} ~.}
In this form we see that the \Lcfbp\  is equivalent to a sum of \hlrprtn\   (after the identification of $\chi$ with $\psi_{CF}$) and a decoupled $U(1)_2$ theory of $\tilde c$.  We can now safely drop the $U(1)_2$ decoupled factor to obtain a different mathematically sensible theory.

\listrefs
\end